\documentclass[11pt]{article}

\usepackage[letterpaper,margin=1in]{geometry}
\usepackage[T1]{fontenc}
\usepackage{amsmath,amssymb,mathtools}
\usepackage{amsthm}
\usepackage{thmtools,thm-restate}
\makeatletter
\@ifundefined{newcounteralias}{}{\renewcommand\thmt@autorefsetup{\@xa\def\csname\thmt@envname autorefname\@xa\endcsname\@xa{\thmt@thmname}}}
\makeatother
\usepackage{mathrsfs,dsfont,pifont,stmaryrd,braket}
\SetSymbolFont{stmry}{bold}{U}{stmry}{m}{n}
\usepackage{enumitem,xcolor,needspace}
\usepackage{booktabs,array,tabularx,multirow,graphicx}
\usepackage{etoolbox,environ,framed,appendix}
\usepackage[framemethod=tikz]{mdframed}
\usepackage{tcolorbox}
\usepackage[labelfont=bf]{caption}
\usepackage{titlesec}
\titlespacing*{\paragraph}{0pt}{1ex plus 0.5ex minus 0.2ex}{1em}

\usepackage[algo2e,linesnumbered,ruled,vlined]{algorithm2e}
\let\PreambleAlgoIf\If
\let\PreambleAlgoElse\Else
\let\PreambleAlgoFor\For
\let\PreambleAlgoForAll\ForAll
\let\PreambleAlgoWhile\While
\let\PreambleAlgoRepeat\Repeat
\let\PreambleAlgoReturn\Return
\forcsvlist{\csundef}{If,Else,For,ForAll,While,Repeat,Return}
\usepackage[noend]{algpseudocode}
\newcommand{\PreambleRestoreAlgorithmTwoE}{\let\If\PreambleAlgoIf
  \let\Else\PreambleAlgoElse
  \let\For\PreambleAlgoFor
  \let\ForAll\PreambleAlgoForAll
  \let\While\PreambleAlgoWhile
  \let\Repeat\PreambleAlgoRepeat
  \let\Return\PreambleAlgoReturn
}
\AtBeginEnvironment{algorithm2e}{\PreambleRestoreAlgorithmTwoE}
\AtBeginEnvironment{algorithm2e*}{\PreambleRestoreAlgorithmTwoE}

\usepackage[color,final]{showkeys}

\colorlet{refkey}{orange!20}
\colorlet{labelkey}{blue!30}
\usepackage[colorlinks=true,linkcolor=blue!45!black,citecolor=blue!45!black,
 urlcolor=blue!45!black,pdftitle={Reed-Solomon Subspace Designs: Optimal List Decoding and Proximity Gaps for Explicit RS codes},pagebackref=true]{hyperref}
\usepackage[nameinlink,capitalize,noabbrev]{cleveref}
\crefformat{equation}{#2(#1)#3}
\Crefformat{equation}{#2(#1)#3}
\crefrangeformat{equation}{#3(#1)#4 to~#5(#2)#6}
\Crefrangeformat{equation}{#3(#1)#4 to~#5(#2)#6}
\crefmultiformat{equation}{#2(#1)#3}{ and~#2(#1)#3}{, #2(#1)#3}{ and~#2(#1)#3}
\Crefmultiformat{equation}{#2(#1)#3}{ and~#2(#1)#3}{, #2(#1)#3}{ and~#2(#1)#3}
\AtBeginEnvironment{appendices}{\crefalias{section}{appendix}}
\Crefname{subsection}{Subsection}{Subsections}
\Crefname{subsubsection}{Subsubsection}{Subsubsections}
\crefname{algocf}{algorithm}{algorithms}
\Crefname{algocf}{Algorithm}{Algorithms}
\newcounter{PreambleAlgorithmic}
\AtBeginEnvironment{algorithmic}{\stepcounter{PreambleAlgorithmic}}
\makeatletter
\def\theHALG@line{\arabic{PreambleAlgorithmic}.\arabic{ALG@line}}
\makeatother

\allowdisplaybreaks[1]
\setlist{topsep=4pt,itemsep=3pt,parsep=0pt}
\numberwithin{equation}{section}

\theoremstyle{plain}
\newtheorem{theorem}{Theorem}[section]

\newtheorem{lemma}[theorem]{Lemma}
\newtheorem{proposition}[theorem]{Proposition}
\newtheorem{corollary}[theorem]{Corollary}
\newtheorem{claim}[theorem]{Claim}

\newtheorem*{theorem*}{Theorem}

\theoremstyle{definition}
\newtheorem{definition}[theorem]{Definition}
\newtheorem{remark}[theorem]{Remark}

\newtheorem*{question*}{Question}
\newtheorem*{definition*}{Definition}
\newtheorem*{notation}{Notation}
\theoremstyle{plain}
\Crefname{fact}{Fact}{Facts}
\crefname{claim}{Claim}{Claims}
\Crefname{claim}{Claim}{Claims}
\newcommand{\Real}{\mathbb{R}}

\newcommand{\Natural}{\mathbb{N}}

\newcommand{\Field}{\mathbb{F}}
\newcommand{\F}{\Field}

\newcommand{\K}{\mathbb{K}}
\renewcommand{\L}{\mathbb{L}}
\newcommand{\E}{\mathbb{E}}
\newcommand{\N}{\Natural}

\newcommand{\R}{\Real}

\newcommand{\mb}[1]{\mathbb{#1}}

\newcommand{\wt}{\widewidetilde}
\newcommand{\eps}{\varepsilon}
\newcommand{\RS}{\operatorname{RS}}

\newcommand{\Span}{\operatorname{span}}
\newcommand{\Gal}{\operatorname{Gal}}
\newcommand{\Tr}{\operatorname{Tr}}
\newcommand{\mult}{\operatorname{mult}}

\newcommand{\poly}{\operatorname{poly}}

\newcommand{\Enc}{\mathsf{Enc}}

\newcommand{\smcdef}{\sum_{j=0}^{\ell}c_j\alpha^j}
\newcommand{\smudef}{\sum_{j=0}^{\ell}u_j\alpha^j}

\DeclareSymbolFont{symbolsC}{U}{pxsyc}{m}{n}
\SetSymbolFont{symbolsC}{bold}{U}{pxsyc}{bx}{n}
\DeclareFontSubstitution{U}{pxsyc}{m}{n}
\DeclareMathSymbol{\medcircle}{\mathbin}{symbolsC}{7}

\let\originalleft\left
\let\originalright\right
\renewcommand{\left}{\mathopen{}\mathclose\bgroup\originalleft}
\renewcommand{\right}{\aftergroup\egroup\originalright}

\title{Explicit Reed-Solomon Subspace Designs: \\
Optimal List and Line Decoding}
\author{Omar Alrabiah\thanks{Department of Electrical Engineering and Computer Sciences, UC Berkeley, Berkeley, CA 94709, USA. Email: \href{mailto:oalrabiah@berkeley.edu}{\texttt{oalrabiah@berkeley.edu}}. Research supported in part by a Saudi Arabian Cultural Mission (SACM) Scholarship,  ONR grant N00014-24-1-2491, and V.~Guruswami's Simons Investigator Award.} \and Rohan Goyal\thanks{Massachusetts Institute of Technology, Cambridge, MA, USA. Email: \href{mailto:rohan_g@mit.edu}{\texttt{rohan\_g@mit.edu}}.  Research supported by Yael Tauman Kalai's NSF Award 6956054.} \and Venkatesan Guruswami\thanks{Department of Electrical Engineering and Computer Sciences, Department of Mathematics, and the Simons Institute for the Theory of Computing, UC Berkeley, Berkeley, CA 94709, USA. Email: \href{mailto:venkatg@berkeley.edu}{\texttt{venkatg@berkeley.edu}}. Research supported in part by a Simons Investigator Award, a UC Noyce initiative award, DOD Advanced
Research Projects Agency grant HR0011262E031, and NSF grant 2211972.}}
\date{}
\newcommand{\cL}{\mathcal L}
\newcommand{\Ready}{\mathop{\mathit{ready}}}
\renewcommand{\wt}{\operatorname{wt}}
\begin{document}

\maketitle
\thispagestyle{empty}

\begin{abstract}
For any desired rate $R \in (0,1)$, and gap to list-decoding capacity $\epsilon > 0$, we give an explicit $n^{O_{\eps}(1)}$ time construction of Reed-Solomon codes over extension fields of size polynomial in the block length $n$ that can be list-decoded from a fraction $1-R-\epsilon$ of errors with the optimal list-size of $O(\tfrac{1-R}{\epsilon})$. We also give a simpler, strongly explicit construction of evaluation points for which the associated Reed-Solomon code can be list-decoded from $1-R-\eps$ errors with list size $(1/\eps)^{O(1/\eps)}$. We provide a deterministic list-decoding algorithm for both constructions running in time $O_{\eps}(1)\poly(n)$.

In fact, our Reed-Solomon codes are near-optimal subspace-design codes over a subfield. Via known results, this directly implies the combinatorial list-decoding guarantee, as well as near-optimal list-recovery bounds. We also present a new reduction showing that the subspace-design guarantee also implies line decoding over the extension field itself, which then yields proximity gaps and mutual correlated agreement up to proximity parameter approaching the optimal $1-R$ radius.

\smallskip Compared to the recent algorithm of Brakensiek, Chen, Putterman, Zhang, and Zheng (2026), which list decodes every Reed–Solomon code over prime fields up to capacity, our approach achieves constant (in fact, near-optimal) list size rather than a polynomially growing one, at the cost of carefully choosing the code's evaluation points. In a companion paper, we show that these respective restrictions are unavoidable.

\end{abstract}

\clearpage
\tableofcontents
\thispagestyle{empty}
\clearpage

\setlength{\parskip}{3pt}
\section{Introduction}
\label{sec:intro}

Reed-Solomon (RS) codes~\cite{ReedS1960} are among the most fundamental and widely deployed error-correcting codes. At their core, they encode messages by evaluating low-degree polynomials over a finite field. Specifically, given a a finite field $\F$ and $n$ distinct evaluation points $(\alpha_1,\dots,\alpha_n)$ in $\F$, the Reed-Solomon code of dimension $k$ is defined as
\[
 \RS_{\F}((\alpha_1, \dots, \alpha_n),k)\coloneqq \bigl\{(f(\alpha_1),\dots,f(\alpha_n)) : f\in \F[X],\ \deg f<k\bigr\}\subseteq \F^n,
\]
and has rate $R=k/n$. Since any two distinct polynomials of degree less than $k$ agree at at most $k-1$ points, the code has minimum distance $n-k+1$, attaining the Singleton bound. Thus, Reed–Solomon codes are canonical examples of maximum distance separable (MDS) codes. Beyond their optimal distance properties, Reed–Solomon codes also admit efficient decoding algorithms. Classical decoders due to Peterson, Berlekamp-Massey and Berlekamp-Welch~\cite{Peterson1960,Berlekamp1968,Massey1969,WelchB1986} can correct any fraction of errors approaching $(1-R)/2$, which is the most that \emph{unique} decoding can achieve at rate $R$.

Beyond half the minimum distance, a decoder can no longer always identify the transmitted codeword, but it can output the \emph{list} of all codewords within the given radius. This notion, known as \emph{list decoding}~\cite{Elias1957,Wozencraft1958}, is useful when the output list remains small. By the Johnson bound~\cite{Johnson1962}, \emph{every} code of relative distance $1-R$ has polynomially bounded list sizes up to $1-\sqrt R$, which exceeds $(1-R)/2$ for every rate $R\in(0,1)$ (the $1-\sqrt{R}$ threshold is therefore referred to as the \emph{Johnson radius}). Just as in the unique-decoding setting, Reed–Solomon codes admit efficient decoding algorithms for list decoding. In particular, the algorithms of Sudan~\cite{Sudan1997} and Guruswami and Sudan~\cite{GuruswamiS1998} showed that Reed-Solomon codes can be efficiently list-decoded all the way up to the Johnson radius.  

\paragraph{List-decoding capacity.}
While this result remained the state of the art for explicit Reed-Solomon codes till very recently, the Johnson radius is far from the ultimate limit of list decoding. For every rate $R\in(0,1)$ and $\eps>0$, there are codes of rate $R$ that are list-decodable from a fraction $1-R-\eps$ of errors with list-size bounded by $O(1/\eps)$, whereas list decoding beyond radius $1-R$ requires exponentially large lists~\cite{ZyablovP1982,Elias1991}. \emph{Variants} of Reed--Solomon codes were shown to achieve list-decoding capacity explicitly and efficiently: building on Parvaresh and Vardy~\cite{ParvareshV2005}, Guruswami and Rudra~\cite{GuruswamiR2008} showed that \emph{folded} Reed--Solomon codes can be efficiently list-decoded from a fraction $1-R-\eps$ of errors, and the same was later shown for univariate \emph{multiplicity codes}~\cite{GuruswamiW2013,Kopparty2015}. While the original list-size guarantee grew as $n^{O(1/\eps)}$, constant list sizes (depending only on $\eps$) were then obtained first for subcodes of these codes~\cite{Guruswami2011,DvirL2012,GuruswamiW2013} and then for the codes themselves~\cite{KoppartyRSW2023,Tamo2024,Srivastava2025}, further improved to $O(1/\eps^2)$ by \cite{Srivastava2025}, and culminating in the breakthrough work of \cite{ChenZ2025} who showed the matching optimal $O(1/\eps)$ bound~\cite{ChenZ2025} for these codes. In fact, the results of~\cite{Srivastava2025, ChenZ2025} held for a broader class of codes called \textit{subspace-design} codes, which are a class of codes typically over a vector-valued alphabet.

None of these codes is a Reed–-Solomon code, however. A folded Reed–Solomon code bundles evaluations at several related points into a single symbol over a larger alphabet, whereas a multiplicity code bundles each evaluation of a polynomial with the evaluations of its derivatives. It is exactly this extra structure that the decoders exploit. Given the centrality of Reed–Solomon codes, this has led to the following prominent question:
\begin{quote}
\emph{Can Reed--Solomon codes themselves, for a suitable choice of evaluation points, achieve list-decoding capacity?}
\end{quote}
\paragraph{Generic Reed--Solomon codes and list-decoding capacity.} To address this question, a large body of work
focused on whether there even \emph{exist} RS codes (even if they are not explicitly constructed) that can achieve list-decoding capacity, or at least beat the Johnson radius, in terms of combinatorially having small lists (even if not accompanied by an efficient list-decoding algorithm). The focus was on ``generic'' Reed--Solomon codes whose evaluation points are sampled randomly from a large enough underlying field. 
After a sequence of works establishing list-decodability of random Reed--Solomon codes beyond the Johnson radius in various regimes~\cite{RudraW2014,ShangguanT2020,GuoLSTW2024,FerberKS2022,GoldbergST2023}, a remarkable work of Brakensiek, Gopi and Makam~\cite{BrakensiekGM2023} showed that generic Reed--Solomon codes achieve list-decoding capacity.
In fact, they showed a more fine-grained statement. Let us say a code is $(\delta,L)$-\emph{list-decodable} if every Hamming ball of relative radius $\delta$ contains at most $L$ codewords. Shangguan and Tamo~\cite{ShangguanT2020} showed that a $(\delta,L)$-list-decodable code of rate $R$ satisfies the \emph{generalized Singleton bound} (GSB)
\begin{equation*}\label{eq:GSB}
 \delta\;\le\;\frac{L}{L+1}\,(1-R),
\end{equation*}
which interpolates between the unique-decoding radius (for $L=1$) and capacity (as $L\to\infty$). \cite{ShangguanT2020} also conjectured that generic Reed--Solomon codes attain the GSB for every fixed $L$, meaning that they can be list-decoded up to radius $\tfrac{L}{L+1}(1-R)$ with list-size $L$ and proved tightness for $L=2,3$.  This conjecture was established in \cite{BrakensiekGM2023}, using the theory of \emph{higher-order MDS codes}~\cite{BrakensiekGM2022,Roth2022}and the GM-MDS theorem~\cite{Lovett2018,YildizH2019}. The field size in the result of \cite{BrakensiekGM2023} grows exponentially in the block length $n$, and this is in fact necessary for codes meeting the GSB exactly~\cite{BrakensiekDG2024}.However, allowing a small gap with respect to the GSB limit, one can get much smaller alphabet sizes: Guo and Zhang~\cite{GuoZ2023} showed that randomly punctured Reed--Solomon codes over fields of quadratic size are $(\frac{L}{L+1}(1-R)-\eps,\,L)$-list-decodable for every fixed $L$ and $\eps>0$, and subsequent work~\cite{AlrabiahGL2024-RandomRS,AlrabiahGGLZ2025} reduced the field size to linear in $n$.

Despite this progress, the existence of good evaluation points is known only through the probabilistic method. A uniformly random set works with high probability, but no deterministic construction was known, nor any efficient way to certify that a given set works. In particular, an explicit construction of Reed--Solomon evaluation points approaching the generalized Singleton bound (GSB) or having constant list-sizes approaching capacity has remained open.\footnote{The earlier mentioned variants of RS codes, like folded Reed-Solomon and univariate multiplicity codes, do approach the GSB, as shown in \cite{ChenZ2025}, by virtue of their having the subspace design property~\cite{GuruswamiK2016}.} As a spoiler, this is the question that the present work resolves. Our codes are constructed over extension fields and have a stronger subspace-design property that yields list-decodability approaching the GSB as a corollary. Further, we also give an efficient algorithm to perform the guaranteed list decoding. 

\paragraph{Subspace-Design Codes.}
Subspace-designs were first introduced by Guruswami and Xing \cite{GuruswamiX2013} as a way to precode Reed-Solomon codes given by evaluating over a subfield in order to make them list-decodable with small lists, and they were first explicitly constructed by Guruswami and Kopparty \cite{GuruswamiK2016} using folded Reed-Solomon and Univariate Multiplicity codes. 

In a recent breakthrough~\cite{ChenZ2025}, it was demonstrated that any code with a near-optimal subspace-design property is optimally list-decodable in the strong sense of approaching the generalized Singleton bound. Subspace-designs have also formed the backbone of many breakthroughs in coding theory recently, such as~\cite{Srivastava2025, AshvinkumarHS2026, GoyalG2025-Proximity, GoyalG2025-Structure, BrakensiekCDZ2025-BrascampLieb, BrakensiekCDZ2025-Explicit, GoyalGSW2026, GoyalGH2026, GuoRSZ2026, CohenDG2026, Liu2026}. They have also found further applications in derandomization, leading to the breakthrough of \cite{ChatterjeeGGRT2026} showing that bipartite matching is in $\mathsf{NC}$.

Informally, \cite{GuruswamiX2013} defined a collection $\{H_1,\dots,H_n\}$ of linear subspaces in $\K^k$ to be a subspace-design if any low-dimensional subspace $A \subseteq \K^k$, the average intersection $\E_{i}\dim(H_i \cap A)$ is much smaller than $\dim(A)$. For a $\K$-additive code $C \subseteq (\K^s)^n$, we say that it is a subspace design code if the collection $\{C_i\}_{i\in [n]}$, where $C_i = \{x\in C \mid x_i=0\}$, forms a subspace design (see \Cref{def:SubspaceDesigns}). We will use the subspace-design-code terminology coined by \cite{GoyalG2025-Proximity}, but an earlier definition was made by Chen and Zhang in \cite{ChenZ2025} who called it a subspace-designable code.

\cite{BrakensiekCDZ2025-Explicit} showed that in fact, subspace-design codes have all \textit{local-coordinatewise-linear (LCL)} properties of random linear codes. LCL properties are a broad class of properties defined by \cite{LeviMS2025} to capture list-decoding and list-recovery, and further shown to capture line and curve-decoding by \cite{GoyalG2025-Proximity, GoyalGSW2026}. Thus, this leads to a natural question in our quest for finding explicit Reed-Solomon codes approaching the generalized Singleton bound: can we construct subspace-design Reed-Solomon codes? 

This question as is seems ill-posed as we require the alphabets to be low-dimensional vectors to get the subspace-design property. We can thus modify the question to ask:

\begin{quote}
\emph{Can explicit Reed-Solomon codes over \textit{extension-fields} have the subspace-design property when viewed as an additive code over a subfield?}
\end{quote}

This is the key question we tackle in this paper and answer it in the affirmative. This is the first construction of a near-optimal subspace-design code that is also linear over its alphabet. The previous near-optimal constructions with $\tau(r)\leq R+\eps$ are due to \cite{GuruswamiK2016} and \cite{GoyalGH2026} who gave the first constant alphabet construction of subspace-design codes building on the ideas of \cite{AlonEL1995,SrivastavaT2025,JeronimoS2026} using the AEL transform.

\paragraph{Decoding every Reed-Solomon code over prime fields.}
Our work comes on the heels of a very recent breakthrough by Brakensiek, Chen, Putterman, Zhang and Zheng~\cite{BrakensiekCPZZ2026}, who gave an efficient algorithm that list-decodes \emph{every} Reed--Solomon code of rate $R$ over a prime field up to capacity. For every prime $p \ge O_\eps(n)$ and every set of $n$ evaluation points in $\F_p$, the associated Reed-Solomon code can be list-decoded from a fraction $1-R-\eps$ of errors in polynomial time, with list-size $n^{O_{\eps}(1)}$. (See also the recent exposition by \cite{HarshaKS2026}.) Thus, over prime fields, \emph{all} Reed-Solomon codes, and not just generic ones, achieve list-decoding capacity! These codes, however, do not approach the GSB and in fact list-size grows with $n$; we give a detailed comparison between our results and \cite{BrakensiekCPZZ2026} after \Cref{thm:intro-ld}.

\paragraph{Proximity gaps.}
In addition to its intrinsic interest in coding theory and pseudorandomness, the question of list decoding Reed-Solomon codes up to capacity has gained recent prominence due to its connection with practical proof systems through the problem of \textit{proximity-gaps}. 

A code $C\subseteq \F^n$ has a proximity gap at proximity parameter $\delta$ if for every affine line $\{u+\alpha v:\alpha\in \F\}$ in $\F^n$, either \emph{all} of its points are $\delta$-close to $C$, or only a negligible fraction are \cite{AmesHIV2017,BenSassonKS2018}. This dichotomy underlies the soundness of interactive oracle proofs \cite{ReingoldRR2021, BenSassonCS2016} and the succinct efficient arguments systems built on them, such as FRI, SNARKs, STARKs and their descendants~\cite{BenSassonBHR2018-Reed-a,BenSassonBHR2019,ArnonCFY2024-STIR-a,ArnonCFY2025-WHIR}: the prover claims that several words are close to a Reed--Solomon code, and the verifier tests a random linear combination. Larger proximity parameter $\delta$ and better error guarantees lead to shorter proofs.

\cite{BenSassonCIKS2023} proved proximity gaps for Reed--Solomon codes up to the Johnson radius and conjectured that they extend up to $1-R$. This challenge is the basis of Ethereum's proximity prize~\cite{Ethereum2025}, and has attracted intense recent attention~\cite{ DiamondG2025,CritesS2025,BenSassonCHKS2025,GoyalG2025-Proximity,BordageCGM2025, Jeronimo2026}.
Our work constructs Reed-Solomon codes which are subspace-designs over a subfield. 

This automatically bestows upon them a ``line decoding'' guarantee that implies proximity gaps (and stronger correlated agreement guarantees) up to proximity parameter approaching $1-R$~\cite{GoyalG2025-Proximity} for affine lines over a subfield of the code. In \cref{sec:ImprovedExtensionFieldConsequences}, we extend \cite{GoyalG2025-Proximity}'s techniques to also apply over extension fields even when the subspace-design guarantee is only over a subfield. This lets us get the first explicit construction of codes over an alphabet $\Sigma$ with proximity gaps with error $\frac{\poly(1/\eps)n}{|\Sigma|}$.

\subsection{Our results}\label{sec:intro-results}

We now state our results more precisely. As mentioned, our list-decodability and proximity gap guarantees both stem from a stronger subspace-design property of our Reed-Solomon construction, which we now describe.

Our codes are constructed over an extension field $\F/\K$, and therefore their symbols can be viewed as vectors over $\K$.  Slightly more formally, we say a $\K$-linear code $C \subseteq (\K^s)^n$ of rate $R$ is a $\tau$-\emph{subspace-design code} for $\tau: \N\mapsto \R$,if for all $\K$-subspaces $A$ of $C$ of dimension at most $r$, if we consider the subspace $A_i$ of codewords in $A$ which are $0$ in the $i$'th coordinate, i.e., $A_i = \{ a  \in A \mid a_i=0\}$, then on average, for a random $i \in [n]$, $\dim A_i \le \tau(r) \dim A$. See \cref{def:SubspaceDesigns} for a formal definition.

A subspace-design code is near optimal up to dimension $r$ if $\tau(r)\leq R+\eps$. That is, the dimension shrinks by a factor of about $R$ (which is best possible) when a random codeword position is set to $0$. When $r=1$, this property is the same as the code's relative distance being at least $1-R-\eps$, and thus the code being near-MDS. The subspace-design guarantee is thus a higher-dimensional generalization of this near-MDS property.\footnote{Since $\dim A_i \ge \dim A - s$, one needs $s$ growing with $r$ in order for the subspace design property to hold up to dimension $r$; in particular for RS codes, one has to work over extension fields.} And just as the MDS property implies optimal unique decoding, due to~\cite{ChenZ2025}, the subspace-design property implies optimal list-decoding approaching the GSB (for list-sizes up to $r$).

\begin{restatable}[Explicit RS codes that are near-optimal subspace-design codes]{theorem}{mainresult}
\label{thm:Main}
For every integer $r\ge2$, every $0<\varepsilon<1$, all sufficiently
large $n$, and every finite field $\K$ with $|\K|\ge n$, one can
deterministically construct in time $|\K|^{O_{r,\varepsilon}(1)}$
an extension $\F/\K$ of degree $O_{r,\varepsilon}(1)$ and evaluation points
$A=(\alpha_1,\ldots,\alpha_n)\in\F^n$ with the following properties.
For every $1\le k\le n$, the Reed--Solomon code
$C=\RS_\F(A,k)$, of rate $R=k/n$, is a $\tau$-subspace-design code over every subfield
$\K_0\subseteq\K$, where
\[
  \tau(d)=\min\{1,R+\varepsilon/4\}\quad(1\le d\le r),
  \qquad \tau(d)=1\quad(d>r).
\]
Moreover, for every integer $L\ge1$, $C$ is
$(\delta_L,L)$-average-radius list-decodable, where
\[
  \delta_L=\frac{L}{L+1}(1-R-\varepsilon).
\]
Given any received word, the lists within relative distances
$\delta_L$ and $1-R-\varepsilon$ can be found deterministically in
time $O_{r,\varepsilon}(1)\poly(n,\log|\F|)$, with sizes at most
$L$ and $O((1-R)/\eps)$ respectively.
\end{restatable}

We in fact show that the subspace-design property holds over $\K$ and prove that the subspace-design property transfers to subfields generically. This is closely related to a question raised by \cite{GoyalGH2026} in their paper about the behavior of local properties under field extensions.

We next state our main result for list-decoding more explicitly. The combinatorial guarantee of approaching the GSB follows, via~\cite{ChenZ2025}, from the subspace-design property above; we additionally develop an efficient list-decoding algorithm.

\begin{theorem}[Explicit RS codes approaching GSB and efficiently list-decodable up to capacity (informal)]
\label{thm:intro-ld}
Fix $\eps\in (0,1)$. For all sufficiently large $n$, and an arbitrary field $\K$ of size at least $n$, one can deterministically construct in time $|\K|^{O_{\eps}(1)}$, an extension field $\F$ over $\K$ with $[\F:\K]=O_{\eps}(1)$ and evaluation points $A=(\alpha_1, \dots, \alpha_n) \in \F^n$, such that for every $1\le k \le n$ the Reed-Solomon code $\RS_{\F}(A,k)$ of rate $R = k/n$ has the following properties:

\begin{enumerate}[label=(\alph*)]
\item \textup{(Generalized Singleton bound.)} For every list-size $L$, it is $(\tfrac{L}{L+1}(1-R-\eps),L)$-list-decodable.

\item  \textup{(Efficient list decoding up to capacity.)} There is a deterministic algorithm that, given $y\in \F^n$, outputs all Reed-Solomon codewords within relative distance $1-R-\eps$ of $y$ in time $O_{\eps}(1)\poly(n, \log |\K|)$.\end{enumerate}
\end{theorem}

Additionally, we give a weaker construction achieving $O_{\eps}(1)$ list-size instead of the near-optimal $O(1/\eps)$ that is simpler and strongly explicit.

\begin{restatable}[Strongly explicit list-decodable Reed-Solomon codes]{theorem}{stronglyexplicit}
\label{cor:StronglyExplicitListDecoding}
Fix $\eps\in (0,1)$. For all sufficiently large $n$, and an arbitrary field $\K$ of size at least $n$, there is a strongly explicit construction of an extension field $\F$ over $\K$ with $[\F:\K]=O_{\eps}(1)$ and evaluation points $A=(\alpha_1, \dots, \alpha_n) \in \F^n$, such that for every $1\le k \le n$ the Reed-Solomon code $\RS_{\F}(A,k)$ of rate $R = k/n$ is
\[
  \bigl(1-R-\varepsilon,\,(1/\varepsilon)^{O(1/\varepsilon)}\bigr)
\]
list-decodable, and the list can be found deterministically in time $O_{\eps}(1)\poly(n,\log|\F|)$. Given an index $i\in[n]$, $\alpha_i$ can be computed in time $\poly(\log|\F|)$.
\end{restatable}

In fact, the algorithm used for both constructions is the same. A simpler version of the same idea also lets us get a more efficient deterministic list-decoding algorithm for folded Reed-Solomon codes that was previously known. We present this algorithm and the corresponding result in \Cref{thm:rs-subspace-decoding} and \cref{cor:frs}, using it as a warm-up to our main list-decoding algorithm. 

\smallskip
\paragraph{Comparison with \cite{BrakensiekCPZZ2026}.} 
Both \Cref{thm:intro-ld} and~\cite{BrakensiekCPZZ2026} give efficient list decoding of Reed--Solomon codes up to capacity, but the results apply under incomparable settings.
\begin{itemize}
\item \textbf{Underlying field.} The algorithm of \cite{BrakensiekCPZZ2026} applies to prime fields, whereas our codes are constructed over extension fields. The base field $\K$ that is extended can be arbitrary, and in particular we can construct RS codes over fields of characteristic $2$, perhaps the most traditional use case of RS codes in practice. In terms of field size,  we need fields of size $n^{O_\eps(1)}$, whereas the result of~\cite{BrakensiekCPZZ2026} applies to fields of size $O_\eps(n)$.\footnote{This can be improved to any prime at least $n$, see \cite{Jeronimo2026,HarshaKS2026}.}

\item \textbf{List size.} We achieve constant, in fact optimal, list size that approaches the generalized Singleton bound. In contrast, the list size in \cite{BrakensiekCPZZ2026} is a polynomial in $n$ with exponent growing doubly exponentially in $1/\eps$.\footnote{If the goal is just to decode past the Johnson radius, the list-size can be improved to $O(n)$~\cite{Gabizon2026, DaoKT2026}.} 

\item \textbf{Evaluation points.} The result of~\cite{BrakensiekCPZZ2026} works for \emph{arbitrary} evaluation points over a prime field. In contrast, our evaluation points are carefully chosen.
\end{itemize}

We now elaborate on the respective restrictions within which our result and that of \cite{BrakensiekCPZZ2026} operate, and why they are unavoidable.

\smallskip
\paragraph{Necessity to choose field or evaluation points carefully.}
Our list decoding result (\Cref{thm:intro-ld}) above \emph{cannot} be established for arbitrary evaluation points. 
Likewise, the restriction to prime (or large characteristic) fields is necessary for the result in \cite{BrakensiekCPZZ2026} which applies to arbitrary evaluation points.
Known limitations of list decodability of RS codes over fields of small characteristic show that the list-size can grow linearly in the block length at the Johnson radius~\cite{JustesenH2001}.In fact, just past the Johnson radius, the list-size can grow faster than any fixed polynomial, for small enough (but constant) rates~\cite{BenSassonKR2010}.
See ~\cite{GuruswamiR2006,ChengW2007} for related limitations.

In an upcoming companion paper, we in fact show that the list-size can grow super-polynomially just beyond the Johnson radius~\cite{AlrabiahGG2026}. (This was also concurrently shown in \cite{Zhu2026, DaoKTZ2026}.)
So without restrictions on the field (as in~\cite{BrakensiekCPZZ2026}) or the evaluation points (as in \Cref{thm:intro-ld}), the Johnson radius is the limit of list decoding Reed-Solomon codes with polynomial-sized lists.

\paragraph{Necessity of growing lists for the prime field result of \cite{BrakensiekCPZZ2026}.}
In the same paper~\cite{AlrabiahGG2026}, we also show that over prime fields, list decoding up to capacity is in general \emph{not} possible with constant-size lists: in particular for any $\eps>0$, there are certain full-length Reed--Solomon codes of constant rate $R=R(\eps)$ over prime fields, for which some Hamming ball of relative radius bounded away from capacity ($1-R-\Omega(\eps^2)$) contains $n^{1/\eps}$ codewords.\footnote{This also refutes a conjecture in the S-two white paper~\cite{CarmonGHLLPS2026} which said that every prime-field RS code can be list decoding up to capacity with list-size independent of the block length.} Consequently, a result like~\cite{BrakensiekCPZZ2026}, which is oblivious to the evaluation points, cannot establish a list-size bound independent of the block length.

\smallskip

\paragraph{Implication for proximity gaps/correlated agreement.} Finally, we state our result for proximity gaps of Reed-Solomon codes. Towards this, we simply leverage the subspace-design property proved for our Reed-Solomon codes together with the line decodability (and more generally, low-degree curve decodability) of subspace-design codes established in  \cite{GoyalG2025-Proximity}, and its downstream implication of proximity gaps and mutual correlated agreement up to capacity. 

Informally, $(\ell,\delta,a,b)$-curve decoding provides the guarantee that amongst any $a$ codewords that are $\delta$-close to points on a degree $\ell$ curve, there are $b$ that are themselves on a degree $\ell$ curve.  As is, this approach will give a result for lines parameterized by the field $\K$ and not the extension field $\F$. We present these weaker consequences from \cite{GoyalG2025-Proximity} in \cref{sec:ProofMain}.

However, we also show that one can lift the guarantee to one for lines parameterized by $\F$ with a modest loss in parameters. Here we state one variant for line decoding; see \Cref{sec:ImprovedExtensionFieldConsequences} for more details and precise implications for proximity gaps and (mutual) correlated agreement (the formal definitions of these notions appear in~ \Cref{subsec:SubspaceDesignCodes}). As far as we are aware, these are the first explicit constructions of codes with proximity gaps with soundness $\frac{\poly(1/\eps)}{\gamma |\Sigma|}$ at distance $\delta = 1-R-\eps$ and proximity loss $\gamma$. In particular, previous subspace-design constructions are not linear over the alphabet, and thus have soundness worse than $\frac{\poly(1/\eps)}{|\K|}$ but the alphabet is $\K^s$ for a large constant $s$, typically at least $1/\eps^2$.

\begin{theorem}[Explicit RS codes line-decodable up to capacity]
\label{thm:intro-capacity-line-decoding}
Fix $0<R<1$ and $0<\varepsilon<1$.
For all sufficiently large $n$ and every finite field $\K$ with
$|\K|\ge n$, there is an explicit construction of an extension field $\F$ with $[\F:\K]=O_{\eps}(1)$ along with explicit evaluation points $\alpha_1, \dots, \alpha_n \in \F$ in time $|\K|^{O_{\eps}(1)}$ such that for all $t\ge 1$, $\RS((\alpha_1, \dots, \alpha_n), Rn)$ is
\[
  \left(1-R-\eps,O\!\left(\frac{t}{\varepsilon}
    +\varepsilon^{-3}\right),t\right)
  \text{-\(\F\)-line-decodable}.
\]
\end{theorem}

\begin{corollary}[Improved proximity gaps and correlated agreement]
\label{cor:intro-extension-field-agreement}
Let $\F$, $\alpha_1, \dots, \alpha_n$ and $\varepsilon$ be as in
\Cref{thm:intro-capacity-line-decoding}.
Then, over $\F$, the code $\RS((\alpha_1, \dots, \alpha_n), Rn)$ has:
\begin{enumerate}[label=(\roman*),leftmargin=2em]
  \item For every real $m>1$,
    \[
      \left(1,1-R-\eps,O\!\left(
        \frac{m/\varepsilon+\varepsilon^{-3}}{|\F|}\right),\frac1m\right)
    \]
    correlated agreement and proximity gap, 
  \item and
    \[
      \left(1,1-R-\eps,O\!\left(
        \frac{n/\varepsilon+\varepsilon^{-3}}{|\F|}\right)\right)
    \]
    mutual correlated agreement.
\end{enumerate}
\end{corollary}

\subsection{Technical overview}\label{sec:intro-overview}

\paragraph{Reed-Solomon codes over extension fields that are subspace designs.} 
Interestingly, our starting point is the subfield-evaluation construction from the work of Guruswami and Xing~\cite{GuruswamiX2013, GuruswamiX2022} which also introduced subspace-designs.

In their proof that folded Reed-Solomon codes yield subspace-designs, the central observation that Guruswami and Kopparty~\cite{GuruswamiK2016} used is that when polynomials $f_1(X), \dots, f_r(X)$ are linearly independent, the Wronskian matrix given by $W(f_1, \dots, f_r)$ given by $\{f_i(X\gamma^{j-1})\}_{i, j \in [r]}$ is full rank where $\gamma$ is any fixed element of sufficiently high order. They then used failure of the subspace-design property for folded Reed-Solomon codes to show that $\det W(f_1, \dots, f_r)$ must have too many roots for a non-zero polynomial.

The natural analog of the above Wronskian for extension field Reed-Solomon codes is the Moore/Galois determinant. In particular, if $\K\subseteq \F$ is a subfield and $f_1, \dots, f_r \in \F[X]$ are linearly independent over $\K(X)$, then the matrix $G(f_1, \dots, f_r)=\{\sigma(f_i(X))\}_{i \in [r], \sigma\in \Gal(\F/\K)}$ is full rank. Here,  $\Gal(\F/\K)$ is the group of all automorphisms on $\F$ fixing $\K$, and we let $\sigma$ act on $\F[X]$ coefficient-wise fixing the variable $X$.

The first issue with direct root counting methods is that if we select a full rank submatrix $M$ of $G$ corresponding to automorphisms $\omega_1, \dots,\omega_r$, then it is possible that $\omega_i(\alpha)\ne \alpha$ for an evaluation point $\alpha$. Thus $\alpha$ may not correspond to a root of $\det M$ at all. 

In \cite{GuruswamiX2013}, the evaluation points were all chosen to belong to the subfield $\K$ itself. This runs into the secondary issue that the independence assumption for the Galois matrix to be full rank is over $\K[X]$. Thus, the space could be rank $1$ while being all of $\K[X]$. In particular, each coordinate $\alpha_i$ is now not a root directly of $\det M$ but of the determinant of $MT_i$ where $T_i$ is a low-rank matrix with coefficients in $\K[X]$, which introduces the possibility that $\alpha_i$ is a root of $\det T_i$. Thus, we could face the issue that each $\alpha_i$ is simply a root of the determinant of the $T_i$ chosen for that coordinate.

In order to circumvent the second issue, it would be convenient to take all $\alpha_i$ to be of sufficiently high extension degree so that $T_i$ cannot introduce new zeroes. This would work, but now we are back to the possibility that $\sigma(\alpha)\ne \alpha$. 

We get around both issues by allowing $\alpha_i$ to be of high extension degree but picking them so that there is always a choice of $\mathcal B = \{\omega_1, \dots, \omega_r\} \subseteq \Gal(\F/\K)$ such that $M$ given by picking rows corresponding to $\mathcal B$ is full rank, and moreover $|\{i\mid \omega(\alpha_i)\ne \alpha_i\}|<\eps n$ for any $\omega \in \mathcal B$.  
A combination of these two ideas immediately results in a natural construction with $[\F:\K]=n^{O_{r, \eps}(1)}$. This is, however, unsatisfactory since the resulting field size is then exponential. 

Thus, we observe that the determinants $\det T_i$ can be further restricted to a small class of polynomials given by the span of minors of a low-dimensional space, if we further assume that we are interested only in low-dimensional $\K$-spaces and not low-dimensional $\K(X)$ spaces. For a formalization of this idea, see \cref{def:EffectivePoints} for \textit{effective evaluation points} and the corresponding proof of the subspace-design property. We note that the final proof, once we capture both of these ideas in our definition of effective evaluation points, is rather clean and short. We then present a simple construction of such points in \cref{sec:ExplicitInstantiation}.

We additionally give a simpler construction of evaluation points, that does not have the full property of controlling how many roots they contribute to $\det T_i$, but with the points still of sufficiently high degree, and show that this suffices for list-decoding with size $O_{\eps}(1)$. The advantage of this construction is that it is strongly-explicit, i.e., every evaluation point can be output in time $\poly(\log |\F|)$.

The broad construction idea works as follows. We pick pairwise non-conjugate $\alpha_1, \dots, \alpha_n$ as coming from extension fields of $\K$ in $g =  O((\log r)/\eps)$ different directions, i.e., we pick co-prime integers $d_1, \dots, d_g$ and pick roughly $n/g$ evaluation points from each of $\K^{d_1}, \dots, \K^{d_g}$. This lets us naturally pick the automorphisms as $\sigma_j:x \mapsto x^{|\K|^{\left(\prod_{i=1}^g d_i\right)/d_j}}$. They generate the Galois group since the $d_i$ are co-prime, and there is a choice of $\mathcal B$ as before such that each $\omega \in \mathcal B$ is supported on at most $\log r$ of these $\sigma_i$, giving us the required guarantee that $|\{i\mid \omega(\alpha_i)\neq \alpha_i\}|\leq n(\log r)/g$ (see \cref{lem:GaloisRowsAndMovement}).

For the strongly explicit construction, it suffices to pick $d_i$ to be distinct sufficiently large primes co-prime to $|\K|$. For the full subspace-design construction, the construction begins with the same idea, but we need to ensure that the determinants $\det T_i$ belong to a small space, and they do not automatically vanish for our choice of $\alpha_i$. This requires a brute force search in each coordinate for a specific evaluation point that is not a root of any bounded-degree sparse polynomial, which takes time $|\K|^{O_{r, \eps}(1)}$

\paragraph{List-decoding Algorithm,}

List decoding algorithms for subspace-design codes have proceeded along the following general blueprint:

Given a $\mathbb K$-subspace design code $C \subseteq \left(\K^s\right)^n$ and a received word $y\in \left(\K^s\right)^n$, do the following two steps:

\begin{itemize}
    \item Find a low-dimensional subspace $A\subseteq C$ such that all close-by codewords to $y$ in $C$ also lie in $A$.
    \item Find all close-by codewords in the low-dimensional space. 
\end{itemize}

The first step is usually a code-specific procedure, and we do not know how to implement it generically for all subspace-design codes. In particular, \cite{GuruswamiW2013} showed that the first step can indeed be implemented for the folded Reed-Solomon and univariate Multiplicity codes. This argument can be viewed as an algorithmization of the Wronskian proof used by \cite{GuruswamiK2016} to show that in fact folded Reed-Solomon codes and Univariate Multiplicity codes form subspace-designs.

The second step is known to work generically for all linear codes. In particular, \cite{KoppartyRSW2023,Tamo2024} showed that given any $d$-dimensional affine space $A\subset C$ with distance $\delta$, and $y \in \Sigma^n$, we have that $\left|\{a\in A\mid \Delta(a,y)<(\delta-\eps)\}\right|\leq \eps^{-d}$. They also provided a randomized \textit{pruning} procedure to implement this step.

In a recent work, \cite{AshvinkumarHS2026} showed how to improve the randomized algorithm if $C$ in fact forms a near optimal-subspace design code, resulting in lists of size $O(\tfrac{d}{\eps} + 1)$. They also showed that for general codes, they can implement \cite{KoppartyRSW2023, Tamo2024}'s procedure deterministically but in time $(1/\eps)^{\exp(d)}\cdot \poly(n)$.

For us, we follow a variant of this blueprint. In particular, our codes are linear over an extension field $\mathbb F$ and thus $\mathbb K^s = \mathbb F$. Even though we know the combinatorial fact that there is a low-dimensional $\K$ space containing all close by codewords, we do not know a direct algorithmic way of finding such a space. This is primarily since our proof of subspace-design first proved a variant of the subspace-design property over spaces of bounded $\mathbb K(X)$-dimension, and then applied the \textit{transfer} result \Cref{thm:SubspaceDesignTransfer}. There is no natural algorithmization of the latter, thus we focus on the first step and in fact do recover a bounded $\K(X)$ dimension-space.

Now, the goal has shifted from pruning a bounded $\K$-dimension space to pruning a bounded $\K(X)$ dimension space. This cannot be done generically for all choices of evaluation points, as all of $\mathbb K[X]$ has dimension $1$. Even with stronger assumptions, we do not see a natural way of adapting \cite{KoppartyRSW2023, Tamo2024, AshvinkumarHS2026} to this setting. We therefore present a variant. 

We first revisit the procedure in the aforementioned papers. The idea is to prune by randomly sampling a coordinate $i\in [n]$ and restricting $A$ to a subspace $A_i$, and repeating until the resulting space has a unique element. This process is then shown to find any given codeword with constant probability. We view this procedure broadly as \textit{repeatedly imposing linear constraints on $A$ until exactly one codeword remains}. This process can only find a single codeword at a time, so we can perform a branching process where we sample a few coordinates  at each step.

We implement a version of this algorithm via the following process:
\begin{quote}
Repeatedly impose $\textit{low-degree}$ constraints on $A$ until exactly one codeword remains.    
\end{quote}

As a warmup we implement this procedure to improve the deterministic list-decoding of low-dimensional subspaces of Reed-Solomon codes. Specifically, this yields a more efficient (in the sense of improved dependence on $\eps$, as \cite{GoyalHKS2024, AshvinkumarHS2026} both give near-linear time decoders in terms of $n$) deterministic list-decoder of folded Reed-Solomon codes than was known via \cite{AshvinkumarHS2026}, while almost matching the best list-size known for this task due to \cite{KoppartyRSW2023,Tamo2024}. 

Moreover, we show that the above can be adapted to bounded $\mathbb K(X)$-dimensional spaces as long as all evaluation points $\alpha_i$ are pairwise non-conjugate and have comparable and sufficiently large (but still bounded by $O_{d, \eps}(1)$) extension-degrees.

\paragraph{Line decoding and proximity gaps.}
As discussed, our codes are linear over an extension field $\F$ but the subspace-design property is only when viewed as an additive code over a subfield $\K$. The results of~\cite{GoyalG2025-Proximity} thus imply line-decoding and proximity-gaps results, but only over $\K$-affine lines, and a naive lift to $\F$ by partitioning into $\K$ lines is rather lossy. We nevertheless present this black-box argument in \cref{sec:ProofMain}.

In \cref{sec:ImprovedExtensionFieldConsequences}, we show that the methods of~\cite{GoyalG2025-Proximity}, which were inspired by the pruning step we described for list-decoding as in \cite{AshvinkumarHS2026}, are nevertheless useful. In particular, the sampling-based pruning method, that proceeds via sampling one coordinate at a time and pruning the space to agree with the received word on that coordinate till we are left with a zero-dimensional space,
is also meaningful when we stop the pruning at an earlier point. In fact, it can achieve a better success probability than what can be proven when we keep pruning all the way down to a $0$-dimensional space.

In particular, for near-optimal subspace-design codes, the works \cite{AshvinkumarHS2026, GoyalG2025-Proximity} showed that starting from an $d$-dimensional linear space $A$, there is an algorithm that samples a set $S\subseteq [n]$ such that $\{a\in A \mid a_{S}=0\}=\{0\}$, and for any $T \subseteq [n]$ with $|T|\ge (R+\eps)n$ has the property that $S\subseteq T$ with probability at least $\frac{\eps}{d+\eps}$. \cite{GoyalG2025-Proximity} then uses various sets $T$ as the agreement positions of different codewords, i.e., $T_{\alpha}=\{i\mid c_{\alpha, i}=u_i+\alpha v_i\}$, to conclude  that many must have agreed on $S$. This implies that for many $\alpha$, there is a common set $S\subseteq T_\alpha$ and $\{a\in A \mid a_{S}=0\}=\{0\}$. This implies that all such $c_{\alpha}$ must themselves have been collinear.

We use the same pruning method but show the following slightly more general conclusion: for any $D$-dimensional space $A$ and integer $d$, there is an algorithm that samples a set $S\subseteq [n]$ such that $\dim \{a\in A \mid a_{S}=0\}< d$ and for any $T$ as above, $\Pr[S\subseteq T]\ge \frac{\eps}{D/d + \eps}$. 

Now we need to choose what space $A$ to apply this algorithm on, and what choice of $d$ to use. To mimic \cite{GoyalG2025-Proximity} where $D=O(1/\eps)$ and they stop at $d=1$, we use $d= \Omega(\eps D)$.

We show that if $C$ is an $\F$-linear code which is a near-optimal subspace-design over a subfield $\K$, then given any $c_{\alpha_1}, \dots, c_{\alpha_a}$ such that $\Delta(c_{\alpha_i}, u+\alpha_iv)<1-R-\eps$ for all $i\in [a]$,  we can find a $\K$-linear space $V\subseteq \F$ of $O_{\eps, a}(1)$ dimension such that if \[V_1 = \mathsf{span}_{\K}\{\alpha_iV\}_{i \in [a]}\qquad V_2 = \mathsf{span}_{\K}\{\alpha_i\alpha_jV\}_{i,j \in [a]}.\] 
Then we have $\dim V_2 \leq 2 \dim V$ (this is for our special choice of V).

Using a dimension counting argument, this lets us upper bound the dimension of $W = \mathsf{span}_{\K}\{c_{\alpha_i}V_1\}_{\alpha_i}$ by $O((\dim V)/\eps)$. This $W$ plays the role of the space $A$ described above. 

 Now, we use $D = \dim W$ and $d = \dim V$ and sample the set $S$ as before. We then  observe that if $\alpha_1, \alpha_2, \alpha_3$ are such that $S\subseteq T_{\alpha_1}, T_{\alpha_2}, T_{\alpha_3}$, then for $z = (\alpha_3 - \alpha_2)c_{\alpha_1}+(\alpha_1-\alpha_3)c_{\alpha_2}+(\alpha_2-\alpha_1)c_{\alpha_3}$, we must have $z_{\mid S} = 0$. If $z\neq 0$, then $Vz \subseteq W$ but also $Vz_{\mid S}=0$. In particular, $d>\dim Vz = \dim V = d$ which is a contradiction. By the sampling bound, some $S$ is contained in $T_\alpha$ for $\Omega(\eps^2a)$ of the assigned codewords, which are then guaranteed to  lie on a common codeword line.

\subsection{Organization} 
We begin with the necessary definitions in \Cref{sec:Preliminaries}. \Cref{sec:ProofMain} proves \Cref{thm:Main} and gives its combinatorial consequences. The subspace-design property (\Cref{thm:SubspaceDesign}) is established in \Cref{sec:SubspaceDesigns}. We then turn to the list-decoding algorithm which is presented in \Cref{sec:EfficientListDecoding}, with the crucial pruning step established in \Cref{sec:ExtractingList}. \Cref{sec:ExplicitInstantiation} constructs evaluation points satisfying both sets of hypotheses. Finally, the improved line-decoding and proximity-gap results are established in \Cref{sec:ImprovedExtensionFieldConsequences}. The algorithmic list-decoding result uses a root-finding step that follows from standard reductions combined with the recent \cite{OpenAI2026} deterministic polynomial factorization algorithm, and this reduction is presented in Appendix~\ref{app:computing-common-zero}.

\label{sec:Introduction}

\subsection{AI Disclosure}
Our key results, and the paper writing were substantially aided by the use of ChatGPT 5.6 Sol and ChatGPT 6 Astra.

The question of constructing explicit list-decodable Reed-Solomon near the generalized Singleton bound via extension field subspace-designs has been a central goal for the authors for the past year. Consequentially, we queried ChatGPT 5.6 Sol to construct explicit near generalized Singleton-bound codes, and it found one of the central ideas in the paper of replacing Guruswami-Xing's subfield evaluation and frobenius automorphism method with a small family of automorphisms generating the Galois group and fixing all but $\eps$-fraction of evaluation points. The initial construction by GPT only worked for list-decoding and had $O(nr)$ extension degree for each evaluation points leading to a construction growing as $\exp(n^{O(\log(1/\eps)/\eps)})$.

GPT 5.6 Sol over many rounds of prompting was able to extend this to an algorithm for list-decoding as well. The authors on understanding the ideas knew that it should extend to subspace-designs due to the known connection with Galois/Moore determinants and this was then achieved over many rounds of discussion and prompting with GPT 5.6. 

The authors then worked extensively with ChatGPT 6 Astra and simplified, reworked, and developed the techniques to further improve the results, and lead to clearer exposition. Definitions such as effectiveness and generalizations such as to deterministic subspace-decoders for standard Reed-Solomon codes and faster deterministic decoders of folded Reed-Solomon codes were then discovered by the human authors once the algorithmic aspects of GPT's construction were cleared up, and this was then further used to simplify our main algorithm and results. Similar processes happened for the constructions of our choice of evaluation points.

The improvements for line-decoding and proximity-gaps in section 8 were found completely by GPT 6 Astra on being asked to extend \cite{GoyalG2025-Proximity}'s results when only subspace-design guarantees over a subfield are provided in a way independent of the extension-field's extension order.

AI tools also assisted with the exposition, under detailed guidance of the authors. The authors assume responsibility for the correctness of the proofs and the quality of exposition.

\section{Preliminaries}\label{sec:Preliminaries}
We write $[n]=\{1,\ldots,n\}$ and $\N=\{1,2,\ldots\}$.
All logarithms are to base two unless a base is displayed.
For a finite field $\K$, a \emph{$\K$-additive code}
is a $\K$-linear subspace $C\subseteq\Sigma^n$ with alphabet
$\Sigma=\K^s$ for some $s\ge1$. In this paper, we typically use
$\Sigma=\F$, where $\F$ is a finite extension of $\K$.
The rate of $C$ is $\log_{|\Sigma|}|C|/n$, and the relative Hamming
distance between $x,y\in\Sigma^n$ is $\Delta(x,y)=\frac{|\{i\in[n]\mid x_i\ne y_i\}|}{n}$.

For distinct $\alpha_1,\ldots,\alpha_n\in\F$ and $1\le k\le n$,
the Reed--Solomon code is
\[
 \RS_\F((\alpha_1,\ldots,\alpha_n),k)
 =\{(f(\alpha_1),\ldots,f(\alpha_n)):f\in\F[X]_{<k}\}.
\]
It has dimension $k$ over $\F$, rate $k/n$, and distance $1-(k-1)/n$.

\subsection{Subspace-design codes and decoding consequences}
\label{subsec:SubspaceDesignCodes}
\label{subsec:PrelimList}\label{subsec:PrelimLines}
Guruswami and Xing~\cite{GuruswamiX2013} introduced subspace designs
to restrict the message spaces of certain Reed-Solomon and algebraic-geometric codes and obtain small lists in list decoding. Informally, a subspace design is a family of subspaces such that every low-dimensional space meets only a few members nontrivially. Their constructions were random; Guruswami and Kopparty~\cite{GuruswamiK2016} gave explicit constructions from folded Reed--Solomon codes. Constant-alphabet subspace-design codes were constructed in~\cite{GoyalGH2026} using the Alon--Edmonds--Luby transformation~\cite{AlonEL1995}.

We follow the terminology for subspace-design codes in~\cite{GoyalG2025-Proximity}. A similar definition first appears in~\cite{ChenZ2025} under the name \emph{subspace-designable codes}.

\begin{definition}[Subspace-design codes]\label[definition]{def:SubspaceDesigns}
    For a nondecreasing function $\tau: \mathbb Z_{\ge1}\rightarrow [0,1]$, a $\K$-additive code $\mathcal C\subseteq\Sigma^n$ is a $\tau$-\textit{subspace-design code over $\K$} if for every integer $r\ge1$ and every $\K$-linear subspace $\mathcal A\subseteq\mathcal C$ of dimension at most $r$,
    \[\frac{\sum\limits_{i=1}^n \dim_\K \mathcal A_i}{n}\le \dim_\K(\mathcal A)\cdot \tau(r) \]
    where $\mathcal A_i = \{a\in \mathcal A\mid a_i=0\}$.
\end{definition}

For Reed--Solomon codes, evaluation identifies $\mathcal A$ with a
$\K$-subspace $U\subseteq\F[X]_{<k}$ and $\mathcal A_i$ with
$U_i=\{f\in U:f(\alpha_i)=0\}$.

\begin{definition}[List decoding]\label[definition]{def:ListDecoding}
Let $\delta<1$ and let $L\ge1$ be an integer. A code $C\subseteq \Sigma^n$ is \emph{$(\delta,L)$-list-decodable} if, for every $y\in\Sigma^n$, \[|\{c\in C:\Delta(c,y)\le\delta\}|\le L.\] 
\end{definition}

\begin{definition}[Average-radius list decoding]\label[definition]{def:AverageRadius}
Let $\delta<1$ and let $L\ge1$ be an integer. A code $C\subseteq \Sigma^n$ is \emph{$(\delta,L)$-average-radius list-decodable} if, for every $y\in\Sigma^n$ and every set of $L+1$ distinct codewords $c_1, \dots, c_{L+1}$, \[\frac{\sum\limits_{i=1}^{L+1}\Delta(c_i, y)}{L+1}>\delta.\]
\end{definition}

The following list-decoding statement is due to \cite{ChenZ2025} and the list-recovery statement is due to \cite{BrakensiekCDZ2025-BrascampLieb}. We use the subspace-design formulation of their results as used by \cite{GoyalGH2026}, and presented for list-recovery in \cite{GoyalG2025-Structure}.
\begin{theorem}[Average-radius list decoding~\cite{ChenZ2025}]
\label{thm:AverageRadius}
Every $\tau$-subspace-design code over $\K$ is
$(\delta,L)$-average-radius list-decodable, and hence list-decodable,
for every integer $L\ge1$ and
$\delta<\frac{L}{L+1}(1-\tau(L))$.
\end{theorem}

\begin{corollary}[Generalized Singleton bound for all list sizes]
\label[corollary]{cor:SubspaceDesignAllListSizes}
Let $0<\varepsilon<1$ and $r=\lfloor1/\varepsilon\rfloor$.
Every $\tau$-subspace-design code over $\K$ is
$\left(\frac{L}{L+1}(1-\tau(r)-\varepsilon),L\right)$-average-radius
list-decodable, and hence list-decodable, for every integer $L\ge1$.
\end{corollary}
\begin{proof}
For $L\le r$, the claim follows from \Cref{thm:AverageRadius} and the monotonicity of $\tau$. For $L>r$, fix a received word $y$ and order any $L+1$ distinct codewords by increasing distance to $y$. Their average distance is at least the average of the first $r+1$,
which, by \Cref{thm:AverageRadius}, is at least
\[
  \frac{r}{r+1}\bigl(1-\tau(r)\bigr)
  \ge 1-\tau(r)-\frac1{r+1}
  >1-\tau(r)-\varepsilon.
\]
The claim follows.
\end{proof}

\begin{definition}[List recovery]\label[definition]{def:ListRecovery}
A code $C\subseteq\Sigma^n$ is \emph{$(\delta,\ell,L)$-list-recoverable} if,
for every $S_1,\ldots,S_n\subseteq\Sigma$ with $|S_i|\le\ell$,
\[
 \left|\left\{c\in C:
 |\{i\in[n]:c_i\notin S_i\}|\le\delta n
 \right\}\right|\le L.
\]
\end{definition}

\begin{theorem}[List recovery~\cite{BrakensiekCDZ2025-BrascampLieb}]
\label{thm:BlackboxListRecovery}
For every integer $\ell\ge1$ and $\eta>0$, set $r=\lceil\ell/\eta\rceil$.
Every $\tau$-subspace-design code over $\K$ is
\[
 \left(\delta,\ell,
 \left(\frac{\ell}{\tau(r)+\eta}\right)^{(\tau(r)+\eta)/\eta}\right)
\]
list-recoverable for every $\delta<1-\tau(r)-\eta$.
\end{theorem}

Curve-decoding was first defined in \cite{GoyalG2025-Proximity} as a new notion capturing proximity gaps, correlated agreement, and mutual correlated agreement. The line-decoding case was independently defined by Hab\"ock and is also used in \cite{CarmonGHLLPS2026}.

\begin{definition}[Curve decodability]
\label[definition]{def:CurveDecodability}
    A $\K$-additive code $C\subseteq\Sigma^n$ is $(\ell, \delta, a, b)$-$\K$-curve-decodable if, for every $u_0, u_1,\cdots, u_\ell\in \Sigma^n$ and every function $f: \K \to C$, whenever the set 
    \[ A=\left\{\alpha \in \K : \Delta\left(\smudef, f(\alpha)\right) \le  \delta\right\}\] has at least $a$ elements, there exist $c_0, c_1,\cdots, c_\ell \in C$ such that 
    \[\left|\left\{\alpha \in A  :  f(\alpha)=\smcdef \right\}\right|\ge b \ .  \]
When $\ell=1$, we call this \emph{line decodability}.
\end{definition}

\begin{theorem}[Curve decoding from subspace designs~{\cite{GoyalG2025-Proximity}}]
\label{thm:BlackboxCurve}
Let $C\subseteq\Sigma^n$ be a $\K$-additive
$\tau$-subspace-design code over $\K$.  For positive integers $r,\ell,a$ and
$\eta\ge(\ell+1)/r$, the code $C$ is
\[
 \left(\ell,1-\tau(r)-\eta,a,\frac{\eta a}{r+\eta}\right)
 \text{-\(\K\)-curve-decodable}.
\]
\end{theorem}

Proximity gaps were studied in~\cite{RothblumVW2013,BenSassonKS2018,BenSassonCIKS2023}, with correlated agreement formalized in~\cite{BenSassonCIKS2023}. We use the curve formulations of~\cite{GoyalG2025-Proximity}.

\begin{definition}[Proximity gaps (PG)]\label[definition]{def:CurvePg}
   A $\K$-additive code $C\subseteq\Sigma^n$ has an $(\ell, \delta, \eps, \gamma)$ proximity gap if for all $u_0, u_1, \cdots, u_\ell \in \Sigma^n$, and for all $\delta'\le \delta$, \[\mb{P}_{\alpha\in \K}\left(\Delta\left(\sum_{j=0}^\ell u_j \alpha^j, C\right)\le \delta'\right)>\eps \implies \forall \alpha \in \K, \ \Delta\left(\sum_{j=0}^\ell u_j \alpha^j, C\right)\le \frac{\delta'}{1-\gamma}\ .   \]
   When $\gamma = 0$, we say that $C$ has an $(\ell, \delta, \eps)$ proximity gap.  
\end{definition}

\begin{definition}[Correlated agreement (CA)]\label[definition]{def:CurveCa}
    A $\K$-additive code $C\subseteq\Sigma^n$ has $(\ell, \delta, \eps, \gamma)$ correlated agreement if for all $u_0, u_1, \cdots, u_\ell \in \Sigma^n$ and $\delta'\le \delta$ 
    \[\mb{P}_{\alpha\in \K}\left( \Delta\left(\sum_{j=0}^\ell u_j \alpha^j, C\right)\le \delta'\right)>\eps\implies
    \exists c_0, c_1,\cdots c_{\ell} \in C:
    \left|\left\{i : \bigvee_{j=0}^\ell  c_{j, i}\ne u_{j, i}\right\}\right| \le\frac{\delta' n}{1-\gamma} . \] 
\end{definition}

The parameter $\gamma$ is often called the \emph{proximity loss}~\cite{BenSassonCHKS2025}.

\begin{definition}[Mutual correlated agreement (MCA)~\cite{ArnonCFY2024-STIR-a}]
\label[definition]{def:CurveMCA}
A $\K$-additive code $C\subseteq\Sigma^n$ has
$(\ell,\delta,\varepsilon)$ mutual correlated agreement if, for every
$u_0,\ldots,u_\ell\in\Sigma^n$ and $0\le\delta'\le\delta$,
\[
\begin{split}
 \Pr_{\alpha\in\K}\biggl[\,&\exists S\subseteq[n],\ c\in C,\ j\in\{0,\ldots,\ell\}:\\
 &|S|\ge(1-\delta')n,\quad
 c|_S=\left(\sum_{h=0}^{\ell}\alpha^hu_h\right)\bigg|_S,\\
 &c'|_S\ne u_j|_S\text{ for every }c'\in C\biggr]
 \le\varepsilon.
\end{split}
\]
\end{definition}

Correlated agreement implies proximity gaps with the same parameters;
mutual correlated agreement implies both with $\gamma=0$.

\begin{theorem}[Consequences of curve decodability~{\cite[Theorem~3.6 and Corollary~3.9]{GoyalG2025-Proximity}}]\label{thm:CurveDecodingConsequences}
Let $\ell,L,a,b$ be positive integers with $b>\ell$, and assume $|\K|>(L+1)^2\ell$. If a $\K$-additive code $C\subseteq\Sigma^n$ is \[
\left(\delta\left(1+\frac{\ell}{b-\ell}\right), L\right)\text{ list decodable \quad and \quad } \left(\ell, \delta, a, b\right)\text{-\(\K\)-curve-decodable}
\] 
then $C$ has: \begin{enumerate}
    \item\label{item:CurveAmplification} for any integer $t\ge 1$, $(\ell, \delta, (t-1)L+a, t)$-$\K$-curve decodability,
    \item\label{item:CorrelatedAgreement} for any $m >1$, $\left(\ell, \delta, \frac{\ell m L + a}{|\K|}, \frac{1}{m}\right)$ correlated agreement and proximity gap,
    \item\label{item:MutualCorrelatedAgreement} $(\ell, \delta, \frac{\ell n L+a}{|\K|})$ mutual correlated agreement.
\end{enumerate}
\end{theorem}
For \Cref{item:CorrelatedAgreement}, take $t=\lceil\ell m\rceil$ in the
curve-decoding amplification: $(t-1)L+a\le\ell mL+a$ and
$\ell/t\le1/m$.

\subsection{Polynomial modules and degree}\label{subsec:PrelimPolynomials}

\begingroup
\begin{theorem}[Cauchy--Binet]
\label{thm:CauchyBinet}
Let $\mathbb E$ be a field and $1\le t\le e$.  For
$B\in\mathbb E^{t\times e}$ and $D\in\mathbb E^{e\times t}$,
\[
 \det(BD)=\sum_{\substack{S\subseteq[e]\\|S|=t}}
       \det B_{[t],S}\,\det D_{S,[t]}.
\]
The submatrices use the indicated rows and columns in increasing order.
\end{theorem}
\endgroup

For a polynomial vector $a=(a^{(1)},\ldots,a^{(m)})$, write
$\deg a=\max_{1\le j\le m}\deg a^{(j)}$. We use the conventions
$\deg 0=-\infty$ and $\max\varnothing=-\infty$. For a field $E$ and an
integer $r$, write $E[X]_{<r}=\{f\in E[X]:\deg f<r\}$; in particular,
$E[X]_{<r}=\{0\}$ when $r\le 0$.

A $\K[X]$-module $M\subseteq\K[X]^m$ is an additive subgroup
closed under multiplication by polynomials in $\K[X]$. Its
\emph{rank} is the dimension of $\Span_{\K(X)}M$ over $\K(X)$.

\begin{theorem}[Reduced bases~{\cite{Alekhnovich2005}}]
\label{thm:ReducedBases}
Let $\mathbb{K}$ be a field and let $M\subseteq\mathbb{K}[X]^m$ be a
$\mathbb{K}[X]$-module of rank $d$. Then $M$ has a basis
$a_1(X),\ldots,a_d(X)$ over $\mathbb{K}[X]$ such that
\[
  M=\left\{\sum_{i=1}^d c_i(X)a_i(X)
    \;\middle|\; c_i(X)\in\mathbb{K}[X]\right\}
\]
and, for every $c_1(X),\ldots,c_d(X)\in\mathbb{K}[X]$,
\begin{equation}
\label{eq:PredictableDegree}
  \deg\!\left(\sum_{i=1}^d c_i(X)a_i(X)\right)
  =\max_{1\le i\le d}\bigl(\deg a_i+\deg c_i\bigr).
\end{equation}
We call such a basis \emph{reduced}.
Given $s$ polynomial generators of $M$ of degree at most $D$, a reduced
basis can be computed deterministically using $\poly(m,s,D+1)$
arithmetic operations over $\K$.
\end{theorem}

\begin{corollary}
\label[corollary]{cor:ReducedPolynomialBasis}
Let $\mathbb{K}\subseteq\mathbb{F}$ be finite fields, let
$\mathbb{L}=\mathbb{K}(X)$, and let $A\subseteq\mathbb{F}(X)$ be an
$\mathbb{L}$-linear space of dimension $d$. Set
\[
  \mathcal{M}=A\cap\mathbb{F}[X],
  \qquad
  \mathcal{M}_k=A\cap\mathbb{F}[X]_{<k}
  \quad (k\in\mathbb{Z}_{\ge 0}).
\]
Then $\mathcal{M}$ has rank $d$ over $\mathbb{K}[X]$ and has a reduced
$\mathbb{K}[X]$-basis $a_1(X),\ldots,a_d(X)$ with respect to ordinary
polynomial degree. In particular, for every integer $k\ge 0$,
\[
  \mathcal{M}_k
  =\left\{\sum_{i=1}^d a_i(X)c_i(X)
    \;\middle|\;
    c_i(X)\in\mathbb{K}[X]_{<k-\deg a_i}\right\}.
\]
\end{corollary}

\subsection{Construction of Field Extensions and Deterministic Factorization}

\begin{theorem}[{OpenAI~\cite[Theorem~1.1]{OpenAI2026}}]
\label{thm:PrimeFieldFactorization}
Given a prime $p$ and a nonzero polynomial $f\in\F_p[X]$ of
degree $d$, there is a deterministic algorithm that computes the
leading coefficient of $f$, its distinct monic irreducible factors,
and their multiplicities in time $\poly(d,\log p)$.
\end{theorem}

Berlekamp~\cite{Berlekamp1970} gives a deterministic polynomial-time
reduction from factorization over an arbitrary finite field to
root finding over its prime subfield. Thus
\Cref{thm:PrimeFieldFactorization} also yields deterministic
factorization over non-prime finite fields.

\begin{corollary}[Factorization over finite fields]
\label{cor:FiniteFieldFactorization}
Given a finite field $\F$ and a nonzero polynomial
$f\in\F[X]$ of degree $d$, there is a deterministic algorithm
that computes its polynomial factorization into irreducibles in time
$\poly(d,\log |\F|)$.
\end{corollary}

Shoup~\cite{Shoup1990} gave a deterministic polynomial-time reduction from constructing irreducible polynomials over a finite field to factoring polynomials over its prime subfield. Combining this reduction with
\Cref{thm:PrimeFieldFactorization} gives the following immediate corollary.

\begin{corollary}[Construction of field extensions]
\label{cor:IrreduciblePolynomialConstruction}
Given a finite field $\K$ and an integer $d\ge1$,
one can deterministically construct an irreducible polynomial
$f\in\K[X]$ of degree $d$ and the extension $\K[X]/(f)$
in time $\poly(d,\log |\K|)$.
\end{corollary}

We note that the above results were known to work in run time $\poly(\mathsf{char}(\F), \log |\F|, d)$ earlier and the characteristic requirement is now removed due to OpenAI's result.

\subsection{Independent conjugates and coordinate movement}
\label{subsec:PrelimGalois}
For a field extension $\F/\K$, we extend $\sigma\in\Gal(\F/\K)$ to $\F(X)$ by applying it to coefficients and fixing $X$. 

\begin{lemma}[A downward-closed set of independent conjugates]
\label{lem:GaloisRowsAndMovement}
Let $\mathbb{K}\subseteq\mathbb{F}$ be finite fields, let
$\sigma_1,\ldots,\sigma_g$ generate
$G=\operatorname{Gal}(\mathbb{F}/\mathbb{K})$, and let
$u_1,\ldots,u_d\in\mathbb{F}[X]$ be linearly independent over
$\mathbb{K}(X)$, where $g,d\ge 1$. For
$\omega\in\mathbb{Z}_{\ge 0}^g$ write
$\sigma^\omega=\sigma_1^{\omega_1}\cdots\sigma_g^{\omega_g}$.
There is a set $\mathcal{B}\subseteq\mathbb{Z}_{\ge 0}^g$ of size $d$,
downward closed under the coordinatewise order, with the following
properties:
\begin{enumerate}[label=(\roman*),leftmargin=2em]
\item The polynomial
\[
  P(X)=\det\!\bigl(\sigma^\omega(u_a)(X)\bigr)_{
    \omega\in\mathcal{B},\,1\le a\le d}
\]
is nonzero.

\item For any distinct points $\alpha_1,\ldots,\alpha_n\in\F$,
the following bound holds:
\begin{equation}
\label{eq:Movement}
  \sum_{i=1}^n
  |\{\omega\in\mathcal B:\sigma^\omega(\alpha_i)\ne\alpha_i\}|
  \le d\log_2d\,
  \max_{\sigma\in\{\sigma_1,\ldots,\sigma_g\}}
  |\{\alpha_i:\sigma(\alpha_i)\ne\alpha_i\}|.
\end{equation}
\end{enumerate}
Such a set $\mathcal B$ can be computed deterministically in polynomial time.
\end{lemma}

\begin{proof}
Put $E=\F(X)$. The matrix $M=(\sigma(u_a))_{\sigma\in G,\,a\in[d]}$ has rank $d$ over $E$. Indeed, otherwise choose a nonzero kernel vector $c$ of minimal support and normalize $c_j=1$. For every $\tau\in G$, reindexing rows gives $\tau(c)\in\ker M$. Minimality forces $c-\tau(c)=0$, since its $j$-th coordinate is zero. Thus $c\in\K(X)^d$, and the identity row contradicts the independence of the $u_a$.

Write $r_\omega=(\sigma^\omega(u_1),\ldots,\sigma^\omega(u_d))$. Process the indices in graded lexicographic order, retaining $r_\omega$ whenever it is independent over $E$ of the previously retained rows. Let $\mathcal B$ be the retained indices. The family $(r_\omega)_\omega$ contains every row of $M$, so $|\mathcal B|=d$ and the retained determinant is nonzero. If $\omega\notin\mathcal B$, then
\[
  r_\omega=\sum_{\nu\prec\omega}c_\nu r_\nu
  \quad\Longrightarrow\quad
  r_{\omega+e_j}
  =\sum_{\nu\prec\omega}\sigma_j(c_\nu)r_{\nu+e_j}.
\]
The generators commute, and the order is preserved by addition. Hence $\omega+e_j\notin\mathcal B$, proving downward closure.

To compute $\mathcal B$, start with the index $0$ and process indices in this order, adding unseen successors $\omega+e_1,\ldots,\omega+e_g$ only when $\omega$ is already in $\mathcal B$. Independence is tested when each index is processed. Downward closure ensures that every element of $\mathcal B$ is reached. Since at most $\poly(g,d)$ indices are checked in the greedy process, the algorithm runs in polynomial time.

Finally, let $\omega\in\mathcal B$ have support $S$. Observe that
\[
  |\{i\in[n]:\sigma^\omega(\alpha_i)\ne\alpha_i\}|
  \le |S|\max_{j\in[g]}
  |\{i\in[n]:\sigma_j(\alpha_i)\ne\alpha_i\}|.
\]
By downward closedness, $d=|\mathcal B|\ge 2^{|S|}$. Thus $|S|\le\log d$, and summing over $\mathcal B$ proves the result.
\end{proof}

\section{Main theorem and combinatorial consequences}
\label{sec:ProofMain}

We prove \Cref{thm:Main} by combining the subspace-design bound proved in \Cref{thm:SubspaceDesign}, the list-decoding algorithm as in \Cref{thm:FindingList}, and the construction of evaluation points from \Cref{thm:explicit-effective}. We use the list-decoding algorithm to find an $O_{\eps}(1)$ sized-list containing all nearby codewords, and the subspace-design property gives the requisite near-optimal list-sizes, and we just check all elements in the list outputted by our algorithm.

\mainresult*

\begin{proof}
We may assume $r\ge\lceil4/\varepsilon\rceil$.
Let $\K'/\K$ be a quadratic extension, so $|\K'|>nr$ for
sufficiently large $n$. Apply \Cref{thm:explicit-effective} over
$\K'$ with $d=\lceil C(\varepsilon/2)\rceil$ and sufficiently large
$g=O_{r,\varepsilon}(1)$, so that the points satisfy the hypotheses
of \Cref{thm:FindingList} with gap $\varepsilon/2$ and
\[
  \frac{\lceil n/g\rceil}{n}\log_2r\le\varepsilon/4.
\]
The construction takes time $|\K|^{O_{r,\varepsilon}(1)}$ and gives
$[\F:\K]=O_{r,\varepsilon}(1)$.
Thus, \Cref{thm:SubspaceDesign} gives the claimed subspace-design
bound over $\K'$, which also holds over every
$\K_0\subseteq\K$ by \Cref{lem:RestrictionOfScalars}.

Write $R=k/n$. Since $\lfloor2/\varepsilon\rfloor\le r$ and
$\tau(\lfloor2/\varepsilon\rfloor)\le R+\varepsilon/4$,
\Cref{cor:SubspaceDesignAllListSizes} with gap $\varepsilon/2$
gives $(\delta_L,L)$-average-radius list decodability for every
integer $L\ge1$.

Now, run \Cref{thm:FindingList} with gap $\varepsilon/2$ and
retain the codewords within distance $\delta_L$ or $1-R-\varepsilon$,
respectively. For the latter, let $L_0=\lceil2(1-R)/\varepsilon\rceil\le r$.
Since $\tau(L_0)\le R+\varepsilon/4$ and
\[
  \frac{L_0}{L_0+1}(1-R-\varepsilon/4)
  >1-R-\varepsilon,
\]
\Cref{thm:AverageRadius} gives a list of size at most $L_0$.
The initial list has size $(1/\varepsilon)^{O(1/\varepsilon)}$,
and the enumeration and filtering run deterministically in time
$O_{r,\varepsilon}(1)\poly(n,\log|\F|)$.
\end{proof}

\begin{remark}
We expect our algorithmic techniques to extend naturally to list-recovery
and curve-decoding. We restrict ourselves here to the combinatorial
consequences for these tasks.
\end{remark}

We now apply the results in \Cref{sec:Preliminaries} to the codes of
\Cref{thm:Main} to obtain various combinatorial consequences. Throughout, $n$ is sufficiently large in the fixed
parameters and $\K$ is a finite field with $|\K|\ge n$.

\subsection{List recovery}
\label{subsubsec:ListRecovery}

\begin{theorem}[Capacity list recovery]\label{thm:ListRecoveryConsequence}
Fix $0<R<1$, $0<\varepsilon<1-R$, and an integer $\ell\ge2$, and let
\[
 r=\left\lceil\frac{2\ell}{\varepsilon}\right\rceil,\qquad
 L=\left\lceil
 \left(\frac{\ell}{R+\varepsilon/2}\right)^{3+2R/\varepsilon}
 \right\rceil.
\]
Let $\F,A$ be supplied by \Cref{thm:Main} with parameters
$r,\varepsilon$, and put $C=\RS_\F(A,\lfloor Rn\rfloor)$.
Then $C$ is
$(1-R-\varepsilon,\ell,L)$-list-recoverable.
\end{theorem}
\begin{proof}
\Cref{thm:Main} gives $\tau(r)\le R+\varepsilon/4$.
Apply \Cref{thm:BlackboxListRecovery} to $C$ with
$\eta=\varepsilon/2$ and design bound $R+\varepsilon/4$.
At radius $1-R-\varepsilon$, the list size is at most
\[
  \left(\frac{\ell}{R+3\varepsilon/4}\right)^{2R/\varepsilon+3/2}
  \le\left(\frac{\ell}{R+\varepsilon/2}\right)^{2R/\varepsilon+3}
  \le L.\qedhere
\]
\end{proof}

\subsection{Curve decoding}
\label{subsubsec:CurveDecoding}

\begin{theorem}[Base-field capacity curve decoding]\label{thm:KCurveDecoding}
Fix $0<R<1$, $0<\varepsilon<1-R$, and an integer $\ell\ge1$.  Set
\[
 L=\left\lceil\frac{8}{\varepsilon}\right\rceil,
 \quad b_0=\ell+\left\lceil\frac{8(\ell+1)}{\varepsilon}\right\rceil,
 \quad r=\left\lceil\frac{8(\ell+1)}{\varepsilon}\right\rceil,
 \quad a_0=\left\lceil\frac{2b_0(r+\varepsilon/2)}{\varepsilon}\right\rceil.
\]
Let $\F,A$ be supplied by \Cref{thm:Main} with parameters
$r,\varepsilon$, and put $C=\RS_\F(A,\lfloor Rn\rfloor)$.
Then $C$ is
\[
 \bigl(\ell,1-R-\varepsilon,a_0,b_0\bigr)
 \text{-\(\K\)-curve-decodable}.
\]
\end{theorem}
\begin{proof}
\Cref{thm:Main} gives $\tau(d)\le R+\varepsilon/4$ for $d\le r$.
\Cref{thm:BlackboxCurve}, applied over $\K$ with
$\eta=\varepsilon/2$, gives the displayed $(a_0,b_0)$ statement: our
choices ensure $\eta\ge(\ell+1)/r$ and
$\eta a_0/(r+\eta)\ge b_0$.
\end{proof}

\begin{theorem}[Extension-field capacity curve decoding]
\label{thm:CurveDecodingConsequence}
Under the hypotheses and parameter choices of
\Cref{thm:KCurveDecoding}, assume also
$|\F|>(L+1)^2\ell$, as holds for sufficiently large $n$, and put
\[
 a_\F=\left\lceil\frac{|\F|}{|\K|}a_0\right\rceil.
\]
Then, for every integer $t\ge1$, the code $C$ is
\[
 \bigl(\ell,1-R-\varepsilon,(t-1)L+a_\F,t\bigr)
 \text{-\(\F\)-curve-decodable}.
\]
\end{theorem}
\begin{proof}
Partition $\F$ into affine lines over $\K$. Among $a_\F$ good parameters,
some line contains at least $a_0$. Applying
\Cref{thm:KCurveDecoding} after translation gives
$(\ell,1-R-\varepsilon,a_\F,b_0)$-$\F$-curve decodability.
The result follows from \Cref{thm:CurveDecodingConsequences}, using
the $(1-R-\varepsilon/2,L)$-list-decoding bound from
\Cref{thm:AverageRadius}.
\end{proof}

\subsection{Line decoding}
\label{subsubsec:LineDecoding}

\begin{theorem}[Capacity line decoding]\label{thm:LineDecodingConsequence}
Fix $0<R<1$ and $0<\varepsilon<1-R$.  Set
\[
 L=\left\lceil\frac{8}{\varepsilon}\right\rceil,
 \quad r=\left\lceil\frac{16}{\varepsilon}\right\rceil,
 \quad b_0=1+\left\lceil\frac{16}{\varepsilon}\right\rceil,
 \quad a_0=\left\lceil\frac{2b_0(r+\varepsilon/2)}{\varepsilon}\right\rceil.
\]
Let $\F,A$ be supplied by \Cref{thm:Main} with parameters
$r,\varepsilon$, and put $C=\RS_\F(A,\lfloor Rn\rfloor)$.
Then, for every integer $t\ge1$, $C$ is
\[
 \bigl(1,1-R-\varepsilon,(t-1)L+
 \lceil |\F|a_0/|\K|\rceil,t\bigr)
 \text{-\(\F\)-curve-decodable}.
\]
\end{theorem}
\begin{proof}
This is \Cref{thm:CurveDecodingConsequence} with $\ell=1$.
\end{proof}

\subsection{Proximity gap, correlated agreement, and mutual correlated agreement}
\label{subsubsec:ProximityGapCorrelatedAgreementAndMutualCorrelatedAgreement}

Let $C,L,a_\F$ be as in \Cref{thm:CurveDecodingConsequence}, and put
\[
 \delta=1-R-\varepsilon.
\]
We now apply \Cref{thm:CurveDecodingConsequences} over $\F$
to the list-decoding and curve-decoding guarantees proved above.

\begin{corollary}[Capacity correlated agreement and proximity gap]
\label{cor:CapacityCa}\label{cor:CapacityPg}
For every real $m>1$, $C$ has
\[
 \left(\ell,\delta,\frac{\ell mL+a_\F}{|\F|},\frac1m\right)
\]
correlated agreement and proximity gap over $\F$.
\end{corollary}

\begin{corollary}[Capacity mutual correlated agreement]
\label{cor:CapacityMutualCorrelatedAgreement}
$C$ has
\[
 \left(\ell,\delta,\frac{\ell nL+a_\F}{|\F|}\right)
\]
mutual correlated agreement over $\F$.
\end{corollary}

\begin{remark}
The affine-line reduction introduces a factor $|\F|/|\K|$
in the curve-decoding threshold. For $\ell=1$,
\Cref{sec:ImprovedExtensionFieldConsequences} gives a better
result under a stronger subspace-design assumption.
By assuming the same design bound through a larger dimension,
we obtain a line-decoding threshold without the $|\F|/|\K|$ factor.
\end{remark}

\section{Reed-Solomon Subspace Designs}
\label{sec:SubspaceDesigns}

In this section, we detail sufficient properties of evaluation points so that the resulting Reed-Solomon code forms a subspace-design code over a subfield. We prove that these properties suffice, and an explicit construction matching these properties is given in \Cref{sec:ExplicitInstantiation}.

\subsection{Assumptions and the subspace-design theorem}
\label{subsec:AssumptionsAndTheSubspaceDesignTheorem}

\begin{definition}[Properties required of the evaluation points]\label{def:EffectivePoints}
Let $\K\subseteq\F$ be finite fields, fix integers $n,g\ge1$,
$0\le b\le n$, and $r\ge2$.

A set of $n$ distinct points
$A=(\alpha_1,\ldots,\alpha_n)\in\F^n$ is
\emph{$(n,g,b,r)_{\F,\K}$-effective} if:
\begin{enumerate}[label=(\roman*),leftmargin=2em]
\item There are automorphisms $\sigma_1,\ldots,\sigma_g$ generating $\operatorname{Gal}(\F/\K)$ such that
\[
 \bigl|\{i\in[n]:\sigma_j(\alpha_i)\ne\alpha_i\}\bigr|\le b
 \qquad(j\in[g]).
\]
These automorphisms commute, since the Galois group of a finite-field
extension is cyclic and generated by the Frobenius automorphism $x\mapsto x^{|\K|}$.

\item For every nonzero $\K$-subspace
$\mathcal Q\subseteq \K[X]_{<nr}$ of dimension at most $2^r$, then
\begin{equation}\label{eq:MultiplicityAssumption}
 \sum_{i=1}^n\max_{0\ne q\in\mathcal Q}\mult_{\alpha_i}q\le \max_{0\ne q\in\mathcal Q}\deg q.
\end{equation}
Here $\mult_{\alpha}q$ is the multiplicity of $\alpha$ as a root of $q$.
\end{enumerate}
\end{definition}

\begin{restatable}[Subspace-design bound]{theorem}{subspacedesign}\label{thm:SubspaceDesign}
Let $\K\subseteq\F$ be finite fields, and let
$A=(\alpha_1,\ldots,\alpha_n)$ be $(n,g,b,r)_{\F,\K}$-effective. 
For every $1\le k\le n$, the code $C=\operatorname{RS}_\F(A,k)$ is a $\tau$-subspace-design code over $\K$, with
\begin{equation}\label{eq:TauBound}
 \tau(d)=\min\left\{1,\frac{k-1}{n}+\frac{b}{n}\log_2d\right\}
 \qquad(1\le d\le r),
\end{equation}
with $\tau(d)=1$ for $d>r$.
\end{restatable}

The proof here is inspired by Guruswami and Kopparty's Wronskian proof of the subspace-design bound for folded Reed-Solomon codes \cite{GuruswamiK2016}. It is adapted to suit working over extension field Reed-Solomon codes and automorphisms using Moore/Galois determinants as in the work of Guruswami and Xing. The key innovation here in comparison to \cite{GuruswamiX2013} is that we allow a small family of automorphisms generating the Galois group each fixing most evaluation points. In comparison, Guruswami and Xing used the Frobenius automorphism to generate the group. Their argument requires this automorphism to fix all evaluation points, so the evaluation points lie in the subfield $\K$. This variation allows us to have the second property, as we can now consider evaluation points to all be non-conjugate and of high degree. 

\subsection{From effective evaluation points to subspace designs}
\label{subsec:SubspaceDesignTransfer}

Our claimed subspace-design property is a property of spaces with bounded $\K$-dimension. We first prove a bound in terms of $\K(X)$-ranks and then use a transfer theorem to convert it into a bound on $\K$-dimensions. Moreover, we show that the same \textit{transfer theorem} (\cref{thm:SubspaceDesignTransfer}) transfers the subspace-design property to subfields of $\K$.

\begin{proposition}[Rational ranks of the coordinate kernels]\label{prop:CoordinateKernelRanks}
Let $\K\subseteq\F$ be finite fields, let $1\le k\le n$, and let
$A=(\alpha_1,\ldots,\alpha_n)$ be $(n,g,b,r)_{\F,\K}$-effective.
Let
$\{0\}\ne W\subseteq \F[X]_{<k}$ with $\dim_{\K} W = e\leq r$, and $\dim_{\K(X)}\Span_{\K(X)}(W)=t\leq e$.  Set $W_i=\{f\in W:f(\alpha_i)=0\}$ and $t_i=\dim_{\K(X)}\Span_{\K(X)}(W_i)$.
Then
\[
 \sum_{i=1}^n t_i\le t(k-1)+bt\log_2t.
\]
\end{proposition}
\begin{proof}
Put $\L=\K(X)$, and choose generators $\sigma_1,\ldots,\sigma_g$ witnessing the effectiveness of $A$. Apply \Cref{cor:ReducedPolynomialBasis} to $\Span_\L W$ and choose its reduced polynomial basis $u_1,\ldots,u_t$, and let
\[
 \nu=\sum_{a=1}^t\deg u_a,\qquad A_W=t(k-1)-\nu.
\]
Write a $\K$-basis of $W$ as
\[
 (w_1,\ldots,w_e)=(u_1,\ldots,u_t)C(X),
 \qquad C(X)\in \K[X]^{t\times e}.
\]
By~\Cref{eq:PredictableDegree}, the degree of row $a$ of $C$ is at most $k-1-\deg u_a$.  Let $\mathcal Q_W$ be the $\K$-span of the $t\times t$ minors of $C$.  The matrix has rank $t$, so $\mathcal Q_W\ne0$.  Moreover,
\[
 \dim_\K\mathcal Q_W\le\binom et\le 2^r,
 \qquad \deg q\le A_W\le r(n-1)<rn
 \quad(q\in\mathcal Q_W).
\]
Apply \Cref{lem:GaloisRowsAndMovement} to $u_1,\ldots,u_t$ to obtain a row index set $\mathcal B\subseteq\mathbb Z_{\ge0}^g$ of size $t$ such that
\[
 P(X)=\det\!\bigl(\sigma^\omega(u_a)(X)\bigr)_{\omega\in\mathcal B,\,a\in[t]}\ne0.
\]
Expanding the determinant gives
$\deg P\le\sum_a\deg u_a=\nu$.

Fix a coordinate $i$, and set
\[
 b_i=\bigl|\{\omega\in\mathcal B:\sigma^\omega(\alpha_i)\ne\alpha_i\}\bigr|.
\]
Choose $h_1,\ldots,h_{t_i}\in W_i$ independent over $\L$, and extend them,
using elements of $W$, to an $\L$-basis $h_1,\ldots,h_t$ of
$\Span_\L W$.  There is a matrix $T_i\in\K^{e\times t}$ such that
\[
 (h_1,\ldots,h_t)=(u_1,\ldots,u_t)C(X)T_i.
\]
By the Cauchy-Binet formula for determinant (\Cref{thm:CauchyBinet}),
\[
 q_i:=\det(C(X)T_i)
 =\sum_{\substack{S\subseteq[e]\\|S|=t}}
     \det C_{[t],S}(X)\,\det(T_i)_{S,[t]}
 \in\mathcal Q_W\setminus\{0\}.
\]
Here the coefficients $\det(T_i)_{S,[t]}$ lie in $\K$, and $q_i\ne0$
because the $h_j$ are independent over $\L$.  Since every Galois row
fixes the entries of $C(X)T_i$, the determinant for this adapted basis
is $q_i(X)P(X)$.

In every row whose automorphism fixes $\alpha_i$, the first $t_i$
entries vanish at $\alpha_i$.  Each determinant term uses a moving row
in at most $b_i$ of these columns.  Hence
\begin{align*}
 t_i-b_i
 &\le\mult_{\alpha_i}(q_iP)\\
 &=\mult_{\alpha_i}q_i+\mult_{\alpha_i}P\\
 &\le \max_{0\ne q\in \mathcal Q_W}\mult_{\alpha_i}q+\mult_{\alpha_i}P.
\end{align*}
Summing and using distinctness of the points,
\Cref{eq:MultiplicityAssumption}, and~\Cref{eq:Movement}, we obtain
\begin{align*}
 \sum_i t_i
 &\le\sum_i b_i+\sum_i\mult_{\alpha_i}P
                  +\sum_i \max_{0\ne q\in \mathcal Q_W}\mult_{\alpha_i}q\\
 &\le bt\log_2t+\deg P+A_W\\
 &\le bt\log_2t+\nu+A_W
 =t(k-1)+bt\log_2t \ . \qedhere
\end{align*}
\end{proof}

\begin{theorem}[From extension-field ranks to dimensions]
\label{thm:SubspaceDesignTransfer}
Let $\E_0\subseteq \E_1$ be fields, let $C$ be an $\E_0$-subspace of an $\E_1$-vector space, and let $H_1,\ldots,H_n$ be $\E_0$-subspaces of $C$. Let $r\ge1$ and $\tau\ge0$.
Suppose every nonzero $\E_0$-subspace $W\subseteq C$ with $\dim_{\E_0} W\le r$ satisfies
\begin{equation}
\label{eq:RankSubspaceDesign}
  \sum_{i=1}^n\dim_{\E_1}\Span_{\E_1}(W\cap H_i)
  \le n\tau\dim_{\E_1}\Span_{\E_1} W.
\end{equation}
Then every $\E_0$-subspace $U\subseteq C$ of dimension
$1\le d\le r$ satisfies
\[
  \sum_{i=1}^n\dim_{\E_0}(U\cap H_i)\le n\tau d.
\]
\end{theorem}
We prove \Cref{thm:SubspaceDesignTransfer} in \Cref{subsec:ProofSubspaceDesignTransfer} by breaking any $\E_0$-linear space into a chain of spaces on which the hypothesis applies and then taking a nonnegative linear combination of the resulting inequalities. Before proving the result, we use it to obtain both the Reed-Solomon subspace-design guarantee of \Cref{thm:SubspaceDesign} and additionally prove that the same transfers to subfields of $\K$ as below:

\begin{lemma}[Subspace-design transfers to subfields]
\label{lem:RestrictionOfScalars}
Let $\K_0\subseteq\K$ be finite fields. If a $\K$-additive code
$C\subseteq\Sigma^n$ is a $\tau$-subspace design over $\K$, then it is also a $\tau$-subspace design over $\K_0$.
\end{lemma}

\begin{proof}[Proof of \Cref{thm:SubspaceDesign,lem:RestrictionOfScalars}]
Both statements follow from \Cref{thm:SubspaceDesignTransfer}. For \Cref{thm:SubspaceDesign}, let $\E_1=\K(X)$, and $\E_0=\K$. For \Cref{lem:RestrictionOfScalars}, let $\E_1 = \K$ and $\E_0 = \K_0$. 

Observe that the hypothesis in the first case is true by \Cref{prop:CoordinateKernelRanks}, and the conclusion is exactly the guarantee of \Cref{thm:SubspaceDesign}. 

Similarly, if $C$ is a $\tau$-subspace design over $\K$, then the hypothesis of \Cref{thm:SubspaceDesignTransfer} holds with $\K = \E_1$, and the conclusion is exactly that $C$ is also a $\tau$ subspace-design over $\K_0$.

\end{proof}

\subsection{Proof of the transfer theorem}
\label{subsec:ProofSubspaceDesignTransfer}

\begin{lemma}[Dimension comparison]\label{lem:DimensionComparison}
  Let $\E_0\subseteq \E_1$ be fields, and let $U$ be a finite-dimensional $\E_0$-subspace of an $\E_1$-vector space.  For $S\subseteq U$, write $\rho(S)=\dim_{\E_1}\operatorname{span}_{\E_1} S$.  There are spaces
  \[
   0=U_0\subsetneq U_1\subsetneq\cdots\subsetneq U_h=U, ~~~~ U_i = U\cap \operatorname{span}_{\E_1} U_i ~\forall i \in \{0,1,\dots, h\}
  \]
  and nonnegative real numbers $a_1,\ldots,a_h$ such that, for every $\E_0$-subspace $S\subseteq U$,
  \[
   \dim_{\E_0}S\le\sum_{\ell=1}^h a_\ell\rho(S\cap U_\ell),
   \qquad
   \sum_{\ell=1}^h a_\ell\rho(U_\ell)=\dim_{\E_0}U.
  \]
\end{lemma}
\begin{proof}
If $U=0$, take $h=0$. Otherwise, starting with $U_0=0$, choose a
$\E_0$-subspace $U_{\ell-1}\subsetneq U_\ell\subseteq U$ maximizing
\[
 \lambda_\ell=
 \frac{\dim_{\E_0}U_\ell-\dim_{\E_0}U_{\ell-1}}
 {\rho(U_\ell)-\rho(U_{\ell-1})}.
\]
Inductively, $U_{\ell-1}=U\cap\Span_{\E_1}U_{\ell-1}$, so every
denominator is positive. A maximizer exists because there are only
finitely many dimension--rank pairs. Replacing $U_\ell$ by
$U\cap\Span_{\E_1}U_\ell$ preserves its rank and would increase
its dimension unless the two spaces coincide. Thus the required
intersection property persists, and strict dimension growth gives a
sequence of spaces ending at $U$.

The slope from $U_{\ell-1}$ to $U_{\ell+1}$ is a weighted average of
$\lambda_\ell$ and $\lambda_{\ell+1}$. Maximality of $\lambda_\ell$
therefore gives $\lambda_{\ell+1}\le\lambda_\ell$. Set
$\lambda_{h+1}=0$ and $a_\ell=\lambda_\ell-\lambda_{\ell+1}\ge0$.

For a $\E_0$-subspace $S\subseteq U$, put $S_\ell=S\cap U_\ell$.
Maximality applied to $S_\ell+U_{\ell-1}$ and the dimension formulas
over $\E_0$ and $\E_1$ give
\begin{align*}
 \dim_{\E_0}S_\ell-\dim_{\E_0}S_{\ell-1}
 &\le\lambda_\ell\bigl(
 \rho(S_\ell+U_{\ell-1})
 -\rho(U_{\ell-1})\bigr)\\
 &\le\lambda_\ell\bigl(
 \rho(S_\ell)-\rho(S_{\ell-1})\bigr).
\end{align*}
Here the second inequality uses
$\Span_{\E_1}S_{\ell-1}\subseteq
\Span_{\E_1}S_\ell\cap\Span_{\E_1}U_{\ell-1}$;
if $S_\ell\subseteq U_{\ell-1}$, both increments are zero.

Summing these inequalities and then summing by parts yields
\begin{align*}
 \dim_{\E_0}S
 &\le\sum_{\ell=1}^h\lambda_\ell\bigl(
       \rho(S_\ell)-\rho(S_{\ell-1})\bigr)\\
 &=\sum_{\ell=1}^h a_\ell\rho(S\cap U_\ell).
\end{align*}
For $S=U$, every increment inequality is an equality by the definition
of $\lambda_\ell$, giving the required normalization.
\end{proof}

\begin{proof}[Proof of \Cref{thm:SubspaceDesignTransfer}]
Fix a $\E_0$-subspace $U\subseteq C$ of dimension $1\le d\le r$.
Write $\rho(S)=\dim_{\E_1}\Span_{\E_1} S$.
Apply \Cref{lem:DimensionComparison} with scalar field $\E_0$ and
rank field $\E_1$, obtaining spaces $U_\ell$ and weights $a_\ell$.
Each $U_\ell$ has $\E_0$-dimension at most $r$, so
\Cref{eq:RankSubspaceDesign} gives
\begin{align*}
  \sum_i\dim_{\E_0}(U\cap H_i)
  &\le\sum_\ell a_\ell\sum_i\rho(U_\ell\cap H_i)\\
  &\le n\tau\sum_\ell a_\ell\rho(U_\ell)
   =n\tau d.
\end{align*}
The equality is the normalization in \Cref{lem:DimensionComparison}.
\end{proof}

\section{Efficient List Decoding}
\label{sec:EfficientListDecoding}

As discussed in \Cref{sec:intro-overview}, our efficient decoder follows the following blueprint, given our constructed Reed-Solomon code $C\subseteq \mathbb F^n$ with $\K \subseteq \F$.
\begin{itemize}
    \item Finds a low-dimensional $\K(X)$ space of $\F(X)$ containing all close by message polynomials. (\Cref{thm:FindingLinearSpace})
    \item \textit{Prunes} a low-dimensional $\K(X)$ space via low-degree constraints on the space. (\Cref{thm:ExtractingList})
\end{itemize}

We first state our combined result and the required properties of the evaluation points.

\begin{theorem}[List decoding near capacity]
\label{thm:FindingList}
For every $0<\varepsilon<1/2$, there exists a constant
\[
  C(\varepsilon)
  =\exp\!\left(O\!\left(\varepsilon^{-3}\log(1/\varepsilon)\right)\right)
\]
with the following property. Let $\mathbb{K}\subseteq\mathbb{F}$ be finite
fields, and let $\alpha_1,\ldots,\alpha_n\in\mathbb{F}$,
where $n$ is sufficiently large in terms of $\varepsilon$. Write
$\deg_{\mathbb{K}}\alpha=[\mathbb{K}(\alpha):\mathbb{K}]$.
Suppose that the evaluation points satisfy the following conditions:
\begin{enumerate}
  \item The points $\alpha_1,\ldots,\alpha_n$ are pairwise nonconjugate
    over $\mathbb{K}$.

  \item There is a positive integer
    \[
      g\ge \frac{12\log_2(1/\varepsilon)}{\varepsilon}
    \]
    and a generating set $\sigma_1,\ldots,\sigma_g$ of
    $\operatorname{Gal}(\mathbb{F}/\mathbb{K})$ such that, for every
    $1\le \ell\le g$,
    \[
      \left|\left\{j\in\{1,\ldots,n\}:
        \sigma_\ell(\alpha_j)\ne\alpha_j\right\}\right|
      \le \left\lceil\frac{n}{g}\right\rceil.
    \]

  \item For every $1\le i,j\le n$,
    \[
      C(\varepsilon)\le\deg_{\mathbb{K}}\alpha_i
      \le 2\deg_{\mathbb{K}}\alpha_j.
    \]
\end{enumerate}
Then, for every integer $1\le k\le n$, the Reed--Solomon code
$\RS_\F((\alpha_1,\ldots,\alpha_n),k)$ evaluated at
$\alpha_1,\ldots,\alpha_n$ admits a deterministic list-decoding algorithm with the
following guarantee. Given the evaluation points, the generators, and
a received word $y=(y_1,\ldots,y_n)\in\mathbb{F}^n$, the algorithm outputs all
$f\in\mathbb{F}[X]_{<k}$ satisfying
\[
  \left|\left\{j\in\{1,\ldots,n\}:f(\alpha_j)\ne y_j\right\}\right|
  \le n-k-\varepsilon n.
\]
The output list has size at most
\[
  (1/\varepsilon)^{O(1/\varepsilon)}.
\]
and the algorithm runs in time $O_{g,\eps}(1)\poly(n, \log |\F|)$.\footnote{We use the deterministic finite-field factorization algorithm of \Cref{cor:FiniteFieldFactorization}.}
\end{theorem}

Combining \Cref{thm:FindingList} with
\Cref{thm:strongly-explicit-points} gives the following as a corollary.

\stronglyexplicit*
\begin{proof}
We can apply \Cref{thm:strongly-explicit-points} with $g=\lceil12\log_2(1/\varepsilon)/\varepsilon\rceil$ and $d=\lceil C(\varepsilon)\rceil$, and the result follows using \Cref{thm:FindingList} with gap $\varepsilon$.
\end{proof}

\subsection{The two-step list-decoding theorems}

The two theorems below give the two steps of the list-decoding algorithm. Interestingly, the two assumptions of the choice of automorphisms $\sigma_i$, and the requirement of sufficiently high degree of $\alpha_i$ are used in the two steps separately and do not interact with each other.

\begin{restatable}[Finding a low-dimensional $\K(X)$-space]{theorem}{linearspacereduction}
\label{thm:FindingLinearSpace}
Fix $0<\varepsilon<1/2$. Let $\mathbb{K}\subseteq\mathbb{F}$ be finite fields,
and consider a Reed--Solomon code of length $n$ and dimension $k$ over
$\mathbb{F}$, where $n$ is sufficiently large in terms of $\varepsilon$,
with distinct evaluation points
$\alpha_1,\ldots,\alpha_n\in\mathbb{F}$.
Suppose $\operatorname{Gal}(\mathbb{F}/\mathbb{K})$ has $g\ge 1$ generators
$\sigma_1,\ldots,\sigma_g$ with
\[
  g\ge\frac{12\log_2(1/\varepsilon)}{\varepsilon},
\]
such that each $\sigma_\ell$, we have that $|\{\alpha_i\mid \sigma_\ell (\alpha_i)\neq \alpha_i\}|\leq \lceil n/g\rceil$.

Given the evaluation points, these generators, and any received word
$y=(y_1,\ldots,y_n)\in\mathbb{F}^n$, one can compute in time $\poly(n, g, \log |\F|)$ a $\K(X)$-basis consisting of polynomials in $\F[X]_{<k}$ of a space $A\subseteq\mathbb{F}(X)$ with
\[
  \dim_{\mathbb{K}(X)}A\le\frac{2}{\varepsilon}
\]
that contains every polynomial $f\in\mathbb{F}[X]$ of degree less than
$k$ satisfying
\[
  \bigl|\{j\in\{1,\ldots,n\}:f(\alpha_j)=y_j\}\bigr|
  \ge k+\varepsilon n.
\]
\end{restatable}

\begin{restatable}[Pruning a low-dimensional $\K(X)$-space]{theorem}{listextraction}
\label{thm:ExtractingList}
For every integer $d\ge 1$ and every $0<\varepsilon<1/2$, there is a
constant
\[
  C(d,\varepsilon)
  =\exp\!\left(O\!\left(d^3\log(1/\varepsilon)\right)\right)
\]
with the following property.
Let $\mathbb{K}\subseteq\mathbb{F}$ be finite fields, and consider a
Reed-Solomon code of length $n$ and dimension $k$ over $\mathbb{F}$,
where $n$ is sufficiently large in terms of $d$ and $\varepsilon$.
Suppose its evaluation points $\alpha_1,\ldots,\alpha_n\in\mathbb{F}$
are pairwise nonconjugate over $\mathbb{K}$ and satisfy
\[
  C(d,\varepsilon)\le\deg_{\mathbb{K}}\alpha_i
  \le 2\deg_{\mathbb{K}}\alpha_j
  \qquad(1\le i,j\le n),
\]
where $\deg_{\mathbb{K}}\alpha=[\mathbb{K}(\alpha):\mathbb{K}]$.

Given the evaluation points, a received word
$y=(y_1,\ldots,y_n)\in\mathbb{F}^n$, and a
$\mathbb{K}(X)$-linear space $A\subseteq\mathbb{F}(X)$ of dimension at
most $d$, given by a basis over $\K(X)$ consisting of polynomials in
$\F[X]_{<k}$, one can output all polynomials $f\in A\cap\mathbb{F}[X]$
of degree less than $k$ satisfying
\[
  \bigl|\{j\in\{1,\ldots,n\}:f(\alpha_j)=y_j\}\bigr|
  \ge k+\varepsilon n.
\]
The list has size at most
\[
  {(O(1/\varepsilon))^{d}},
\]
and can be computed in time $O_{d, \eps}(1)\poly(n, \log |\F|)$.
\end{restatable}

\begin{proof}[Proof of \cref{thm:FindingList} assuming \cref{thm:FindingLinearSpace}, \Cref{thm:ExtractingList}]
Apply \Cref{thm:FindingLinearSpace}, followed by
\Cref{thm:ExtractingList} with $d=\lceil2/\varepsilon\rceil$ and
$C(\varepsilon)=C(d,\varepsilon)$.
The bounds in \Cref{thm:ExtractingList} give the claimed degree
threshold and list size.
\end{proof}

We now present the proof of \Cref{thm:FindingLinearSpace}, and defer \Cref{thm:ExtractingList} to the next section.

\subsection{Finding a low-dimensional \texorpdfstring{$\K(X)$}{K(X)}-space (\Cref{thm:FindingLinearSpace})}
\label{subsec:FindingLinearSpace}

To prove \Cref{thm:FindingLinearSpace}, we first prove a lemma showing that we can repeatedly refine any given $\K(X)$-space containing close by codewords to be of lower dimension while continuing to contain all close by codewords until we get to dimension $\leq 2/\eps$.

\begin{lemma}[Dimension reduction]
\label{lem:DimensionReduction}
Fix $0<\varepsilon<1/2$. Let $\mathbb{K}\subseteq\mathbb{F}$ be finite fields and let $\alpha_1,\ldots,\alpha_n\in\mathbb{F}$ be distinct, where $n$ is sufficiently large in terms of $\varepsilon$. Suppose $\sigma_1,\ldots,\sigma_g$ generate $\operatorname{Gal}(\mathbb{F}/\mathbb{K})$, where $g\ge 12\log_2(1/\varepsilon)/\varepsilon$, and for each $\sigma_\ell$, we have $|\{\alpha_i\mid \sigma_\ell (\alpha_i)\neq \alpha_i\}|\leq \lceil n/g\rceil$.

Given these points and generators, a received word $y\in\mathbb{F}^n$, an integer $1\le k\le n$, and a $\mathbb{K}(X)$-linear space $A\subseteq\mathbb{F}(X)$ of dimension greater than $2/\varepsilon$, given by a basis over $\K(X)$ consisting of polynomials in $\F[X]_{<k}$, one can compute in time $\poly(n,g,\log|\F|)$ a strict subspace $A'\subsetneq A$ containing every $f\in A\cap\mathbb{F}[X]_{<k}$ with at least $k+\varepsilon n$ agreements with $y$. Moreover, $A'$ has a basis of polynomials of degree less than $k$.
\end{lemma}

\begin{proof}
Put $\mathbb{L}=\mathbb{K}(X)$ and
$d=\lceil2/\varepsilon\rceil$.
Let $V=A\cap\mathbb{F}[X]_{<k}$ and choose
$u_1,\ldots,u_d$ from the given basis of $A$.
\Cref{lem:GaloisRowsAndMovement} gives automorphisms
$\tau_1,\ldots,\tau_d\in\Gal(\F/\K)$ such that
\[
  \det\!\bigl(\tau_i(u_a)\bigr)_{1\le i,a\le d}\ne 0.
\]

By the support bound in the proof of \Cref{lem:GaloisRowsAndMovement}, if
\[
  S_i=\{j:\tau_i(\alpha_j)\ne\alpha_j\},
  \qquad s_i=|S_i|,
\]
we have:
\[
  s_i\le \left\lceil\frac{n}{g}\right\rceil\log_2 d
  \le \frac{2n\log_2 d}{12\log_2(1/\varepsilon)/\varepsilon}
  \le \frac{\varepsilon n}{2}
\]
for sufficiently large $n$. Let $h=\lceil\varepsilon n\rceil$ and define
\[
  E_i(X)=\prod_{j\in S_i}(X-\alpha_j).
\]
Find polynomials $Q_i=E_iR_i$ of degree at most $h$, imposing the constraints
\begin{equation}
\label{eq:AutomorphismInterpolation}
  \sum_{i=1}^d Q_i(\alpha_j)\tau_i(y_j)=0
  \qquad (1\le j\le n).
\end{equation}
These are $n$ homogeneous linear equations over $\mathbb{F}$.
Since $s_i\le\varepsilon n/2\le h$, the number of unknown
coefficients in the $R_i$ is
\[
  \sum_{i=1}^d(h-s_i+1)
  \ge d\left(\frac{\varepsilon n}{2}+1\right)>n.
\]
Thus a nonzero tuple $(Q_1,\ldots,Q_d)$ can be found by Gaussian
elimination in time $\poly(n, \log |\F|)$.

Define the $\mathbb{L}$-linear map
\[
  \Psi:\mathbb{F}(X)\longrightarrow\mathbb{F}(X),
  \qquad
  \Psi(f)=\sum_{i=1}^d Q_i\tau_i(f).
\]
Suppose $f\in V$ agrees with $y$ at position $j$.
If $j\notin S_i$, then
\[
  \tau_i(f)(\alpha_j)
  =\tau_i\bigl(f(\alpha_j)\bigr)=\tau_i(y_j).
\]
If $j\in S_i$, the factor $E_i$ gives $Q_i(\alpha_j)=0$.
\Cref{eq:AutomorphismInterpolation} therefore implies
$\Psi(f)(\alpha_j)=0$ in either case, after summing over $i$.
If $f$ has at least $k+\varepsilon n$ agreements, then $\Psi(f)$
has at least $k+h>\deg \Psi(f)$ distinct roots. Hence, $\Psi(f)=0$.

Since the vectors $(\tau_i(u_1),\ldots,\tau_i(u_d))$, $i\in[d]$, are linearly independent over $\F(X)$ and the $Q_i$ are not all zero, $\Psi$ is nonzero on some $u_a\in A$. We may therefore return
\[
  A'=\operatorname{span}_{\mathbb{L}}
     \{f\in V:\Psi(f)=0\}\subseteq A\cap\ker\Psi\subsetneq A.
\]
This space contains every required polynomial and has a basis in
$\mathbb{F}[X]_{<k}$. This can also be computed in time $\poly(n, \log |\F|)$ as desired.
\end{proof}

\begin{proof}[Proof of \Cref{thm:FindingLinearSpace}]
Start with $A=\mathbb{F}(X)$, using a $\K$-basis of $\F$ as a constant basis over $\K(X)$. Its dimension is $[\mathbb{F}:\mathbb{K}]$. While $\dim_{\mathbb{K}(X)}A>2/\varepsilon$, apply \Cref{lem:DimensionReduction} and replace $A$ by $A'$. Every iteration strictly decreases the dimension and preserves all polynomials in the output list. Hence at most $[\mathbb{F}:\mathbb{K}]$ iterations produce the required space and the algorithm runs in time $\poly(n,g,\log|\F|)$.
\end{proof}

\section{Pruning low-dimensional spaces of Reed-Solomon codes}

Our goal is to prove the following theorem.

\label{sec:ExtractingList}
\listextraction*

We first prove the following warmup result.

\begin{theorem}\label{thm:rs-subspace-decoding}
Given $\eps>0$, a finite field $\F$, distinct evaluation points
$\alpha_1,\ldots,\alpha_n\in\F$, $y\in\F^n$, and an $\F$-linear space
$A\subseteq\F[X]_{<k}$ with $\dim A=d$, given by a basis, we give a deterministic
algorithm running in
\[
  \bigl(O(1/\eps)\bigr)^{O(d)}\cdot
  \widetilde O\!\left(n\,\poly(1/\eps,d,\log|\F|)\right)
\]
which outputs
\[
  \mathcal L:=\left\{f\in A:
    \bigl|\{i:f(\alpha_i)=y_i\}\bigr|\ge k+\eps n\right\},
  \qquad |\mathcal L|\le\bigl(O(1/\eps)\bigr)^d.
\]
\end{theorem}

The key ingredients in proving \cref{thm:rs-subspace-decoding}, are then used in the proof of \Cref{thm:ExtractingList} with various modifications.

\cite{KoppartyRSW2023, Tamo2024} already give a randomized algorithm for a more general version of \cref{thm:rs-subspace-decoding};
we do it deterministically. The previous best deterministic algorithm
is due to \cite{AshvinkumarHS2026} (but it is also more general), getting
parameter dependence $(2/\eps)^{O(2^d)} \tilde O(n\,\poly(\log |\F|))$ in comparison to our
$(O(1/\eps))^{O(d)}\tilde O(n\,\poly(\log |\F|))$.

\paragraph{Proof differences between \Cref{thm:rs-subspace-decoding} and \Cref{thm:ExtractingList}.}
The broad idea is similar, essentially replacing $\F$ with
$\mathbb K(X)$. The resulting issue is that coefficients have degree.
We will always control this degree using Nullstellensatz and linear
algebra!

We will use high extension degree to show that any nonzero $\mathbb K[X]$ polynomial with degree $<\eps\,C(d,\eps)\,n$ has at most $\eps n$ roots from our evaluation points. Comparable degrees are for simplicity in an internal interpolation step.

\paragraph{Proof method (describing only Theorem~\ref{thm:rs-subspace-decoding}).}
Choose a basis $f_1,\ldots,f_d$ of the initial linear space
\[
  A=\left\{\sum_i c_i f_i:c_i\in\F\right\}.
\]
Starting with $R=\{0\}$, we maintain a relation space
\[
  R\subseteq\F[Z_1,\ldots,Z_d]_{\le t},
  \qquad t\sim d/\eps.
\]
If $1\in R$, we stop. Thus $1\notin R$ whenever we continue.
We try to find $Q\in\F[Z_1,\ldots,Z_d]_{\le t}\setminus R$
such that if
\[
  \sum_i c_i f_i\in\mathcal L,
  \qquad R(c_1,\ldots,c_d)=0,
\]
then $Q(c_1,\ldots,c_d)=0$.

There are three failure cases. The first is
$R+\F=\F[Z_1,\ldots,Z_d]_{\le t}$. The second is that at least $k$ evaluation coordinates are
determined: after relabeling them,
\[
  \sum_{i=1}^d c_i f_i(\alpha_j)=y_j
  \qquad\text{for each }j\in[k]
  \text{ and all }(c_1,\ldots,c_d)\in Z(R).
\]
We can then interpolate the unique polynomial from the $k$ determined
coordinates and check whether it belongs to $\Enc(Z(R))\cap\mathcal L$.
In the third case, there exist
\[
  \ell\in\F[Z_1,\ldots,Z_d]_{=1}\setminus R,
  \qquad
  r\in\F[Z_1,\ldots,Z_d]_{\le t-1}\setminus R,
\]
such that $\ell r\in R$. We branch into $R\cup\{\ell\}$ and
$R\cup\{r\}$.

Naively, this gives an output-size bound of
$\exp(\dim\F[Z_1,\ldots,Z_d]_{\le t})$. The first observation is
that adding $\ell$ removes at least one remaining variable, which
strengthens the bound to
\[
  \bigl(\dim\F[Z_1,\ldots,Z_d]_{\le t}\bigr)^d.
\]
The second observation is that the failure condition is stronger:
there are savings in $R\cup\{r\}$ as well. This leads to
\[
  2^d\cdot\dim\F[Z_1,\ldots,Z_d]_{\le t}.
\]

As a corollary of our subspace-decoding of Reed-Solomon codes, we also improve the best known deterministic list-decoding algorithm run-time for folded Reed-Solomon codes.
\newcommand{\polylog}{\mathsf{polylog}}
\begin{corollary}[Deterministic FRS decoding]\label[corollary]{cor:frs}
Let $0<\eps<1-R$, put $m=\lceil2/\eps\rceil$ and $s=m^2$,
and consider the $s$-folded Reed-Solomon code. There is a deterministic algorithm that runs
in time
\[
  (2/\eps)^{O(1/\eps)}\cdot\widetilde O\!\left(n\,\polylog(q)\right)
\]
and, given a received word in $(\F_q^s)^n$, outputs all messages agreeing with it on at least $(R+\eps)n$ coordinates.
\end{corollary}
\begin{proof}
We use the standard linear-interpolation stage for FRS codes,
with its fast implementation in \cite{GoyalHKS2024}.
For parameters $m\le s$, this stage computes a linear space of
message polynomials of dimension at most $m$ containing every
message within relative distance
\[
  1-R-\eps\leq \frac{m}{m+1}\left(1-\frac{sR}{s-m+1}\right)
\]
of the received word due to the choice of $m, s$. 
Thus this linear space contains the entire desired folded list and has dimension $d=O(1/\eps)$.

Unfold the received word into $N=sn$ scalar coordinates. Every desired message agrees with it on at least
\[
  s(R+\eps)n=k+\eps N
\]
scalar coordinates. Apply \Cref{thm:rs-subspace-decoding} to this unfolded word within the computed linear space and compute all close by codewords. Finally, retain only the returned messages agreeing on at least $(R+\eps)n$ coordinates of the original folded Reed-Solomon code. 
\end{proof}

\subsection{Pruning for $\F$-subspaces of Reed-Solomon codes (\cref{thm:rs-subspace-decoding})}

We now follow the proof strategy described above. But, we first set up notation.

Fix a basis $f_1,\ldots,f_d$ of $A\subseteq\F[X]_{<k}$ and distinct evaluation points $\alpha_1,\ldots,\alpha_n\in\F$. We maintain an $\F$-linear relation space $R\subseteq\F[Z_1,\ldots,Z_d]_{\le t}$, initially $R=\{0\}$.

For $0\le s\le t$, let
$M_s=\{Z^a:\wt(a)\le s\}$ and $S_s=\Span(M_s)$,
where $\wt(a)=a_1+\cdots+a_d$ is the total degree.

\begin{definition}
We say that $R\subseteq S_t$ is \textit{ready} if
\[
  (R\cap S_{t-1})Z_a\subseteq R
  \qquad\text{for every }a\in[d].
\]

Write $\Ready(R) = R'$ for the space obtained by closing the space of relations $R$ under the above operation.
All the added relations in $R'$ also vanish on $Z(R)$, and hence $Z(R)=Z(R')$.
\end{definition}

\begin{notation}
The relation space $R$ is stored as a matrix in
$\F^{\dim R\times M_t}$, where the $M_t$ monomials are ordered by
decreasing total degree, with arbitrary tie-breaking. The pivot
monomials of $R$ are the $\dim R$ monomials with a $1$ entry in
the corresponding relation and no entry in any preceding monomial.
The remaining monomials are
$P_t=M_t\setminus\{\text{pivots}\}$. For $0\le s\le t$, write
$P_s=P_t\cap M_s$ and $W_s=\Span(P_s)$. Row reduction gives
$S_s=(R\cap S_s)+W_s$, with $(R\cap S_s)\cap W_s=\{0\}$.
\end{notation}

\begin{definition}
Let $\ell_j$ be the relation
\[
  \ell_j=\left(\sum_{i=1}^d f_i(\alpha_j)Z_i\right)-y_j.
\]
A coordinate $j\in[n]$ is determined if $\ell_j\in R$.
Let $J=\{j\in[n]:\ell_j\notin R\}$ be the undetermined
coordinates.
\end{definition}

\begin{notation}
$Z(R)$ denotes the common zeros of $R$ in $\F^d$, and
$\Enc(c_1,\ldots,c_d)=\sum_i c_i f_i$.
\end{notation}

\begin{claim}\label[claim]{clm:counting}For every \textit{ready} $R$,
\[
  \frac{d|P_t|}{d+t}\ge |P_t|-|P_{t-1}|.
\]
\end{claim}
\begin{proof}
For every $s\le t$, the number of remaining monomials of degree at
most $s$ is
\[
  |P_s|=\dim S_s-\dim(R\cap S_s).
\]
These counts do not depend on how ties are broken. For this proof
only, break ties lexicographically.

If a pivot monomial has degree less than $t$, multiplying its
corresponding relation by any $Z_j$ gives another relation in $R$,
since $R$ is \textit{ready}. Its pivot is the original pivot
multiplied by $Z_j$. Thus every divisor of a remaining monomial
is also remaining.

For $2\le s\le t$, let $N_s$ be the number of triples $(Z^{\bar a}, Z_i, b_i)$ such that $Z^{\bar a} \in P_s\setminus P_{s-1}$, $Z_i^{b_i}\mid Z^{\bar a}$ and $1\le b_i\le a_i$. Note that $N_s = s(|P_s|-|P_{s-1}|)$.

We can also count $N_s$ starting from $Z^{\bar a}/Z_i \in P_{s-1}\setminus P_{s-2}$ and then choosing $Z_i, b_i$. Observe that there are at most $s+d-1$ choices for the pair $Z_i, b_i$ given $Z^{\bar a}/Z_i$. Hence
\[
  s\bigl(|P_s|-|P_{s-1}|\bigr)
  \le(s+d-1)\bigl(|P_{s-1}|-|P_{s-2}|\bigr).
\]
At $s=1$, the same count gives
$|P_1|-|P_0|\le d|P_0|$. It follows that
\[
  \frac{|P_s|-|P_{s-1}|}{\binom{d+s-1}{s}}
  \ge\frac{|P_t|-|P_{t-1}|}{\binom{d+t-1}{t}}
  \qquad(1\le s\le t),
\]
and $|P_0|$ is at least the expression on the right. Therefore
\begin{align*}
  |P_t|
  &=|P_0|+\sum_{s=1}^t\bigl(|P_s|-|P_{s-1}|\bigr)\\
  &\ge\frac{|P_t|-|P_{t-1}|}{\binom{d+t-1}{t}}
    \left(\sum_{s=0}^t\binom{d+s-1}{d-1}\right) = \frac{|P_t|-|P_{t-1}|}{\binom{d+t-1}{t}}
    \binom{d+t}{d}\\
  &=\frac{d+t}{d}\bigl(|P_t|-|P_{t-1}|\bigr) \ . \qedhere
\end{align*}
\end{proof}

\subsubsection{Interpolate a new relation or branch}
The goal is now to interpolate a new relation $Q\notin R$.

\begin{lemma}\label[lemma]{lem:refinement}
Let $t=\lfloor2d/\eps\rfloor$, and let $R\subseteq S_t$ be
\textit{ready}. Then one of the following holds:
\begin{itemize}
  \item We find $Q\in S_t\setminus R$ such that
  $Q(c_1,\ldots,c_d)=0$ whenever
  $(c_1,\ldots,c_d)\in Z(R)$ and $\Enc(c_1,\ldots,c_d)\in\cL$.
  \item $\F+R=S_t$.
  \item There are at least $k$ determined coordinates.
  \item There exists $j\in J$ such that
  \[
    \dim(\ell_j W_{t-1}\cap R)\ge\frac{\eps|P_t|}{2}.
  \]
\end{itemize}
\end{lemma}

\begin{proof}
First check the two terminal cases. If $\F+R=S_t$, the linear
relations fix every coordinate of a possible common zero. Solve
these relations and test the resulting tuple against $R$ and
the agreement requirement. If at least $k$ coordinates are determined,
interpolate the unique polynomial of degree less than $k$ agreeing
there, then check whether it belongs to $\Enc(Z(R))\cap\cL$.
Either case returns at most one message polynomial.

Otherwise, let $D=\lfloor\eps n\rfloor+1$ and interpolate a
nonzero polynomial
\[
  E(X,Z_1,\ldots,Z_d)=\sum_{m\in P_t}E_m(X)\,m,
\]
such that $\deg_X E_m<D$ for all $m\in P_t$, with constraints
\[
  \text{for every }j\in J,\qquad
  E(\alpha_j,\vec Z)\in(\ell_j W_{t-1}+R)\cap W_t.
\]
The polynomials $E_m$ have $D|P_t|$ unknown field coefficients.
The membership conditions give a homogeneous linear system with
\[
  |J|\,|P_t|
  -\sum_{j\in J}\dim\bigl((\ell_j W_{t-1}+R)\cap W_t\bigr)
\]
equations. For $j\in J$, the polynomial $\ell_j$ is nonzero. Multiplication
by $\ell_j$ is therefore injective, and reduction along $R$
into $W_t$ gives
\[
  \dim\bigl((\ell_j W_{t-1}+R)\cap W_t\bigr)
  =\dim W_{t-1}-\dim(\ell_j W_{t-1}\cap R).
\]
Thus the number of equations is at most
\begin{align*}
  &n\bigl(|P_t|-|P_{t-1}|\bigr)
    +\sum_{j\in J}\dim(\ell_j W_{t-1}\cap R)\\
  &\qquad\le n\,\frac{d}{d+t}|P_t|
    +\sum_{j\in J}\dim(\ell_j W_{t-1}\cap R).
\end{align*}

Our choice $t=\lfloor2d/\eps\rfloor$ gives
$d/(d+t)<\eps/2$.

\paragraph{Case 1.}
\[
  \frac{\eps n}{2}|P_t|
  >\sum_{j\in J}\dim(\ell_j W_{t-1}\cap R).
\]
The number of equations is strictly less than $\eps n|P_t|$,
whereas the number of unknown field coefficients is
\[
  D|P_t|>\eps n|P_t|.
\]
We can therefore find a nonzero $E$ by solving this homogeneous
linear system.
If $(c_1,\ldots,c_d)\in Z(R)$ satisfies
\[
  \sum_i c_i f_i(\alpha_j)=y_j
  \qquad\text{and}\qquad \ell_j\notin R,
\]
then $E(\alpha_j,c_1,\ldots,c_d)=0$.
Since fewer than $k$ coordinates are determined, either
\begin{align*}
  &\left|\{j\in[n]:\ell_j\in R\}
    \cup\{j\in[n]:E(\alpha_j,c_1,\ldots,c_d)=0\}\right|\\
  &\qquad\le |\{j\in[n]:\ell_j\in R\}|+(D-1)\\
  &\qquad\le k+D-2<k+\eps n,
\end{align*}
or $E(X,c_1,\ldots,c_d)=0$ identically.
Therefore, for every $f\in\Enc(Z(R))\cap\cL$,
$E(X,c_1,\ldots,c_d)=0$.
Write
\[
  E=\sum_{j=0}^{D-1}r_jX^j.
\]
Every coefficient $r_j$ vanishes at every $c\in Z(R)$ with
$\Enc(c)\in\cL$. Moreover, $r_j\in W_t$, and at least one $r_j$ is
nonzero because $E\ne0$. Since $W_t\cap R=\{0\}$, any such
nonzero coefficient is a new relation outside $R$. Replace $R$ with
\[
  \Ready\bigl(\Span(R\cup\{r_0,\ldots,r_{D-1}\})\bigr).
\]

\paragraph{Case 2.}
The Case~1 inequality fails. Since $|J|\le n$, averaging gives
$j\in J$ such that
\[
  \dim(\ell_j W_{t-1}\cap R)\ge\frac{\eps|P_t|}{2}.
\]
Branch into
\[
  \Ready\bigl(\Span(R\cup\{\ell_j\})\bigr)
\]
and
\[
  \Ready\bigl(\Span(R\cup\{p\in W_{t-1}:p\ell_j\in R\})\bigr).
\]
These branches cover every $c\in Z(R)$. If $\ell_j(c)=0$,
$c$ lies in the first branch. Otherwise, each relation
$p\ell_j\in R$ forces $p(c)=0$, so it lies in the second branch.
Taking spans and making the spaces \textit{ready} preserves these
conclusions.
\end{proof}

\subsubsection{Bounding the Output List-Size}
\begin{claim}\label[claim]{clm:list}
There are $|P_1|-1$ remaining variables, and
\[
  |\cL\cap\Enc(Z(R))|\le 2^{|P_1|-1}|P_t|.
\]
\end{claim}
\begin{proof}
Since $1\notin R$, the constant monomial is remaining. Row reduction
of the affine linear relations expresses the other pivot monomials
of degree one in terms of the remaining variables. Thus their number
is $|P_1|-1$.

We induct on $|P_t|$. The terminal cases return at most one
message polynomial, which satisfies the claimed bound. In particular, if
$|P_t|=1$, then $R+\F=S_t$, so this is a terminal case.
If a new relation is found, adding it strictly decreases $|P_t|$
and cannot increase the number of remaining variables. The induction
hypothesis then gives the bound.

Otherwise, use the two branches from Case~2. The first adds the
independent affine linear relation $\ell_j\notin R$, so it removes
at least one remaining variable unless it contains $1$ and is
discarded. Once made \textit{ready}, it contains
$R+\ell_jW_{t-1}$. Its remaining dimension is therefore at most
\[
  |P_t|-|P_{t-1}|+\dim(\ell_jW_{t-1}\cap R).
\]
The space $\{p\in W_{t-1}:p\ell_j\in R\}$ has dimension
$\dim(\ell_jW_{t-1}\cap R)$, because multiplication by $\ell_j$
is injective. It meets $R$ only in zero. Thus the second branch
has remaining dimension at most
\[
  |P_t|-\dim(\ell_jW_{t-1}\cap R).
\]
Both dimensions are strictly less than $|P_t|$: the first branch
adds $\ell_j\notin R$, and the second adds a space of positive
dimension disjoint from $R$.

By the branch coverage proved above and the induction hypothesis,
the total list size is at most
\begin{align*}
  &2^{|P_1|-2}
    \bigl(|P_t|-|P_{t-1}|+\dim(\ell_jW_{t-1}\cap R)\bigr)\\
  &\quad+2^{|P_1|-1}
    \bigl(|P_t|-\dim(\ell_jW_{t-1}\cap R)\bigr)\\
  &=2^{|P_1|-1}|P_t|\\
  &\quad+2^{|P_1|-2}
    \bigl(|P_t|-|P_{t-1}|-\dim(\ell_jW_{t-1}\cap R)\bigr)\\
  &\le2^{|P_1|-1}|P_t|,
\end{align*}
since Claim~\ref{clm:counting} and the choice of $t$ give
\[
  |P_t|-|P_{t-1}|
  \le\frac{d}{d+t}|P_t|
  <\frac\eps2|P_t|
  \le\dim(\ell_jW_{t-1}\cap R) \ . \qedhere
\]
\end{proof}

Initially $R=\{0\}$ and $P_t=M_t$, so the list size is at most
\[
  2^d|M_t|=2^d\binom{d+t}{d}
  \le\bigl(O(1/\eps)\bigr)^d.
\]

\paragraph{Running time.}
Observe that checking whether $j\in J$ by row reduction takes
$\poly(|M_t|,\log|\F|)$ time, given the values $f_i(\alpha_j)$.
Thus computing $J$ takes $n\,\poly(|M_t|,\log|\F|)$ time.
Similarly, for each $j\in J$, we can compute
$\dim(\ell_jW_{t-1}\cap R)$ and a basis for
$\{p\in W_{t-1}:p\ell_j\in R\}$ in
$\poly(|M_t|,\log|\F|)$ time. Replacing $R$ with
$\Ready(R)$ has the same cost.
The basis evaluations, terminal interpolation, and agreement checks
take $\widetilde O(n)\poly(|M_t|,\log|\F|)$ time using fast
univariate evaluation and interpolation.

It remains to interpolate $E$ within this bound when
\[
  \frac{\eps n}{2}|P_t|
  >\sum_{j\in J}\dim(\ell_jW_{t-1}\cap R).
\]
We use the following specialization of the fast interpolation theorem.

\begin{theorem}[Fast interpolation~{\cite[Theorem~1.5]{JeannerodNSV2017}}]
\label{thm:FastLinearInterpolation}
Let $\F$ be a finite field, let $u\ge1$ and $v\ge0$ be integers,
and let $\beta_1,\ldots,\beta_v\in\F$ and
$a_1,\ldots,a_v\in\F^u$. The points $\beta_h$ need not be distinct.
There is a deterministic algorithm that finds a nonzero tuple
$(q_1,\ldots,q_u)\in\F[X]^u$ satisfying
\[
  \sum_{i=1}^u a_{h,i}q_i(\beta_h)=0
  \qquad(h\in[v])
\]
and minimizing $\max_i\deg q_i$ among all such tuples, using
$\widetilde O\bigl(u^2(v+u)\bigr)$ field operations.
\end{theorem}

For each $j\in J$, row reduction writes the membership condition
\[
  E(\alpha_j,\vec Z)\in(\ell_jW_{t-1}+R)\cap W_t
\]
as $c_j\le|P_t|$ equations
\[
  \sum_{m\in P_t}a_{j,h,m}E_m(\alpha_j)=0
  \qquad(h\in[c_j]),
\]
with coefficients $a_{j,h,m}\in\F$.
Apply \Cref{thm:FastLinearInterpolation} with $u=|P_t|$,
$v=\sum_{j\in J}c_j\le n|P_t|$, and one copy of $\alpha_j$
for each of its $c_j$ equations. Case~1 of Lemma~\ref{lem:refinement}
gives a nonzero solution with $\deg E_m<D$, so the minimum-degree
solution has this property. The cost is
$\widetilde O(n|P_t|^3)$ field operations.
Extracting the coefficient relations and adjoining them to $R$
also takes $n\,\poly(|M_t|,\log|\F|)$ time.

The induction in Claim~\ref{clm:list} also bounds the number of terminal
nodes by $2^d|M_t|$, whether or not they return a message polynomial;
children containing $1$ are discarded at their parent. Every path
has at most $|M_t|$ strict enlargements of $R$, so there are at
most $2^d|M_t|^2$ steps. Since $|M_t|=(O(1/\eps))^d$, the total
running time is
\[
  \bigl(O(1/\eps)\bigr)^{O(d)}\cdot
  \widetilde O\!\left(n\,\poly(1/\eps,d,\log|\F|)\right).
\]
This proves Theorem~\ref{thm:rs-subspace-decoding}.

\subsection{Pruning $\mathbb K(X)$-subspaces (\cref{thm:ExtractingList})}

We follow the proof of \Cref{thm:rs-subspace-decoding}, now over
$\mathbb L=\K(X)$. The coefficients $c_i(X)$ representing a
message polynomial lie in $\K[X]$, and the relations have coefficients in
$\mathbb L$. We still try to add a new relation, and branch when
interpolation fails. The additional work is to control coefficient
degrees and to pass from relations at an evaluation point to
relations over $\mathbb L$.

A space of rational dimension zero contains only the zero polynomial. We use $d$
for the actual rational dimension in this proof; applying the bound
with this dimension implies the bound in the theorem.

In the previous subsection, choosing a basis gave constant
coefficients $c_i\in\F$. Here we use a reduced polynomial basis
to obtain coefficients $c_i(X)\in\K[X]$ with controlled
degrees. To do this, first work in the polynomial module
\[
  \mathcal A=A\cap\F[X].
\]

\begin{lemma}[Reduced polynomial basis (\Cref{cor:ReducedPolynomialBasis})]\label[lemma]{lem:rational-basis}
The module $\mathcal A$ has a
$\K[X]$-basis $f_1,\ldots,f_d$ such that
\[
  \deg\!\left(\sum_{i=1}^d c_i(X)f_i(X)\right)
  =\max_{i\in[d]}\bigl(\deg c_i+\deg f_i\bigr).
\]
In particular,
\[
  \mathcal A\cap\F[X]_{<k}
  =\left\{\sum_{i=1}^d c_i(X)f_i(X):
    c_i\in\K[X]_{<k-\deg f_i}\right\}.
\]

Additionally, given a basis $g_1, \dots, g_d$ of $A$, $f_1, \dots, f_d$ can be computed in $\poly(d, \max\deg g_i, n, \log |\F|)$ time. 
\end{lemma}
\begin{proof}
The basis properties follow from \Cref{cor:ReducedPolynomialBasis}.
For computation, put $D=\max_i\deg g_i$ and identify $\F[X]$
with $\K[X]^{[\F:\K]}$. Linear algebra over $\K(X)$, followed
by comparison of coefficients, computes a $\K$-basis of
$A\cap\F[X]_{\le D}$. The predictable-degree property implies
that every element of a reduced module basis has degree at most
$D$, since the $g_i$ span $A$ over $\K(X)$. Thus the computed
polynomials generate $\mathcal A$ over $\K[X]$, and we apply
\Cref{thm:ReducedBases}.
\end{proof}

\begin{notation}
For $c=(c_1,\ldots,c_d)\in\mathbb L^d$, write
$\Enc(c)=\sum_i c_i f_i$. If $\Enc(c)\in\F[X]_{<k}$, then
$c_i\in\K[X]_{<k-\deg f_i}$; in particular, every $c_i$
has degree less than $k$. Set
\[
  t=\lfloor16d/\eps\rfloor,\qquad
  M_s=\{Z^a:\wt(a)\le s\},\qquad
  S_s=\Span_{\mathbb L}(M_s).
\]
We maintain an $\mathbb L$-linear relation space $R\subseteq S_t$,
starting with $R=\{0\}$, and discard a branch if $1\in R$.
Here $Z(R)$ denotes the common zeros of $R$ in $\mathbb L^d$.
For a polynomial tuple $c(X)$, membership in $Z(R)$ means that
each relation vanishes identically after substituting $Z_i=c_i(X)$.
\end{notation}

The definition of \textit{ready} is unchanged, with spans now
taken over $\mathbb L$. We use the same monomial order, pivot
monomials, and remaining monomials $P_s$, and put
$W_s=\Span_{\mathbb L}(P_s)$. In particular,
\[
  S_s=(R\cap S_s)+W_s,\qquad (R\cap S_s)\cap W_s=\{0\}.
\]
\Cref{clm:counting} used only row reduction and closure under
multiplication. Its proof therefore applies over $\mathbb L$ and gives
\begin{equation}\label{eq:rational-counting}
  |P_t|-|P_{t-1}|
  \le\frac{d}{d+t}|P_t|<\frac\eps{16}|P_t|.
\end{equation}

\begin{definition}
Store $R$ using a row-echelon basis over $\mathbb L$, with
denominators cleared in each row. Thus its stored matrix has
entries in $\K[X]$, and its pivot entries need not be $1$.
Write $\deg R$ for the largest $X$-degree of the entries in this
representation, with $\deg\{0\}=0$.
We say that $R$ is \textit{$H$-prepared} if it is
\textit{ready} and has such a representation of degree at most $H$.
\end{definition}

Preparation plays the role of making the relation space
\textit{ready} in the previous subsection, but also controls the
degrees needed for evaluation. It only needs to preserve the
polynomial tuples under consideration, rather than every zero in
$\mathbb L^d$.

\begin{lemma}[Preparation]\label[lemma]{lem:rational-preparation}
Let $B(d,t)=|M_t|t$. Given $R\subseteq S_t$, we can compute
$R'\supseteq R$ with the following properties:
\begin{itemize}
  \item $R'$ is \textit{$B(d,t)k$-prepared}.
  \item It preserves the polynomial zeros under consideration:
  \[
    Z(R')\cap\bigl(\K[X]_{<k}\bigr)^d
    =Z(R)\cap\bigl(\K[X]_{<k}\bigr)^d.
  \]
  \item The computation takes time
  $\poly(n,\deg R,|M_t|,\log|\F|)$.
\end{itemize}
\end{lemma}
\begin{proof}
For $c\in Z(R)\cap(\K[X]_{<k})^d$ and $r\in R$,
\[
  \sum_{m\in M_t}r_m(X)m(c)=0,
  \qquad \deg m(c)<tk.
\]
Let
\[
  U=\Span_{\mathbb L}\left\{
    u\in(\K[X]_{<tk})^{M_t}:u\cdot r=0
    \text{ for every }r\in R\right\},
  \qquad R'=U^\perp.
\]
Then $R\subseteq R'$. For every such $c$, the vector
$(m(c))_{m\in M_t}$ lies in $U$, so $c\in Z(R')$.
The reverse inclusion follows from $R\subseteq R'$, proving
preservation of the stated polynomial zeros. Computing the
bounded-degree solutions by coefficient comparison and selecting
an $\mathbb L$-basis takes
$\poly(\log|\K|,tk,\deg R,|M_t|)$ time.

Write this basis as the rows of a matrix $V$ with $D=\dim_{\mathbb L}U$
rows and entries of degree less than $tk$. In the fixed monomial
order, choose independent columns greedily from right to left.
Let $\mathcal I$ be their indices and put
$\mathcal J=M_t\setminus\mathcal I$.
For each $j\in\mathcal J$, solve
\[
  V_{\mathcal I}f_j=-V_j,
  \qquad f_j\in\mathbb L^{\mathcal I}.
\]
The vector with coordinate $1$ at $j$, coordinates as $f_j$ on $\mathcal I$, and zero elsewhere lies in $U^\perp$. Every nonzero coordinate of $f_j$ is to the right of $j$, so these vectors, ordered by $j$, form a row-echelon basis of $R'$. Multiplying each row by $\det V_{\mathcal I}$ converts entries from $\K(X)$ elements to $\K[X]$. By Cramer's rule, the resulting entries have degree at most $Dtk\le |M_t|tk$, giving the required degree bound.

If $R'$ is not ready, adjoin a product $Z_i r\notin R'$ of a
stored row $r\in R'\cap S_{t-1}$ and repeat the construction.
This preserves the stated polynomial zeros and strictly increases
the dimension. The added row has $X$-degree at most $B(d,t)k$,
so each repetition takes $\poly(|M_t|,tk,\log|\F|)$ time.
After at most $|M_t|$ repetitions we obtain a $B(d,t)k$-prepared
space in $\poly(n,\deg R,|M_t|,\log|\F|)$ time.
\end{proof}

\begin{notation}
Write
\[
  \delta_j=[\K(\alpha_j):\K],\qquad
  \delta_{\min}=\min_j\delta_j,\qquad
  \delta_{\max}=\max_j\delta_j.
\]
The hypotheses give $\delta_{\min}\ge C(d,\eps)$ and
$\delta_{\max}\le2\delta_{\min}$.
\end{notation}

\begin{proposition}\label[proposition]{clm:rational-roots}
    For every nonzero $g\in\K[X]$,
\[
  \sum_{j:g(\alpha_j)=0}\delta_j\le\deg g.
\]
\end{proposition}

\begin{proof}
If $g(\alpha_j)=0$, the minimal polynomial of $\alpha_j$ over
$\K$ divides $g$. Non-conjugacy means that these minimal
polynomials are distinct irreducible polynomials. Their product
therefore divides $g$, giving the inequality.
\end{proof}

\begin{notation}
Let $R(\alpha_j)$ be the $\K(\alpha_j)$-span of the evaluations
of the stored polynomial rows of $R$, and let
$W_s(\alpha_j)=\Span_{\K(\alpha_j)}(P_s)$.
Let $\mathcal B$ be the set of coordinates at which some pivot
entry of the stored matrix vanishes. Coordinates outside
$\mathcal B$ are called regular.
\end{notation}

\begin{claim}\label[claim]{clm:rational-pivots}
If the stored matrix has degree at most $H$, then
\[
  |\mathcal B|\le\frac{|M_t|H}{\delta_{\min}}.
\]
At every regular coordinate, the pivot monomials are unchanged,
and $W_s(\alpha_j)$ is a complement of
$R(\alpha_j)\cap\K(\alpha_j)[\vec Z]_{\le s}$.
\end{claim}
\begin{proof}
The product of the pivot entries is a nonzero polynomial of
degree at most $(\dim R)H\le|M_t|H$. It vanishes at every
coordinate in $\mathcal B$, so \Cref{clm:rational-roots}
gives the bound. At any other coordinate, the evaluated matrix
is still in row-echelon form with the same pivots. The complement
statement then follows by row reduction, just as before.
\end{proof}

For the estimates below, choose $C(d,\eps)$ large enough that
\begin{equation}\label{eq:rational-degree-choice}
  C(d,\eps)\ge\frac{16|M_t|^2t}{\eps}.
\end{equation}
At a prepared node, $H\le B(d,t)k$, so $k\le n$ gives
$|\mathcal B|\le\eps n/16$.

The single equation $\ell_j$ from the previous subsection is
replaced by a space of affine linear equations. This is because
$c_i(\alpha_j)\in\K(\alpha_j)$, whereas $f_i(\alpha_j)$ and
$y_j$ lie in $\F$.

\begin{definition}
Expand the equation
\[
  \sum_{i=1}^d f_i(\alpha_j)Z_i-y_j=0
\]
in a basis of $\F$ over $\K(\alpha_j)$, and let $L_j$ be the
$\K(\alpha_j)$-span of the resulting affine linear forms. Thus
for $z\in\K(\alpha_j)^d$, all forms in $L_j$ vanish at $z$ if
and only if $\sum_i f_i(\alpha_j)z_i=y_j$.
We say that $j$ is determined if $L_j\subseteq R(\alpha_j)$,
and put $J=\{j:L_j\nsubseteq R(\alpha_j)\}$.
\end{definition}

A polynomial tuple $c\in Z(R)$ specializes to a zero of
$R(\alpha_j)$, so $\Enc(c)$ agrees with $y$ at
each determined coordinate. Thus $k$ determined coordinates
leave at most one message polynomial of degree less than $k$.

\subsubsection{Finding a new relation}
We now repeat the interpolation argument of
\Cref{lem:refinement}. At a regular undetermined coordinate,
choose any $\ell_j\in L_j\setminus R(\alpha_j)$ and put
\[
  a_j=\dim_{\K(\alpha_j)}
    \bigl(\ell_jW_{t-1}(\alpha_j)\cap R(\alpha_j)\bigr).
\]
The interpolation count uses them in exactly the same way as before.

\begin{lemma}[Interpolation]\label[lemma]{lem:rational-interpolation}
Suppose $R$ is $B(d,t)k$-prepared and~\cref{eq:rational-degree-choice}
holds. Then one of the following holds:
\begin{itemize}
  \item We find $E\in S_t\setminus R$ vanishing identically at
  every tuple $c(X)\in Z(R)$ with $\Enc(c)\in\cL$.
  \item $\mathbb L+R=S_t$.
  \item There are at least $k$ determined coordinates.
  \item We have
  \[
    \sum_{j\in J\setminus\mathcal B}a_j
    \ge\frac{\eps n}{8}|P_t|.
  \]
\end{itemize}
\end{lemma}

\begin{proof}
The two terminal cases are handled as in
\Cref{lem:refinement}. If $\mathbb L+R=S_t$, solve for
the unique possible tuple over $\mathbb L$, then check whether it
encodes a nearby message polynomial. If at least
$k$ coordinates are determined, interpolate and test the unique
possible message. Henceforth assume neither case holds.

Set
\[
  D=\left\lfloor\frac{\delta_{\min}\eps n}{2}\right\rfloor+1
\]
and seek a nonzero polynomial
\[
  E(X,\vec Z)=\sum_{m\in P_t}E_m(X)m,
  \qquad E_m\in\K[X]_{<D}.
\]
At every $j\in J\setminus\mathcal B$, impose
\[
  E(\alpha_j,\vec Z)\in
  \bigl(\ell_jW_{t-1}(\alpha_j)+R(\alpha_j)\bigr)
    \cap W_t(\alpha_j).
\]
At every $j\in J\cap\mathcal B$, impose the stronger condition
$E(\alpha_j,\vec Z)=0$. This avoids needing a complement at
the exceptional coordinates.

There are $D|P_t|$ unknown coefficients over $\K$.
At a regular coordinate, multiplication by the nonzero form
$\ell_j$ is injective. As in \Cref{lem:refinement}, the
membership condition imposes
$|P_t|-|P_{t-1}|+a_j$ equations over $\K(\alpha_j)$, hence at
most $\delta_j(|P_t|-|P_{t-1}|+a_j)$ equations over $\K$.
At an exceptional coordinate, there are at most $\delta_j|P_t|$
equations over $\K$. The total is therefore at most
\[
  \delta_{\max}\left(
    n\bigl(|P_t|-|P_{t-1}|\bigr)
    +\sum_{j\in J\setminus\mathcal B}a_j
    +|\mathcal B|\,|P_t|\right).
\]
If the last outcome of the lemma does not hold, this is strictly
less than
\[
  2\delta_{\min}
  \left(\frac{\eps n}{16}+\frac{\eps n}{8}
              +\frac{\eps n}{16}\right)|P_t|
  =\frac{\delta_{\min}\eps n}{2}|P_t|
  <D|P_t|.
\]
Thus the homogeneous linear system has a nonzero solution.

For $c(X)\in Z(R)$ with $\Enc(c)\in\cL$, consider
$E(X,c_1(X),\ldots,c_d(X))\in\K[X]$. Its degree is at most
\[
  D-1+t(k-1)
  \le\frac{\delta_{\min}\eps n}{2}+tn
  \le\frac{3\delta_{\min}\eps n}{4}.
\]
At every agreement coordinate in $J$, this polynomial vanishes:
at a regular coordinate, both $\ell_j$ and $R(\alpha_j)$ vanish
at $c(\alpha_j)$; at an exceptional coordinate, we imposed
$E(\alpha_j,\vec Z)=0$. Fewer than $k$ coordinates are
determined, so there are more than $\eps n$ such roots.
\Cref{clm:rational-roots} forces
$E(X,c_1(X),\ldots,c_d(X))=0$ identically.

Finally, $E$ is a nonzero element of $W_t$, so $E\notin R$.
We add $E$ and prepare the enlarged space using
\Cref{lem:rational-preparation}. Here we add $E$ itself
as a relation over $\mathbb L$. Since the $c_i$ depend on $X$,
its individual coefficients in $X$ need not vanish after
substitution, as they did in the previous subsection.
\end{proof}

\subsubsection{From interpolation failure to branching}
\label{subsubsec:Branching}
We now use the last outcome of Lemma~\ref{lem:rational-interpolation}
to find either a new relation or a branching form over $\mathbb L$.
First, we present two lemmas that will be useful in the proof
of our key theorem.

\begin{lemma}[Effective Nullstellensatz~{\cite[Corollary~1.9]{Kollar1988}}]
\label[lemma]{thm:EffectiveNullstellensatz}
Let $E$ be a field, let $a,D\ge1$, and let
$f_1,\ldots,f_s\in E[Y_1,\ldots,Y_a]$ satisfy
$\deg f_i\le D$. If the $f_i$ have no common zero over
$\overline E$, then there are $g_1,\ldots,g_s\in E[Y_1,\ldots,Y_a]$
such that
\[
  \sum_{i=1}^s g_if_i=1,
  \qquad \deg(g_if_i)\le 2(3D)^a\quad(1\le i\le s).
\]
\end{lemma}

\begin{restatable}[Computing a common zero]{lemma}{computingcommonzero}\label[lemma]{lem:computing-common-zero}
Let $\K$ be a finite field and $\mathbb L=\K(X)$. Given
$f_1,\ldots,f_s\in\K[X,Y_1,\ldots,Y_a]$ with
$\deg_Y f_j\le D$ and $\deg_X f_j\le H$, one can
deterministically decide whether they have a common zero over
$\overline{\mathbb L}$ and, if so, compute a finite extension
$E/\mathbb L$, given by defining polynomials and an
$\mathbb L$-basis, and a common zero $y\in E^a$, in time
\[
  O_{s,D,a}(1)\poly(H,\log|\K|).
\]
\end{restatable}
We prove \Cref{lem:computing-common-zero} in
Appendix~\ref{app:computing-common-zero}.

\begin{theorem}[Finding a relation or a branching form]\label{thm:rational-branching}
Suppose $R\subsetneq S_t$ is $B(d,t)k$-prepared and
\[
  \sum_{j\in J\setminus\mathcal B}a_j
  \ge\frac{\eps n}{8}|P_t|.
\]
Then we can compute one of the following:
\begin{itemize}
  \item A relation $Q\in S_t\setminus R$ vanishing on $Z(R)$.
  \item An affine linear form $\ell\in S_1\setminus R$, over
  $\mathbb L$, such that
  \[
    \dim_{\mathbb L}(\ell W_{t-1}\cap R)
    \ge\frac\eps{16}|P_t|.
  \]
\end{itemize}
\end{theorem}
\begin{proof}
Put $b=|P_t|$, $s=|P_{t-1}|$, and
$r=\lceil\eps b/16\rceil$. First we show that, over some
finite extension $E/\mathbb L$, there is a nonzero
$\ell\in\Span_E P_1$ for which
\begin{equation}\label{eq:rational-extension-kernel}
  \dim_E\{p\in\Span_E P_{t-1}:\ell p\in\Span_E R\}
  \ge r.
\end{equation}

Let
\[
  T=\{j\in J\setminus\mathcal B:a_j\ge r\}.
\]
For $j\in(J\setminus\mathcal B)\setminus T$, we have
$a_j\le r-1<\eps b/16$, and always $a_j\le b$. Hence
\[
  \frac{\eps nb}{8}
  \le\sum_{j\in J\setminus\mathcal B}a_j
  \le\frac{\eps nb}{16}+b|T|,
  \qquad |T|\ge\frac{\eps n}{16}.
\]
In particular $r\le s$.

Without loss of generality, $P_1=\{1,Z_1,\ldots,Z_a\}$. For
$Y=(Y_0,\ldots,Y_a)\in\mathbb L^{a+1}$, let
$A(Y)\in\mathbb L^{b\times s}$ be the matrix, in the bases
$P_{t-1}$ and $P_t$, of the linear map
\[
  A_Y:W_{t-1}\longrightarrow W_t,\qquad
  A_Y(p)=\left(Y_0+\sum_{i=1}^aY_i Z_i\right)p\pmod R.
\]
Here the remainder modulo $R$ is taken in $W_t$, so the entries
of $A(Y)$ are linear in $Y$.
Let $\Pi(X)$ be the product of the pivot entries of the stored
row-echelon matrix for $R$. Cramer's rule shows that
multiplying $A(Y)$ by $\Pi(X)$ clears its denominators. The
entries of $\Pi(X)A(Y)$ lie in $\K[X,Y]$, are linear in $Y$,
and have $X$-degree at most
\[
  |M_t|B(d,t)k=|M_t|^2tk.
\]
At a regular coordinate $j$, $\Pi(\alpha_j)\ne0$, so evaluating
$A(Y)$ at $X=\alpha_j$ gives the multiplication matrix modulo
$R(\alpha_j)$. Moreover, $R(\alpha_j)$ is ready, so replacing
$\ell_j$ by its nonzero remainder in $W_1(\alpha_j)$ does not
change $a_j$.

First assume that there is no nonzero $\ell$ satisfying
\cref{eq:rational-extension-kernel} over any finite extension
of $\mathbb L$. A nonzero coefficient vector $Y$ gives such
a form exactly when $A(Y)$ has rank at most $s-r$, or
equivalently, when all $(s-r+1)\times(s-r+1)$ minors of
$\Pi(X)A(Y)$ vanish. These minors therefore have no common
zero in $\overline{\mathbb L}^{a+1}\setminus\{0\}$. Their $Y$-degree
is at most $|M_t|$ and their $X$-degree is at most
$|M_t|^2B(d,t)k$.

For each $i\in\{0,\ldots,a\}$, set $Y_i=1$ and denote the
resulting nonzero minors by $F_{\lambda,i}$. They have no common zero, so
\Cref{thm:EffectiveNullstellensatz} gives polynomials
$G_{\lambda,i}$ over $\mathbb L$ such that
\[
  \sum_\lambda G_{\lambda,i}F_{\lambda,i}=1,
  \qquad
  \deg_Y\bigl(G_{\lambda,i}F_{\lambda,i}\bigr)
  \le 2(3|M_t|)^d.
\]
Equating coefficients of monomials in $Y$ gives a linear system
for the coefficients of the $G_{\lambda,i}$ over $\mathbb L$,
with at most
\[
  V=\binom{d+2(3|M_t|)^d}{d}
\]
rows and entries in $\K[X]$ of degree at most
$|M_t|^2B(d,t)k$. As in the proof of
Lemma~\ref{lem:rational-preparation}, Cramer's rule gives a solution
with a common nonzero denominator $D_i\in\K[X]$ of degree at most
$V|M_t|^2B(d,t)k$. Clearing denominators gives
$\widetilde G_{\lambda,i}\in\K[X,Y]$ such that
\[
  D_i(X)=\sum_\lambda
       \widetilde G_{\lambda,i}(X,Y)F_{\lambda,i}(X,Y).
\]
Consequently $D=\prod_{i=0}^aD_i$ is nonzero and
\[
  \deg D=\sum_{i=0}^a\deg D_i\le(d+1)V|M_t|^2B(d,t)k.
\]
For every $j\in T$, the remainder of $\ell_j$ is nonzero
and its multiplication matrix has kernel dimension at least
$r$. All these minors therefore vanish at that coefficient
vector. Rescaling a nonzero coefficient $Y_i$ to $1$ gives
$D_i(\alpha_j)=0$, and hence $D(\alpha_j)=0$.

It suffices to take
\begin{equation}\label{eq:rational-specialization-choice}
  C(d,\eps)\ge
  \frac{32|M_t|B(d,t)}{\eps}
    \bigl(1+(d+1)|M_t|V\bigr).
\end{equation}
Indeed, \Cref{clm:rational-roots} and $k\le n$ would then imply
\[
  |T|\le\frac{\deg D}{\delta_{\min}}
       \le\frac{\eps n}{32}.
\]
This contradicts $|T|\ge\eps n/16$. Thus the minor equations
with $Y_i=1$ have a common zero for some $i$. Apply
Lemma~\ref{lem:computing-common-zero} to these systems to compute
a finite extension $E/\mathbb L$ and
$y=(y_0,\ldots,y_a)\in E^{a+1}$ satisfying the minor equations,
with $y_i=1$ for some $i$. Set
\[
  \ell=y_0+\sum_{j=1}^a y_jZ_j.
\]
Then $\ell\ne0$ and satisfies \cref{eq:rational-extension-kernel}.

Let $e_1,\ldots,e_m$ be the computed $\mathbb L$-basis of $E$.
By row reduction, compute an $E$-basis $p_1,\ldots,p_v$ of
the kernel in \cref{eq:rational-extension-kernel}. Write
\[
  \ell=\sum_{u=1}^m e_u\ell_u,\qquad
  p_i=\sum_{u=1}^m e_up_{i,u},\qquad
  \ell_u\in W_1,\quad p_{i,u}\in W_{t-1},
\]
and put $U=\Span_{\mathbb L}\{p_{i,u}:i\in[v],u\in[m]\}$.
Since $\Span_E U$ contains $p_1,\ldots,p_v$,
\[
  \dim_{\mathbb L}U=\dim_E\Span_E U\ge v\ge r.
\]

For every $c\in Z(R)$, we have $\ell(c)p_i(c)=0$ for all $i$.
If $\ell(c)=0$, then $\ell_u(c)=0$ for every $u$.
Otherwise $p_{i,u}(c)=0$ for every $i,u$. Thus every product
$\ell_up_{i,w}$ vanishes on $Z(R)$.

Check by row reduction whether any such product lies outside
$R$; if so, return it as $Q$. Otherwise $\ell_uU\subseteq R$
for every $u$. Choose a nonzero $\ell_u\in W_1$. Then
$\ell_u\notin R$, and multiplication by $\ell_u$ is injective, so
\[
  \dim_{\mathbb L}(\ell_uW_{t-1}\cap R)
  \ge\dim_{\mathbb L}(\ell_uU)
  =\dim_{\mathbb L}U
  \ge r\ge\frac\eps{16}|P_t|.
\]
Thus, we can use $\ell_u$ for branching.

The main cost of the algorithm is computing $\ell$.
Each system has at most $d$ variables and $4^{|M_t|}$ equations,
of $Y$-degree at most $|M_t|$ and $X$-degree at most
$|M_t|^2B(d,t)k$. By Lemma~\ref{lem:computing-common-zero},
we can compute $\ell$ deterministically in time
\[
  O_{d,\eps}(1)\poly(B(d,t)k,\log|\F|).
\]
Thus, the algorithm can be implemented in time
$O_{d,\eps}(1)\poly(n,\log|\F|)$.
\end{proof}

We can now branch into
\[
  R\cup\{\ell_u\}
  \qquad\text{and}\qquad
  R\cup\{p\in W_{t-1}:p\ell_u\in R\}.
\]
We then prepare the resulting spaces and repeat until a single
possible message polynomial remains to be checked, as before.

\subsubsection{Bounding the list and the running time}
The induction in Claim~\ref{clm:list} applies, since at every
branching step,
\[
  |P_t|-|P_{t-1}|
  <\frac\eps{16}|P_t|
  \le\dim_{\mathbb L}(\ell W_{t-1}\cap R).
\]
Preparation only enlarges the relation spaces and preserves the
polynomial tuples under consideration. Thus
\[
  |\cL\cap\Enc(Z(R))|\le 2^{|P_1|-1}|P_t|.
\]
Initially this gives a list of size at most
\[
  2^d|M_t|=2^d\binom{d+t}{d}
  \le\bigl(O(1/\eps)\bigr)^d.
\]

Note that the choice in \cref{eq:rational-specialization-choice} satisfies $C(d,\eps)=\exp(O(d^3\log(1/\eps)))$. The dimension of $R$ strictly increases along each branch, so
the binary tree has depth at most $|M_t|$ and $O_{d,\eps}(1)$ nodes. The main cost at each node is finding a common zero as
in \Cref{thm:rational-branching}. The remaining operations take time $O_{d,\eps}(1)\poly(n,\log|\F|)$. Thus the total running time is
\[
  O_{d,\eps}(1)\poly(n,\log|\F|).
\]
This proves \Cref{thm:ExtractingList}.

\section{Effective Evaluation Points and Explicit Instantiation}
\label{sec:ExplicitInstantiation}

We begin with a simpler construction that can in fact be made strongly-explicit. This construction does not achieve effectiveness \cref{def:EffectivePoints}, but is still sufficient for our list-decoding application with list sizes $(1/\eps)^{O(1/\eps)}$ at a gap of $\eps$ to capacity. 

\begin{theorem}[Strongly explicit evaluation points]\label{thm:strongly-explicit-points}
    For any fixed integers $g, d\in \mathbb N$, for any finite field $\mathbb K$, there exists a deterministic algorithm such that on input $\K$, $g, d$, it runs in time $\poly(g, d, \log |\K|)$ and outputs irreducible polynomials $P_1 (X), \dots, P_g(X)\in \K[X]$ such that:
    \begin{itemize}
        \item $d\leq \deg P_i(X)\leq 2\deg P_j(X)\leq \poly(d,g)$;
        \item $\{\deg P_i\}_{i\in [g]}\cup \{\mathsf{char}(\K)\}$ is a set of $g+1$ distinct primes.
    \end{itemize}

    Moreover, if $n$ is an integer with $\lceil n/g \rceil \leq |\K|$, and $\beta_1, \dots, \beta_g$ are the roots of $P_1, \dots, P_g(X)$, then if we let $\F = \K(\beta_1, \dots, \beta_g)$, and let $S_1, \dots, S_g$ be arbitrary sets of size $\lceil n/g\rceil$ of $\beta_1+\K, \dots, \beta_g+\K$ respectively, and let $\{\alpha_1, \dots, \alpha_n\}$ be an arbitrary subset of $S_1\cup\cdots\cup S_g$, and let $\sigma_i: x\mapsto x^{|\K|^{\frac{\prod_{j=1}^g \deg P_j}{\deg P_i}}}$, 
    \begin{itemize}
        \item $[\F:\K]\leq (d+g)^{O(g)}$.
        \item $d\leq [\K(\alpha_i):\K]\leq 2[\K(\alpha_j):\K]$ for all $i\ne j$
        \item $\alpha_i$ and $\alpha_j$ are non-conjugate.
        \item $\sigma_1, \dots, \sigma_g$ generate $\Gal(\F/\K)$, and \[|\{i \in [n]: \sigma_j(\alpha_i)\neq \alpha_i\}| \leq \lceil n/g \rceil \quad \forall j \in [g]\]
    \end{itemize}
    In particular, given an index $i\in [n]$, $\alpha_i$ can be outputted in time $\poly(\log |\F|)$.
\end{theorem}
\begin{proof}
    We first find via brute force primes $p_1, \dots, p_g$ such that $\mathsf{char}(\K)\neq p_i$ and $d\leq p_i \leq 2p_j$ for all $i\neq j$. Note that such primes exist by the prime number theorem at scale $d+\poly(g)$. This takes time $\poly(d,g)$.

    Now, we can construct $P_1, \dots, P_g$ with $\deg P_i = p_i$ via \cref{cor:IrreduciblePolynomialConstruction} in time $\poly(\log |\K|, d, g)$.

    Now, it suffices to show that the resulting $\alpha_1, \dots, \alpha_n$ satisfy the requisite properties. Observe that $[\F: \K]=\prod_{j=1}^g \deg P_j\leq \poly(d, g)^{g}\leq (d+g)^{O(g)}$.

    $[\K(\beta_i + c_i):\K] = [\K(\beta_i):\K]= p_i$ for any $c_i \in \K$. Thus, the second result follows. Similarly, $\sigma_i$ fixes $(S_1\cup \dots \cup S_g)\setminus S_i$ and $|S_i|\leq \lceil n/g\rceil$. $\sigma_1, \dots, \sigma_g$ clearly generate $\Gal(\F/\K)$ since $\sigma: x\mapsto x^{|\K|}$ is in the group generated by $\sigma_1, \dots, \sigma_g$ since $\gcd\left(\frac{\prod_{j=1}^g p_j}{p_1}, \dots, \frac{\prod_{j=1}^g p_j}{p_g}\right)=1$.

    Now, it suffices to prove that all pairs $\alpha_i, \alpha_j$ are non-conjugate. Write $\Tr=\operatorname{Tr}_{\F/\K}$. Observe that if $\alpha_i$ and $\alpha_j$ are conjugate then they must have the same extension degree. Thus, let $\alpha_1 = \beta_i+c_1$ and $\alpha_2=\beta_i+c_2$. Observe that if $\alpha_1, \alpha_2$ are conjugate then $\Tr(\alpha_1)=\Tr(\alpha_2)\implies \Tr(\beta_i)+\Tr(c_1)=\Tr(\beta_i)+\Tr(c_2)$. But, $\Tr(c_1)-\Tr(c_2)= (c_1-c_2)\prod_{i=1}^g p_i \neq 0$ since $\mathsf{char}(\K)\nmid p_i$ for any $i\in [g]$. Thus, the result follows. 
\end{proof}

Now, we construct the fields and effective evaluation points to prove the full subspace-design condition.

We remind ourselves that the conditions we want are as in \Cref{def:EffectivePoints}.

\begin{theorem}[Explicit effective points]\label{thm:explicit-effective}
    For any fixed integers $r\ge2, g\ge 2^r, d\ge1$, and integer $n\ge g$, for any finite field $\mathbb K$ with $|\K|>nr$, there exists an explicit extension field $\F/\K$ with $[\F:\K]\le(d+g)^{O(g)}$ and points $\alpha_1, \dots, \alpha_n$ with automorphisms $\sigma_1, \dots, \sigma_g$ such that:
    \begin{itemize}
        \item $d\leq [\K(\alpha_i):\K]\leq 2[\K(\alpha_j):\K]$ for all $i\ne j$
        \item $\alpha_i$ and $\alpha_j$ are non-conjugate.
        \item $\sigma_1, \dots, \sigma_g$ generate $\Gal(\F/\K)$, and \[|\{i \in [n]: \sigma_j(\alpha_i)\neq \alpha_i\}| \leq \lceil n/g \rceil \quad \forall j \in [g]\]
        \item For every non-zero $\K$-subspace $\mathcal Q \subseteq \K[X]_{<nr}$ of dimension at most $2^r$, 
        \[\sum_{i=1}^n \max_{0\neq q \in \mathcal Q}\mult_{\alpha_i}q \leq \max_{0\neq q \in \mathcal Q} \deg q\]
    \end{itemize}
    Moreover, such evaluation points can be found in time $|\K|^{O_{d,g}(1)}$.
\end{theorem}

\begin{proof}
    We begin by showing a broad class of explicit constructions that satisfy the first three properties. Let $d_1, \dots, d_g$ be any set of co-prime integers such that $d\leq d_i \leq 2d_j$ for all pairs $i, j$.

    Let $\beta_i$ be such that $[\K(\beta_i):\K]=d_i$. Let $S_i \subseteq \K\cdot \beta_i\setminus \{0\}$ such that $|S_i|=\lceil n/g\rceil$. This is possible since $|\K|>n$. Additionally, we pick $\beta_i$ such that all points in $S_i$ are non-conjugate. Let the set of evaluation points be any $n$-element subset of $S=S_1 \cup S_2 \cup \dots \cup S_g$.

    Now, let $\F = \K(\beta_1, \dots, \beta_g) = \K^{d_1\dots d_g}$. Additionally, let $\sigma_i$ be the map $x \mapsto x^{|\K|^{d_1\dots d_g/d_i}}$. Observe that $\sigma_i$ fixes all of $S\setminus S_i$. Additionally, these $\sigma_i$ always generate $\Gal(\F/\K)$ and thus, the first three constraints hold naturally.

    Now, we construct a special class of $S_i$ to satisfy the final constraint. In particular, we will construct a $\beta_i$ of degree $d_i$ over $\K$ for each $i\in [g]$ and then show that any resulting $S_i$ will be sufficient. Note, that if $d\ge nr$ then all points naturally satisfy $\mult_{\alpha_i}q=0$ since minimal polynomial of each $\alpha_i$ has degree larger than $\deg q$.\footnote{This was the initial construction provided to us by GPT 5.6 Sol, and we since improved it to use degree $O_{d, g}(1)$ as in this proof.}

    We will choose $d_i = p_iq_i$ for distinct primes $p_1,q_1,\ldots,p_g,q_g\in [P, 1.4P]$ for some $P\ge 2^{r+1}g, d$. Note that this is possible for sufficiently large $P=\poly(d,g)$ due to the prime number theorem.

    Let $\gamma_i$ be an element of degree $p_i$ over $\K$, and let $\delta_i$ be an element of degree $q_i$ over $\K$. We will choose $\delta_i$ carefully so that $\beta_i = \gamma_i \delta_i$ suffices. Additionally, we assume that $\gamma_i^{p_i}, \delta_i^{q_i}\notin \K$. Note that this condition suffices to show that $\beta_i$ and $c\beta_i$ are non-conjugate for any $c\ne 1$ since if $\sigma$ is an automorphism such that $\sigma (\beta_i)=c\beta_i$, then $\sigma^{p_i}(\beta_i)=c^{p_i}\beta_i = \gamma_i\bar{\delta_i}$ for some conjugate $\bar{\delta_i}$ of $\delta_i$. Thus, either $\delta_i$ has a conjugate such $\delta_i/\bar\delta_i \in \K$, but this is not true by the assumption on $\delta_i$, or $\sigma(\delta_i)=\delta_i$. Similarly, $\sigma(\gamma_i)=\gamma_i$. Thus, $c=1$!
    
    Thus, let $\mathcal Q\subseteq \K[X]_{<nr}$ be a $\mathbb K$-subspace of dimension $h\leq 2^r$. Let $q_1, \dots, q_h$ be monic polynomials forming a basis of $\mathcal Q$ with $\deg q_1<\deg q_2 <\dots <\deg q_h$. Now, let \[D_i(Y)=\det(\{q_j(\delta_i^{|\K|^{u-1}}Y)\}_{u, j \in [h]}) \in \K(\delta_i)[Y]\]

    We show that there is a choice of $\delta_i$ such that $D_i(Y)$ is a non-zero polynomial, and in particular $\deg D_i(Y)\leq h\deg q_h$. Choosing such a $\delta_i$ is possible due to the following lemma:

    \begin{lemma}\label{lem:sparse-independent-delta}
        Let $\K$ be a finite field and let $n,r\in\mathbb N$ satisfy $|\K|>nr>2^r$, and let $q_i\ge2\cdot4^r$ be a prime. Then there exists a $\delta_i$ with $[\K(\delta_i):\K]=q_i$ such that for any $1\le h\le2^r$ and $0\leq e_1<\dots < e_h<nr$, we have that $\delta_i^{e_1}, \dots, \delta_i^{e_h}$ are linearly independent over $\K$. 

        Such a $\delta_i$ can in fact be computed in $|\K|^{O(q_i)}$ time.
    \end{lemma}

    We prove this lemma at the end, but conclude using it. Observe that $D_i(Y)$ has leading coefficient $\det(\{\delta_i^{(|\K|^{u-1})\deg q_j}\}_{u, j \in [h]}) = \det(\{(\delta_i^{\deg q_j})^{|\K|^{u-1}}\}_{u, j \in [h]})$ which is non-zero since the $\delta_i^{\deg q_j}$ are linearly-independent over $\K$, and thus, by \cref{lem:GaloisRowsAndMovement}, using the fact that $x\mapsto x^{|\K|}$ generates the automorphism group of $\K(\delta_i)$, we get the resulting determinant is also non-zero.

    Moreover, if $(X-c_i\delta_i\gamma_i)^{t}\mid q$ for some $q\in \mathcal Q$, then we also get that all conjugates of $c_i\delta_i\gamma_i$ are also order $t$ roots of $q$. But, observe that $c_i\delta_i^{|\K|^u}\gamma_i$ are all conjugates of $c_i \delta_i \gamma_i$ since $p_i$ and $q_i$ are co-prime. 

    Let $P_i$ be the minimal polynomial of $\gamma_i$ over $\K$.
    Since $p_i$ and $q_i$ are coprime, $P_i$ remains irreducible
    over $\K(\delta_i)$. If $0\ne q\in\mathcal Q$ has multiplicity
    $t$ at $c\gamma_i\delta_i$, taking $q$ as a basis element of
    $\mathcal Q$ shows that $(Y-c\gamma_i)^t\mid D_i(Y)$.
    Thus $P_i(Y/c)^t\mid D_i(Y)$. For distinct $c\in\K^*$,
    these scaled minimal polynomials are pairwise coprime, since
    the corresponding $c\gamma_i$ are non-conjugate. Consequently,
    \[
      \sum_{i=1}^g\sum_{c\in\K^*}
        \max_{0\ne q\in\mathcal Q}\mult_{c\gamma_i\delta_i}q
      \le \sum_{i=1}^g\frac{\deg D_i}{p_i}
      \le \sum_{i=1}^g\frac{h\deg q_h}{p_i}
      \le \deg q_h.
    \]

    Thus, this construction works! In particular, it suffices to just show \cref{lem:sparse-independent-delta}.
    \end{proof}

    \begin{proof}[Proof of \cref{lem:sparse-independent-delta}]
        Observe that any nonzero polynomial of the form $\sum_{i=1}^{s}\eta_iX^{e_i}$ for $s\leq 2^r$ and distinct $0\le e_i<nr$ has at most $nr<|\K|$ roots. There are at most $\binom{nr}{\leq 2^r}|\K|^{2^r}\leq |\K|^{4^r}$ choices of such polynomials. Thus, there is a $\delta_i \in \K^{q_i}\setminus\K$ which is not a root of any such polynomial. Now, we can simply brute force check over all $\delta_i$ in the field and polynomial pairs resulting in a construction in time $|\K|^{O(q_i)}$ as required.

        For the construction above, we additionally discard $\gamma_i$ with $\gamma_i^{p_i}\in\K$ and $\delta_i$ with $\delta_i^{q_i}\in\K$. These exclude at most $p_i|\K|$ and $q_i|\K|$ elements, respectively. Together with the fewer than $|\K|^{4^r+1}$ roots of nonzero sparse polynomials excluded above, these leave valid choices. Exhaustive search for both elements takes $|\K|^{O(p_i+q_i)}=|\K|^{O_{d,g}(1)}$ time.
    \end{proof}

\section{Improved line decoding and proximity gaps}
\label{sec:ImprovedExtensionFieldConsequences}
The following theorem gives line decoding over $\F$ from a
subspace-design bound over $\K$. We write
$(\delta,a,b)$-$\F$-line-decodable for
$(1,\delta,a,b)$-$\F$-curve-decodable, as in
\Cref{def:CurveDecodability}.

\begin{restatable}[Line decoding over the extension field]{theorem}{extensionfieldlinedecoding}
\label{thm:ExtensionFieldLineDecoding}
Let $\K\subseteq\F$ be finite fields, and let
$C\subseteq\Sigma^n$ be an $\F$-additive code, where $\Sigma$ is a
finite-dimensional $\F$-vector space. Suppose $C$ is a
$\tau$-subspace-design code over $\K$.
For an integer $a\ge1$ and $0<\varepsilon<1$, set
\[
  D_a=\left\lceil\frac{2^{5a}}{\varepsilon}\right\rceil.
\]
If $\tau(D_a)+\varepsilon\le1$, then $C$ is
\[
  \left(1-\tau(D_a)-\varepsilon,\ a,
    \frac{a\varepsilon^2}{4}\right)
  \text{-\(\F\)-line-decodable}.
\]
\end{restatable}
We prove this theorem in \Cref{subsec:ExtensionFieldLineDecodingProof}. We will only apply the theorem with alphabet $\Sigma = \F$, but state it in more generality.

\subsection{Improved combinatorial consequences}
\label{subsec:ImprovedCombinatorialConsequences}

We apply \Cref{thm:ExtensionFieldLineDecoding} to the codes of
\Cref{thm:Main}, using a design order that depends only on the gap
to capacity. All line parameters and probabilities in this subsection
are over $\F$.

\begin{corollary}[Improved capacity line decoding]
\label{cor:ImprovedCapacityLineDecoding}
Fix $0<R<1$ and $0<\varepsilon<1-R$, and set
\[
\begin{gathered}
  \delta=1-R-\varepsilon,\qquad
  L=\left\lceil\frac{8}{\varepsilon}\right\rceil,\qquad
  b_0=1+\left\lceil\frac{2}{\varepsilon}\right\rceil,\\
  a_0=\left\lceil\frac{16b_0}{\varepsilon^2}\right\rceil,
  \qquad
  r=\left\lceil\frac{2^{5a_0+1}}{\varepsilon}\right\rceil.
\end{gathered}
\]
For all sufficiently large $n$ and every finite field $\K$ with
$|\K|\ge n$, let $\F$ and $A$ be supplied by \Cref{thm:Main}
with parameters $r,\varepsilon$, and put
$C=\RS_\F(A,\lfloor Rn\rfloor)$.
Then $C$ is $(1-R-\varepsilon/2,L)$-list-decodable and
$(\delta,a_0,b_0)$-$\F$-line-decodable.
Moreover, for every integer $t\ge1$, it is
\[
  \bigl(\delta,(t-1)L+a_0,t\bigr)
  \text{-\(\F\)-line-decodable}.
\]
\end{corollary}
\begin{proof}
The design bound from \Cref{thm:Main} gives
$\tau(d)\le R+\varepsilon/4$ for $d\le r$.
Apply \Cref{thm:ExtensionFieldLineDecoding} with gap
$\varepsilon/2$ and $a=a_0$. Its required order is exactly $r$,
and
\[
  1-\tau(r)-\varepsilon/2\ge1-R-3\varepsilon/4>\delta,
  \qquad \frac{a_0\varepsilon^2}{16}\ge b_0.
\]
This proves the first line-decoding guarantee.
Since $L\le r$ and
\[
  \frac{L}{L+1}(1-R-\varepsilon/4)>1-R-\varepsilon/2,
\]
\Cref{thm:AverageRadius} gives the asserted list bound, including
the closed radius. Also,
\[
  \delta\left(1+\frac{1}{b_0-1}\right)
  \le\delta+\varepsilon/2=1-R-\varepsilon/2.
\]
For sufficiently large $n$, we have $|\F|\ge|\K|\ge n>(L+1)^2$.
Thus \Cref{item:CurveAmplification} of
\Cref{thm:CurveDecodingConsequences}, with scalar field $\F$ and
$\ell=1$, gives the amplified line-decoding guarantee.
\end{proof}

\begin{corollary}[Improved proximity gaps and correlated agreement]
\label{cor:ImprovedExtensionFieldAgreement}
Let $C$, $\F$, $\delta$, $L$, and $a_0$ be as in
\Cref{cor:ImprovedCapacityLineDecoding}.
Then, over $\F$, the code $C$ has:
\begin{enumerate}[label=(\roman*),leftmargin=2em]
  \item For every real $m>1$,
    \[
      \left(1,\delta,\frac{mL+a_0}{|\F|},\frac1m\right)
    \]
    correlated agreement and proximity gap.
  \item
    \[
      \left(1,\delta,\frac{nL+a_0}{|\F|}\right)
    \]
    mutual correlated agreement, and
    \[
      \left(1,\delta,\frac{nL+a_0}{|\F|},0\right)
    \]
    correlated agreement and proximity gap.
\end{enumerate}
\end{corollary}
\begin{proof}
The proof of \Cref{cor:ImprovedCapacityLineDecoding} verifies the
hypotheses of \Cref{thm:CurveDecodingConsequences} with scalar field
$\F$, $\ell=1$, $a=a_0$, and $b=b_0$.
Its \Cref{item:CorrelatedAgreement,item:MutualCorrelatedAgreement}
give the stated bounds. Mutual correlated agreement implies
correlated agreement and proximity gap with $\gamma=0$,
as recorded in \Cref{subsec:PrelimLines}.
\end{proof}

The parameters above satisfy
\[
  L,b_0=O(\varepsilon^{-1}),\qquad
  a_0=O(\varepsilon^{-3}),\qquad
  r=2^{O(\varepsilon^{-3})}.
\]
In particular, the design order and the required extension degree
depend only on $\varepsilon$.

The probability thresholds in the two conclusions are, respectively,
\[
  O\!\left(\frac{m/\varepsilon+\varepsilon^{-3}}{|\F|}\right)
  \qquad\text{and}\qquad
  O\!\left(\frac{n/\varepsilon+\varepsilon^{-3}}{|\F|}\right).
\]
Both estimates have denominator $|\F|$, with no factor
$|\F|/|\K|$ in the line-decoding threshold $a_0$.

\subsection{Proof of the extension-field line-decoding theorem}
\label{subsec:ExtensionFieldLineDecodingProof}

\begin{lemma}[Slow growth of monomial spaces]
\label{lem:MonomialSpaceGrowth}
Let $\K\subseteq\F$ be fields, and let $A\subseteq\F$ have
$a\ge2$ elements, including $0$ and $1$. For $j\ge0$, let
\[
  V_j=\Span_\K\{\alpha_1\cdots\alpha_j:
    \alpha_1,\ldots,\alpha_j\in A\},\qquad V_0=\K.
\]
There is an integer $j\ge0$ such that
\[
  \dim_\K V_{j+2}\le2\dim_\K V_j,
  \qquad \dim_\K V_{j+2}\le32^{a-2}.
\]
\end{lemma}
\begin{proof}
Since $1\in A$, the spaces $V_j$ are nested. Put $m=a-2$.
If $m=0$, take $j=0$. Otherwise, monomials in the $m$ elements of
$A\setminus\{0,1\}$ of total degree at most $h$ span $V_h$, so
\[
  \dim_\K V_h\le\binom{h+m}{m},\qquad
  \dim_\K V_{11m}\le\binom{12m}{m}
  \le\left(\frac{12^{12}}{11^{11}}\right)^m<32^m.
\]
If $\dim_\K V_{j+2}>2\dim_\K V_j$ for every
$j\in\{0,2,\ldots,10m-2\}$, then
$\dim_\K V_{10m}>2^{5m}=32^m$, a contradiction. Any index where
this fails gives both conclusions.
\end{proof}

We use a dimension-threshold version of the weighted pruning
method of~\cite{AshvinkumarHS2026,GoyalG2025-Structure}.

\begin{lemma}[Pruning to a dimension threshold]
\label{lem:DimensionThresholdPruning}
Let $\K$ be a finite field and $C\subseteq\Sigma^n$ a $\K$-additive
$\tau$-subspace-design code over $\K$. Let $D,d\ge1$ be integers,
and let $W\subseteq C$ be a $\K$-subspace of dimension $M\le D$.
Fix $\varepsilon>0$, and set
$\delta=1-\tau(D)-\varepsilon$.
There is a randomized procedure producing $S\subseteq[n]$ such that
\[
  \dim_\K W_S<d,\qquad
  W_S=\{w\in W:w|_S=0\},
\]
and, for every $T\subseteq[n]$ with
$|T|\ge(\tau(D)+\varepsilon)n$,
\[
  \Pr[S\subseteq T]\ge
  \frac{\varepsilon d}{\delta M+\varepsilon d}.
\]
\end{lemma}

\begin{quote}
\textbf{Pruning to a dimension threshold $\mathsf{PruneThreshold}$}
\begin{algorithmic}[1]
\Require $W,d,\delta,\varepsilon$ as in \Cref{lem:DimensionThresholdPruning}.
\State $U\gets W$; $S\gets\varnothing$.
\While{$\dim_\K U\ge d$}
  \State Compute $U_i=\{u\in U:u_i=0\}$ for each $i\in[n]$.
  \State $I\gets\{i\in[n]:U_i\subsetneq U\}$.
  \State Sample $i\in I$ with probability
    $\displaystyle\frac{\delta\dim_\K U_i+\varepsilon d}
    {\sum_{h\in I}(\delta\dim_\K U_h+\varepsilon d)}$.
  \State $S\gets S\cup\{i\}$; $U\gets U_i$.
\EndWhile
\State \Return $S$.
\end{algorithmic}
\end{quote}

\begin{proof}
The invariant is $U=W_S$. At each iteration, the design bound and
$\tau(D)<1$ ensure that $I\ne\emptyset$, and the dimension
strictly decreases. Thus the procedure terminates with
$\dim_\K W_S<d$.

Fix $T$ as in the statement. We prove the probability bound by
induction on $M$. If $M<d$, the output is empty and the bound holds.
Otherwise, write $W_i=\{w\in W:w_i=0\}$ and let $h$ be the
fraction of coordinates with $W_i=W$. The design bound gives
$h\le\tau(D)$ and bounds the total sampling weight by
\[
  Z:=\sum_{i:W_i\subsetneq W}
       (\delta\dim_\K W_i+\varepsilon d)
  \le n\bigl(\delta M(\tau(D)-h)+\varepsilon d(1-h)\bigr).
\]
For each sampled $i\in T$, induction bounds the probability that
all subsequent samples also lie in $T$ by
$\varepsilon d/(\delta\dim_\K W_i+\varepsilon d)$.
These denominators cancel the sampling weights, giving
\begin{align*}
  \Pr[S\subseteq T]
  &\ge\frac{\varepsilon d\,n(\tau(D)+\varepsilon-h)}{Z}\\
  &\ge\frac{\varepsilon d}{\delta M+\varepsilon d}.
\end{align*}
For the last inequality, the difference between
$(\tau(D)+\varepsilon-h)(\delta M+\varepsilon d)$ and
$\delta M(\tau(D)-h)+\varepsilon d(1-h)$ is
$\varepsilon\delta(M-d)\ge0$.
\end{proof}

\begin{proof}[Proof of \Cref{thm:ExtensionFieldLineDecoding}]
Write $\delta=1-\tau(D_a)-\varepsilon$.
Fix a received line $u_0+\alpha u_1$ and a set $A\subseteq\F$
of $a$ parameters, with assigned codewords $c_\alpha\in C$
satisfying $\Delta(u_0+\alpha u_1,c_\alpha)\le\delta$.
The case $a=1$ follows by taking a constant codeword line.
For $a\ge2$, an invertible affine reparameterization lets us assume
$0,1\in A$. Such a reparameterization preserves codeword lines
because $C$ is $\F$-additive.

Choose $j$ and the spaces $V_j,V_{j+1},V_{j+2}$ from
\Cref{lem:MonomialSpaceGrowth}, and put
\[
  d=\dim_\K V_j,\qquad
  W=\Span_\K\{v c_\alpha:v\in V_{j+1},\ \alpha\in A\}
  \subseteq C,\qquad M=\dim_\K W.
\]
We first show that
\begin{equation}
\label{eq:ScaledCodewordDimension}
  M<D_a,\qquad
  \varepsilon M\le\dim_\K V_{j+1}+\dim_\K V_{j+2}\le4d.
\end{equation}
To apply the design hypothesis without assuming $M\le D_a$, choose
$m=\min\{M,D_a\}$ independent elements $v_\ell c_{\alpha_\ell}$
from the generating set of $W$, and let $U$ be their $\K$-span.
At coordinate $i$, let $b_i$ count the selected generators whose
assigned codeword $c_{\alpha_\ell}$ disagrees with the received
line there. Then $\sum_i b_i\le\delta nm$. Every other generator
has its $i$th entry in
\[
  V_{j+1}u_{0,i}+V_{j+2}u_{1,i},
\]
since $\alpha_\ell V_{j+1}\subseteq V_{j+2}$.
Writing $U_i=\{u\in U:u_i=0\}$, we obtain
\[
  m-\dim_\K U_i
  \le\dim_\K V_{j+1}+\dim_\K V_{j+2}+b_i.
\]
Summing and using the design bound on $U$ yields
\[
  \varepsilon m
  \le\dim_\K V_{j+1}+\dim_\K V_{j+2}
  \le2\cdot32^{a-2}<\varepsilon D_a.
\]
Thus $m<D_a$, so $m=M$. The slow-growth of dimensions from  
\Cref{lem:MonomialSpaceGrowth} now gives
\Cref{eq:ScaledCodewordDimension}.

For each $\alpha\in A$, let
$T_\alpha=\{i:c_{\alpha,i}=u_{0,i}+\alpha u_{1,i}\}$.
These sets have size at least $(\tau(D_a)+\varepsilon)n$.
Apply \Cref{lem:DimensionThresholdPruning} to $W$ with $D=D_a$
and threshold $d$. By \Cref{eq:ScaledCodewordDimension},
\[
  \Pr[S\subseteq T_\alpha]
  \ge\frac{\varepsilon d}{\delta M+\varepsilon d}
  \ge\frac{\varepsilon^2}{4\delta+\varepsilon^2}
  \ge\frac{\varepsilon^2}{4},
\]
where the last inequality uses $\delta\le1-\varepsilon$.
By linearity of expectation, some output $S$ satisfies
\[
  \dim_\K W_S<d,\qquad
  |B|\ge\frac{a\varepsilon^2}{4},\qquad
  B=\{\alpha\in A:S\subseteq T_\alpha\}.
\]

If $|B|=1$, a constant codeword line suffices. Otherwise, fix distinct
$\beta,\gamma\in B$. For every $\alpha\in B$, set
\[
  z_\alpha=(\gamma-\beta)c_\alpha
    -(\gamma-\alpha)c_\beta-(\alpha-\beta)c_\gamma.
\]
The three codewords agree with the received line on $S$, so
$z_\alpha|_S=0$. Also, $V_jz_\alpha\subseteq W$, since
$V_j(A-A)\subseteq V_{j+1}$. If $z_\alpha\ne0$, multiplication
by $z_\alpha$ is injective on $V_j$, giving a $d$-dimensional
$\K$-subspace $V_jz_\alpha\subseteq W_S$, a contradiction.
Hence every $z_\alpha=0$, and
\[
  c_\alpha=c_\beta+
    \frac{\alpha-\beta}{\gamma-\beta}(c_\gamma-c_\beta)
  \qquad(\alpha\in B).
\]
This is a line with both coefficients in $C$, proving the required
$\F$-line-decoding guarantee.
\end{proof}

\section{Acknowledgements}

We would like to thank Josh Brakensiek, Yeyuan Chen, Jun-Ting (Tim) Hsieh and Zihan Zhang for various helpful discussions on the problem of constructing explicit list-decodable Reed-Solomon codes and extension-field subspace-designs.

R.G. would also like to thank Prahladh Harsha, Mrinal Kumar, Yihang (Kimi) Sun, Mary Wootters, and Madhur Tulsiani for various discussions and comments on extension field subspace-designs and LCL properties. R.G. thanks Shivam Nadimpalli for raising the question about whether our constructions were strongly explicit. 

Part of this work was done while R.G. was visiting the Simons Institute for the Theory of Computing, and he thanks the institute for its hospitality.

\addcontentsline{toc}{section}{References}
\begingroup
\bibliographystyle{alpha}
\bibliography{coding-theory}
\endgroup

\appendix
\section{Computing a common zero}
\label{app:computing-common-zero}

\computingcommonzero*

We reduce the proof to the following factorization lemma.

\begin{lemma}[Factorization over finite extensions]
\label[lemma]{lem:FunctionFieldFactorization}
Let $\K$ be a finite field and $\mathbb L=\K(X)$. Let
$E/\mathbb L$ be an extension of degree $m$ with a specified
basis $1=e_1,\ldots,e_m$. Suppose we are given its
multiplication table, consisting of rational functions
$c_{ij\ell}(X)\in\mathbb L$ such that
\[
  e_i e_j=\sum_{\ell=1}^m c_{ij\ell}(X)e_\ell.
\]
We are also given a nonzero polynomial $g\in E[T]$ of
degree at most $d$, with each coefficient expressed in this
basis. Suppose all rational functions in these expressions
and in the multiplication table have numerators and
denominators of degree at most $h$.

Then one can deterministically compute the factorization
$g(T)=c\prod_{\nu=1}^r g_\nu(T)^{\mu_\nu}$,
where $c\in E$, the $g_\nu$ are distinct monic irreducible
polynomials over $E$, and the $\mu_\nu$ are their
multiplicities, in time $O_{m,d}(1)\poly(h,\log|\K|)$.
The coefficients of each $g_\nu$ are returned in the given
basis, with numerator and denominator degrees of their
coordinates at most $O_{m,d}(1)h$.
\end{lemma}

We use the following two results.

\begin{theorem}[{von zur Gathen--Kaltofen~\cite[Theorem~4.2(i)]{GathenK1985}}]
\label[theorem]{thm:BivariateFactorization}
Let $\K$ be a finite field. Factoring a polynomial in
$\K[X,T]$ of total degree at most $D$ reduces deterministically
to factoring univariate polynomials of degree at most $D$
over finite extensions of $\K$ of degree $\poly(D)$.
The reduction takes $\poly(D,\log|\K|)$ time.
\end{theorem}

Clearing denominators and applying
\Cref{cor:FiniteFieldFactorization}, we can therefore factor
a degree-$d$ polynomial over $\K(X)$ whose coefficient
numerators and denominators have degree at most $h$
deterministically in time $\poly(d,h,\log|\K|)$.

\begin{theorem}[{Consequence of Jambor~\cite[Theorem~6.6 and Proposition~4.10]{Jambor2014}}]
\label[theorem]{thm:AlgebraDecomposition}
Let $\K$ be a finite field and $\mathbb L=\K(X)$.
Let $E/\mathbb L$ have degree $m$ and be given by a basis
and multiplication table, and let $g\in E[T]$ be nonzero
of degree at most $d$. Suppose the rational functions in
these data have numerator and denominator degrees at most $h$.

Computing the monic irreducible factors of $g$ and their
multiplicities reduces deterministically to $\poly(m,d)$
factorizations of polynomials in $\K[X,T]$ with
\[
  \deg_X\le O_{m,d}(1)h,\qquad \deg_T\le md.
\]
The reduction takes $O_{m,d}(1)\poly(h,\log|\K|)$ time
outside these calls. The returned factor coefficients have
coordinates with numerator and denominator degrees at most
$O_{m,d}(1)h$ in the given basis.
\end{theorem}

\begin{proof}
Apply Jambor's algorithms to $E[T]/(g)$, whose dimension
over $\mathbb L$ is at most $md$. In each resulting residue
field, compute the minimal polynomial of the image of $T$
over $E$. These are the irreducible factors of $g$, and
repeated division gives their multiplicities. The algorithms
use $\poly(m,d)$ operations, including factorizations of
polynomials over $\mathbb L$ of degree at most $md$ and
$p$-th root extractions, where $p=\operatorname{char}(\K)$. Clearing
denominators turns the former into bivariate factorizations.

If $p\le md$, Jambor's coefficientwise root
computation~\cite[Section~4.2]{Jambor2014} takes time polynomial
in $m,d$, the current coefficient degrees, and $\log|\K|$,
using parameter extensions of degree at most $md$.
If $p>md$, no root extractions are needed, since the relevant
polynomials have degree at most $md$.

For coprime $A,B\in\K[U]$, write
$\operatorname{ht}(A/B)=\max(\deg A,\deg B)$.
An arithmetic operation increases the current height bound
by at most a factor of two. Clearing denominators of a
polynomial of degree at most $md$ increases the parameter
degree by at most a factor of $md+1$, and its bivariate
factors have no larger parameter degree. The root computations
and changes of coordinates between the parameter fields also
increase heights by at most $\poly(m,d)$.
Since there are $\poly(m,d)$ operations and factorization
calls, all intermediate and output heights are $O_{m,d}(1)h$.
This gives the stated degrees for the bivariate polynomials,
and the remaining computations take
$O_{m,d}(1)\poly(h,\log|\K|)$ time.
\end{proof}

\begin{proof}[Proof of \Cref{lem:FunctionFieldFactorization}]
Apply \Cref{thm:AlgebraDecomposition,thm:BivariateFactorization}
to reduce to univariate factorization over finite fields.
Write $p=\operatorname{char}(\K)$. If $p\le md$,
Berlekamp's algorithm~\cite{Berlekamp1970} already gives
the required bound: its running time is polynomial in $p$,
the polynomial degree, and the logarithm of the field size.
If $p>md$, use \Cref{cor:FiniteFieldFactorization}.
In either case, each call takes
$O_{m,d}(1)\poly(h,\log|\K|)$ time, giving the claimed
runtime and coefficient-degree bound.
\end{proof}

\begin{proof}[Proof of \Cref{lem:computing-common-zero}]
Put
\[
  Q=2(3D)^a,\qquad
  b=\binom{a+Q+D}{a},\qquad B=Db.
\]
By \Cref{thm:EffectiveNullstellensatz}, a system of polynomials
of degree at most $D$ in at most $a$ variables has no common
zero over the algebraic closure of its coefficient field
exactly when
\[
  \sum_{j=1}^s G_jf_j=1,\qquad \deg G_j\le Q
\]
has a solution. Comparing coefficients gives a linear system
with at most $b$ rows and $sb$ columns. Over $\mathbb L$, this
decides whether the original system has a common zero in time
$\poly(s,b,H,\log|\K|)$. Henceforth assume it does.

Suppose we have chosen $\alpha_1,\ldots,\alpha_{i-1}$ such
that the remaining equations have a common zero over
$\overline E$, where
$E=\mathbb L(\alpha_1,\ldots,\alpha_{i-1})$.
After substituting these values, regard the $f_j$ as
polynomials in $T=Y_i$ and $Z=(Y_{i+1},\ldots,Y_a)$ over $E$.
The identity
\[
  \sum_{j=1}^s G_j(T,Z)f_j(T,Z)=1,
  \qquad \deg_Z G_j\le Q,
\]
gives a linear system $M(T)c=e$ over $E(T)$, where $e$ is
the coefficient vector of $1$. The matrix has at most $b$
rows and $sb$ columns, and its entries lie in $E[T]$ with
degree at most $D$. For every algebraic value $\alpha$ of
$T$, the same degree bound shows that $M(\alpha)c=e$ is
solvable exactly when the equations $f_j(\alpha,Z)=0$ have
no common zero.

If $M(T)c=e$ is solvable, Cramer's rule gives a common
nonzero denominator $h\in E[T]$ of degree at most $B$.
Clearing denominators gives
\[
  h(T)=\sum_{j=1}^s \widetilde G_j(T,Z)f_j(T,Z),
  \qquad \widetilde G_j\in E[T,Z].
\]
Thus the $T$-coordinate of every common zero is a root of
$h$. Factor $h$ using \Cref{lem:FunctionFieldFactorization}.
For each irreducible factor $p$, put $E'=E[T]/(p)$ and
$\alpha=T\bmod p$, and use the same linear-system test to
check whether $f_j(\alpha,Z)=0$ has a common zero over
$\overline{E'}$. At least one factor passes this test, since
the current system has a common zero. Choose such a factor
and set $\alpha_i=\alpha$. This gives
$[E':E]=\deg p\le B$.

Otherwise, put $r=\operatorname{rank}_{E(T)}M(T)$. There is
a nonzero $(r+1)\times(r+1)$ minor $h(T)$ of
$[M(T)\mid e]$, again of degree at most $B$.
Whenever $h(\alpha)\ne0$, the augmented matrix has rank at
least $r+1$, whereas $M(\alpha)$ has rank at most $r$.
Thus $M(\alpha)c=e$ remains unsolvable, and the polynomial
equations have a common zero. Choose $\alpha_i$ from
$1,X,\ldots,X^B$ with $h(\alpha_i)\ne0$; these are $B+1$
distinct elements of $\mathbb L$. In this case the
coefficient field does not change.

Repeating this procedure gives a common zero
$(\alpha_1,\ldots,\alpha_a)$ in an extension of degree at
most $B^a$. The defining polynomials and the power bases of
the successive extensions give the required representation
and $\mathbb L$-basis.

It remains to bound the running time using
\Cref{lem:FunctionFieldFactorization}. At each step there
is at most one factorization of a polynomial of degree at
most $B$, followed by at most $B$ linear-system tests.
Let $h_i$ bound the numerator and denominator degrees of
the coordinates of the current coefficients, the chosen
coordinates, and the entries of the multiplication table;
initially $h_0=H$.
All matrix dimensions and extension degrees are
$O_{s,D,a}(1)$. Cramer's rule and
\Cref{lem:FunctionFieldFactorization} bound the coefficient
degrees after the linear-system computations and factorization
by $O_{s,D,a}(1)h_i$. Reducing powers modulo the chosen
factor constructs the new multiplication table and substitutes
$\alpha_i$ into the remaining equations, with the same bound.
Choosing $\alpha_i$ from $1,X,\ldots,X^B$ contributes at
most $DB$ to these degrees. Consequently, all coefficient
degrees are $O_{s,D,a}(1)\poly(H)$.
Since $B$ depends only on $D,a$, each factorization takes
$O_{D,a}(1)\poly(h_i,\log|\K|)$ time. There are $a$ steps,
so the total running time is
$O_{s,D,a}(1)\poly(H,\log|\K|)$, as required.
\end{proof}

\end{document}